\documentclass{aa}

\usepackage{twoopt,natbib}
\usepackage{graphicx}
\usepackage{xcolor}
\usepackage{txfonts}
\usepackage[breaklinks=true,colorlinks=true,linkcolor=blue,citecolor=blue,filecolor=blue,urlcolor=blue]{hyperref}
\usepackage{multirow}
\usepackage{amssymb}	
\usepackage{textcomp}
\usepackage{amsmath}
\usepackage[normalem]{ulem}
\usepackage{physics}
\usepackage{subcaption}
\usepackage{soul}

\newcommand{\code}[1]{\texttt{#1}}

\newcommand{\mesa}{\code{MESA}}
\newcommand{\cmfgen}{\code{CMFGEN}}

\newcommand{\arepo}{\code{AREPO}}
\usepackage{placeins}

\def\mej{$M_{\rm ej}$}

\def\kms{\mbox{km~s$^{-1}$}}
\def\lsun{\,L$_{\odot}$}
\def\ergs{\,erg s$^{-1}$}

\def\one{{\,\sc i}}
\def\two{{\,\sc ii}}
\def\three{{\,\sc iii}}
\def\four{{\,\sc iv}}
\def\five{{\sc v}}

\def\nad{Na\one\,$\lambda\lambda\,5896,5890$}
\def\heiiuv{He\two\,$\lambda$\,$1640$}
\def\ciiiuv{C\three\,$\lambda$\,$2297$}
\def\civuv{C\four\,$\lambda\lambda$\,$1548,\,1551$}
\def\mgiidoub{Mg\two\,$\lambda\lambda$\,$2795,\,2802$}

\newcommand{\Msol}{\,\mathrm{M}_\odot}
\newcommand{\Rsol}{\,\mathrm{R}_\odot}

\begin{document}

   \title{Light curves and spectra for stellar collisions between main-sequence stars in galactic nuclei}
   \titlerunning{High-velocity stellar collisions}

   \author{
   Taeho Ryu,\inst{\ref{inst1},\ref{inst2},\ref{inst3}}
   Luc Dessart\inst{\ref{inst4}}
	        }

\institute{Max-Planck-Institut für Astrophysik, Karl-Schwarzschild-Straße 1, 85748 Garching bei München, Germany\label{inst1}
\and
JILA, University of Colorado and National Institute of Standards and Technology, 440 UCB, Boulder, 80308 CO, USA\label{inst2}
\and
Department of Astrophysical and Planetary Sciences, 391 UCB, Boulder, 80309 CO, USA\label{inst3}
\and
Institut d'Astrophysique de Paris, CNRS-Sorbonne Universit\'e, 98 bis boulevard Arago, F-75014 Paris, France\label{inst4}
}

   \date{Received XXX; accepted YYY}

  \abstract{
    High-velocity stellar collisions in galactic nuclei produce ejecta that generate potentially observable electromagnetic radiation, making them promising nuclear transients. However, the photometric and spectroscopic properties of these collisions, which would more frequently involve main-sequence stars, remain largely unexplored. Here, using 3D hydrodynamics and 1D radiative-transfer simulations, we investigate the properties and observables of the debris produced in high-velocity collisions between terminal-age main-sequence stars, covering a wide range of collision configurations. The ejecta produce bright ultraviolet (UV) transients with bolometric luminosities typically peaking at $\gtrsim10^{43}$ erg s$^{-1}$, declining steeply as $t^{-2}-t^{-4}$ to reach $\gtrsim10^{41}-10^{42}$ erg s$^{-1}$ at 0.5\,d, and leveling off on a plateau at $10^{39}-10^{41.5}$ erg s$^{-1}$ ($M_V$ between $-$10 to $-$15\,mag) after a few days. The total radiated energy is less than $10^{49}$ ergs, which corresponds to $10^{-3}-10^{-5}$ of the initial collision kinetic energy. Their spectra evolve considerably during the first few days, morphing from UV- to optical-dominated. The UV range shows numerous resonance transitions from metals like C, N, and O, whereas the optical primarily shows H{\,\sc i}\ Balmer lines. These properties are qualitatively similar to those observed, as well as obtained in models of Type II supernovae.  Observables from these events exhibit clear correlations with collision configurations, including impact parameter, relative velocity, and stellar masses. We provide fitting formulae to describe these correlations. Detecting these transients requires sub-day cadence surveys such as ULTRASAT, combined with spectroscopic observations to disentangle degeneracies and infer collision characteristics.}


   \maketitle
%

\begin{figure}
    \centering
    \includegraphics[width=0.98\linewidth]{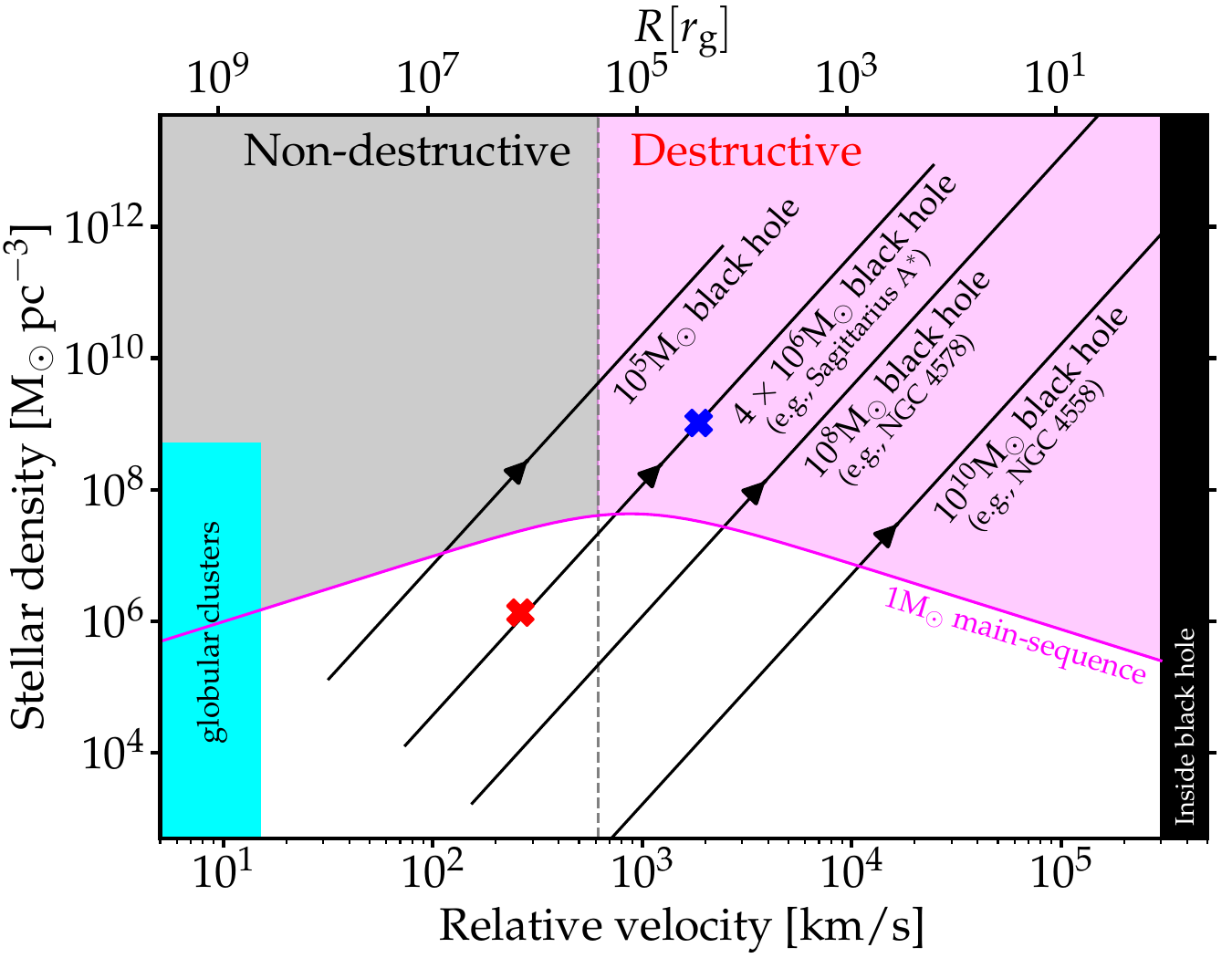}\vspace{-0.1in}
    \caption{Ranges of the relative velocity between stars and stellar density for destructive (magenta-shaded) and non-destructive (grey-shaded) collisions, separated by the surface escape velocity of a $1\Msol$ main-sequence star. The velocity  $v$ and distance $R$ from the black hole are related by $v/c = \sqrt{r_{\rm g}/R}$, with the black-shaded region ("Inside black hole") indicating $v/c > 1$. Here, $r_{\rm g}$ is the gravitational radius. Black diagonal lines indicate nuclear clusters with the same mass as the black hole, assuming a \citet{BahcallWolf1976} profile extending inwards (arrow) from the influence radius until $2\Msol$ is enclosed or the event horizon is reached.}
    \label{fig:parameterrange}
\end{figure}
\section{Introduction} \label{sec:introduction}
Stellar collisions can occur in dense stellar environments such as globular clusters or nuclear stellar clusters, as shown in Fig.~\ref{fig:parameterrange} depicting the parameter ranges of the relative velocity between stars and stellar density for collisions\footnote{This figure resembles Figure 1 in \citet{FreitagBenz2005}, but the primary difference lies in the interpretation of the velocity axis -- $y-$axis in their figure and $x-$axis in ours. In both figures, the velocity is estimated under the same assumption of a Keplerian orbit around the central supermassive black hole. However, we use velocity as a proxy for distance by fixing the black hole mass whereas they use it as a proxy for black hole mass by fixing the distance. As a result, each diagonal line in our figure represents the relation between number density and distance from the galactic center for a given black hole mass, while any curves in theirs show a relation between the black hole mass and number density at a fixed location.}. As indicated by the magenta curve representing the critical density at which the collision timescale equals the lifetime of $1\Msol$ main-sequence stars\footnote{ The critical density is estimated as $n_{\rm cri}\simeq (\Sigma v t_{\rm MS})^{-1}$, where $\Sigma$ is the collisions cross-section accounting for gravitational focusing ($\propto R_{\rm p}/v^{2}$ for $v\ll v_{\rm esc}$ and $\propto R_{\rm p}^{2}$ for $v\gg v_{\rm esc}$), $R_{\rm p}$ is the closest-approach distance yielding a collision, approximately equal to the sum of the radii of the two colliding stars, $t_{\rm MS}$ is the main-sequence lifetime, and $v_{\rm esc}$ is the surface scape velocity of the colliding stars. Thus, $n_{\rm cri}\propto v$ for $v\ll v_{\rm esc}$ and $\propto v^{-1}$ for $v\gg v_{\rm esc}$. The exact value of the critical density depends on the type and mass of the colliding stars. }, stars can undergo multiple collisions in both globular and nuclear stellar clusters, including the Galactic nuclear cluster corresponding to the line for $4\times10^{6}\Msol$ black hole, with the red and blue crosses indicating the observationally constrained density at $0.25$ pc from the Galactic center \citep{Genzel+2010} and the semimajor axis of the S2 star \citep{Gillessen+2009}, respectively.

Collisions can be divided into two regimes based on the collision velocity relative to the escape velocity of two colliding stars (grey vertical line in Fig.~\ref{fig:parameterrange})\footnote{In this figure, the stellar escape velocity for $1\Msol$ main-sequence star is used as a representative value to demarcate the destructive regime, where collision kinetic energy is dominant over stellar binding energy, and the non-destructive regime, where stellar binding energy exceeds collision kinetic energy. While the escape velocity for main-sequence stars itself weakly depends on stellar mass, the exact boundary between the two regimes depend on several factors like impact parameters and stellar structure. }. When the relative velocity between two colliding stars is low enough that the total energy is negative (grey-shaded region), the two stars would merge into a more massive star, which may appear as a blue straggler \citep[e.g.,][]{Sandage1953,BurbidgeSandage1958,Sills+1995,WangRyu2024, Ryu+2024b}. These low-velocity collisions typically occur in globular clusters (cyan region in the lower-left corner, \citealt{BaumgardtHilker2018}), where the stellar velocity dispersion is $\lesssim15$ km s$^{-1}$ \citep{Cohen1983,PryorMeylan1993,Harris+2010}. In contrast, in nuclear stellar clusters surrounding a supermassive black hole (diagonal black lines), the relative velocities between stars near the black hole can reach $\gtrsim O(10^{3})$ km s$^{-1}$. When two stars collide at such high velocities, exceeding their stellar binding energy, they can be fully or partially destroyed (magenta-shaed region) \citep[e.g.,][]{FreitagBenz2005}. This process, which occurs at a rate of $10^{-3}-10^{-7}$ yr$^{-1}$ galaxy$^{-1}$ \citep{Rose+2020,AmaroSeoane2023, Rose+2023, BalbergYassur2023,HuLoeb2024}, can produce bright electromagnetic transients \citep{Balberg+2013, BalbergYassur2023, AmaroSeoane2023, AmaroSeoane2023b,Ryu+2024, Dessart+2024}, making them promising candidates for nuclear transients for current and future surveys.

What do their light curves and spectra look like? This is one of the central questions to be addressed in identifying transients caused by destructive stellar collisions. Previous analytic studies \citep[e.g.,][]{Balberg+2013,AmaroSeoane2023} estimated transient luminosities and durations, assuming photons diffuse out in spherically expanding ejecta. More specifically, \citet{Balberg+2013} made an analytic estimate of the luminosity for the first time, considering a scenario where the emitted energy from the region where the diffusion time equals the expansion time contributes to the luminosity, as done in supernova light curve analysis. On the other hand, \citet{AmaroSeoane2023} attempted to refine the estimate by introducing a factor for the conversion efficiency from kinetic energy to radiation energy and by speculating on the time evolution of the diffusion time. While these estimates already suggested potentially luminous transients, they remained approximate. Recently, \citet{Dessart+2024} performed detailed 1D radiative-transfer calculations of light curves and spectra for high-velocity collisions between giant stars -- these calculations were themselves based on the results from hydrodynamics simulations performed by \citet{Ryu+2024}. The radiative-transfer simulations suggest that the luminosities of giant collisions are on the order of $10^{42}-10^{44}$ erg s$^{-1}$ at 1\,d after the onset of the collision, which is in the ballpark of tidal disruption events \citep[e.g.,][]{Gezari+2021,Wevers+2023} or supernovae. In addition, the collision debris reveal a similar spectral evolution as that of Type II supernovae from blue-supergiant star explosions \citep{Dessart+2024}. 

This paper investigates the observables of high-velocity collisions between realistic main-sequence stars at their terminal age, exploring a wide range of collision configurations, including stellar masses ($M_{1}=1,~3,~10\Msol$ and $M_{2}=0.9M_{1}$), mass ratios ($q=M_{2}/M_{1}\simeq 0.1 - 0.9$), relative collisional velocities ($2\,500-10\,000$ km s$^{-1}$), and impact parameters ranging from nearly head-on to off-axis collisions. Employing the same methodology as in \citet{Dessart+2024}, we first performed hydrodynamics simulations of collisions using the moving-mesh hydrodynamics simulation code {\small AREPO} \citep{Arepo,Arepo2,ArepoHydro}. In most cases, these eventually produce quasi-spherically expanding collision debris. We then conducted 1D, nonlocal thermodynamic equilibrium, time-dependent radiative-transfer calculations with the code \cmfgen\ \citep{HD12} using 1D, angle-averaged profiles built from the 3D hydrodynamics simulations at 0.5\,d.

\begin{figure*}
    \centering
    \includegraphics[width=0.99\linewidth]{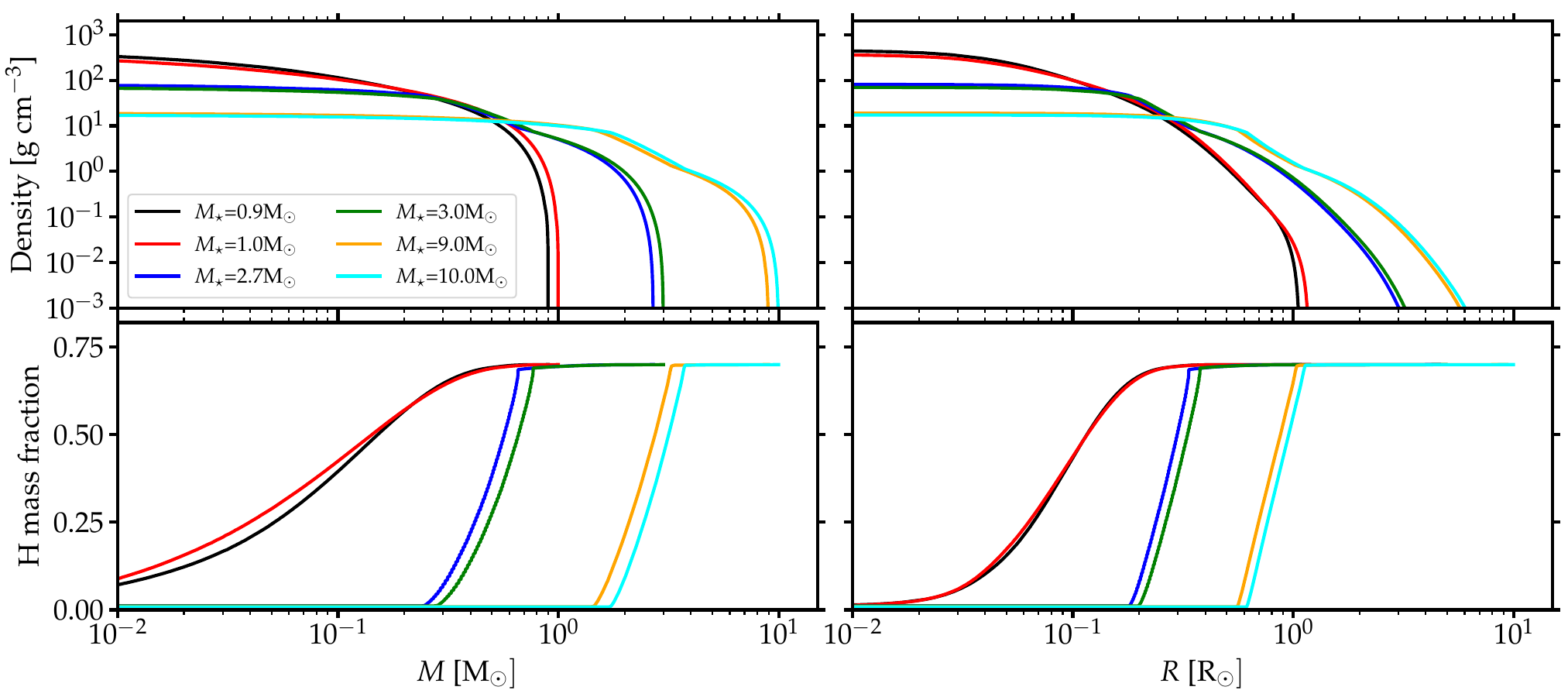}
    \caption{Internal structure of stellar models. Density (top) and H mass fraction (bottom) profiles of our main-sequence stellar models with a metallicity of $Z=0.02$ and at the core H mass fraction of 0.01, as a function of enclosed mass (left) and radius (right). }
    \label{fig:initialprofile}
\end{figure*}

This paper is organized as follows. In \S~\ref{sec:hydro}, we describe our hydrodynamics simulations (\S~\ref{sec:hydro_setup}), followed by a dynamical overview of stellar collisions (\S~\ref{subsec:overview}). In \S~\ref{sec:RT}, we provide a description for the setup of our radiative transfer calculations (\S~\ref{subsec:RT_setup}) and an overview of the photometric and spectroscopic properties of representative models (\S~\ref{subsec:RT_overview}). Using the full set of nearly forty simulations, we discuss correlations between collision parameters and the modeled observables in \S~\ref{sec:correlation}. We present our conclusions in \S~\ref{sec:conclusion}.

\section{Hydrodynamics simulations}\label{sec:hydro}

\subsection{Numerical setup}\label{sec:hydro_setup}
Following \citet{Ryu+2024}, we performed a suite of 3D hydrodynamics simulations of high-velocity collisions between main-sequence stars using the moving-mesh magnetohydrodynamics code \arepo\ \citep{Arepo,ArepoHydro,Arepo2}. We employed an equation of state for a mixture of ideal gas and radiation assuming local thermodynamic equilibrium, in which the total pressure is the sum of the radiation pressure and gas pressure, i.e., $P=aT^{4}/3 + \rho k_{\rm B}T/[\mu m_{\rm p}]$, where $a$ is the radiation constant, $T$ the temperature, $\rho$ the density, $k_{\rm B}$ the Boltzmann constant, $\mu=0.63$ the mean molecular weight assuming a fully ionized plasma with Solar metallicity, defined in terms of the proton mass $m_{\rm p}$. This assumption is valid because the collision product is highly optically thick until the end of simulations. For the initial conditions used in \arepo, we used stellar models evolved at solar metallicity ($Z=0.02$) with \mesa\ \citep[version r24.03.1;][]{Paxton+2011,paxton:13,Jermyn+2023}  We adopted the prescriptions for overshoot, convection, semiconvection, and thermohaline mixing from \cite{Choi+2016}. We used the Ledoux criterion \cite{ledoux_stellar_1947} to identify the boundary of the convective regions. We considered main-sequence stars with six different masses, ranging from 0.9 to $10\Msol$ initially, and evolved them from their original uniform composition until the central hydrogen mass fraction reached 0.01, representing the terminal age of the main sequence. This choice is motivated by the fact that the stellar radius increases toward the terminal age, which will increase the collision probability\footnote{ The stellar radius increases toward the terminal age by almost a factor of two relative to the radius at zero-age main sequence. This, in turn, will increase the collision probability by a factor of four, as the collision rate is proportional to $R^2$, where $R$ is the stellar radius. While the duration of the terminal age may not be the longest main sequence phase, neither zero-age nor middle-age phase would be four times longer than the terminal-age phase, though the duration of each phase depends on its specific definition. For example, the duration during which the core hydrogen mass fraction ($X_{\rm H}$) decreases to 90\% of the initial value is roughly 10 - 20\% of the main sequence lifetime, depending on the mass. On the other hand, the subsequent period until $X_{\rm H}$ = 0.3 lasts 50 - 60\%, while the remaining 25- 40\% corresponds to the time when $X_{\rm H}$ decreases from 0.3 to 0.01.}. At that time, the surface radius of our six models is $1.1\Rsol$ for $0.9\Msol$ star, $1.2\Rsol$ for $1\Msol$ star, $4.6\Rsol$ for $2.7\Msol$ star, $4.9\Rsol$ for $3\Msol$ star, $9.3\Rsol$ for $9\Msol$ star, and $10\Rsol$ for $10\Msol$ star. We show the density and the hydrogen mass fraction versus Lagrangian mass of each stellar model in Fig.~\ref{fig:initialprofile}.

Using the mapping routine of \citet{Ohlmann+2017}, we relaxed each 1D stellar model in 3D using $N=10^{6}$ cells. Then, we conducted collision experiments by placing a pair of such relaxed main-sequence stars $8\,(R_{1}+R_{2})$ apart and with a relative velocity $V_{\rm rel}$ ($R_{1}$ and $R_{2}$ are the surface radii of the colliding stars). Using this numerical setup, we simulated stellar collisions in which were varied the stellar masses $M_{1}$ and $M_{2}$, the relative velocity, and the impact parameter $b$ -- this parameter $b$ will be quoted in units of $(R_{1}+R_{2})$. We considered three suites of (nearly) equal-mass collisions with $q=M_{2}/M_{1}= 0.9$ (i.e., modelsets \# 1, 2, and 3), as well as collisions with widely different mass ratios of 3 to 10 (i.e., modelsets \# 4, 5, and 6). For each modelset, we considered initial configurations with $b = 0.1$, $0.25$, and $0.5$ for $V_{\rm rel}=$\,5\,000\,\kms, and $b=0.1$ and $0.5$ for $V_{\rm rel}=$\,2\,500\,\kms\ and $10^{4}$\,\kms. The complete list of models simulated with \arepo\ is given in Table~\ref{tab:models}. The model nomenclature gives the mass of each stellar component, the relative velocity in units of 1000\,\kms\ and the impact parameter $b$ in units of $(R_1+R_2)$. For example, model ms1/ms0p9/vrel2p5/b0p1 corresponds to a collision with parameters $M_{1}=$\,1\,$\Msol$, $M_{2}=$\,0.9\,$\Msol$, $V_{\rm rel}=$\,2500\,\kms, and  $b=$\,0.1\,($R_{1}+R_{2}$).

We followed the evolution with \arepo up to 0.5\,d after the collision. In all simulations, the fractional errors in total energy remain small ($\lesssim$1\%) until the end of the simulations. To ensure the robustness of our simulations, we conducted resolution tests for an equal-mass collision involving a $10\Msol$ star with $V_{\rm rel}=$\,5\,000\,\kms and $b=0.1$, varying $N$ per star from 1.25 to $20\times10^{5}$. We found that the overall evolution and key quantities, such as the conversion efficiency between kinetic energy and radiation energy or the structure of the debris, are essentially unchanged when $N\gtrsim 5\times 10^{5}$, which is two times lower than the resolution adopted for the entire grid.

\begin{figure*}
    \centering
    \includegraphics[width=0.85\linewidth,height=12cm]{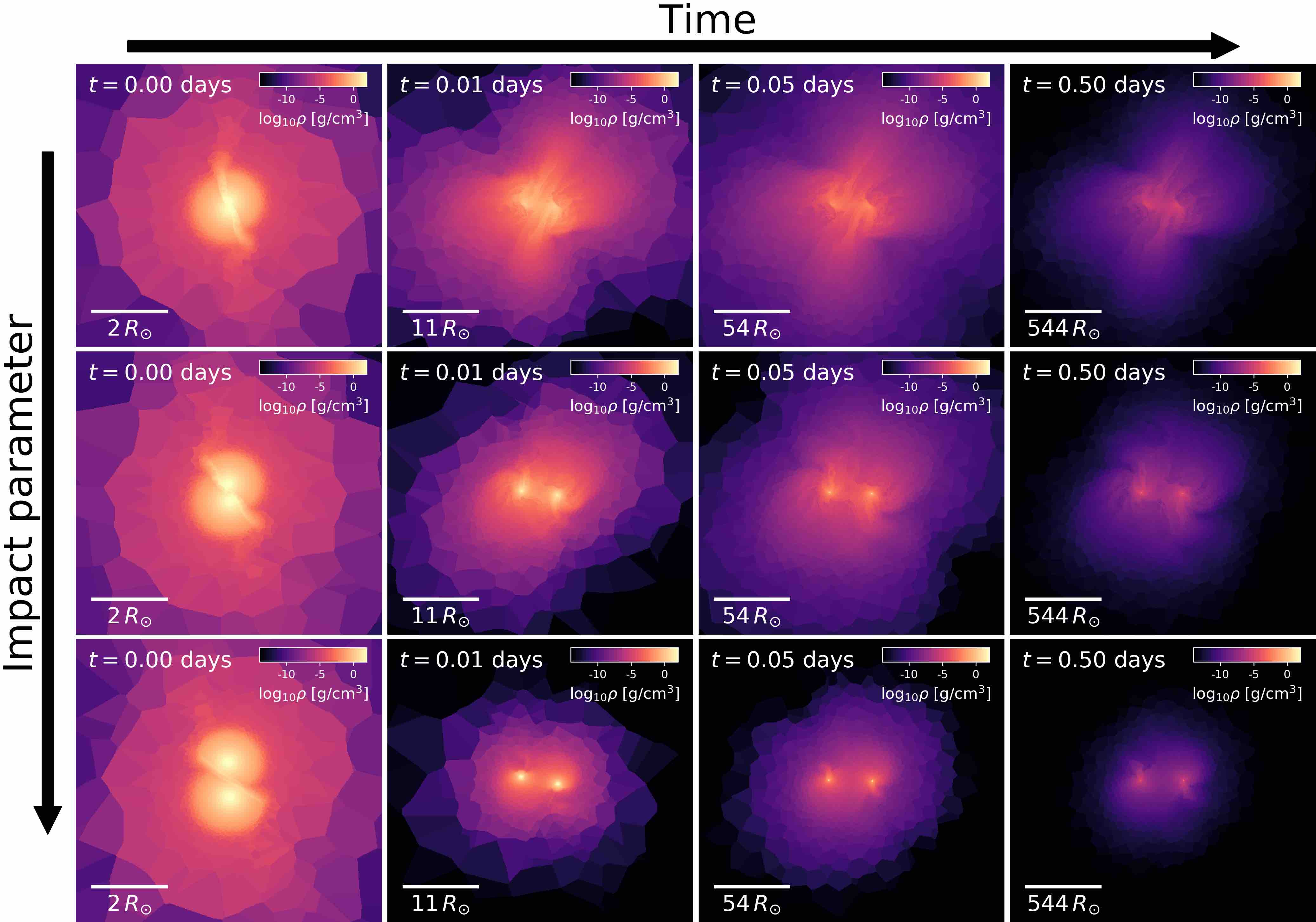}
    \caption{Gas density distribution at successive moments in main-sequence star collisions with $M_{1}=1\Msol$, $M_{2}=0.9\Msol$, $V_{\rm rel}=$\,5\,000\,\kms, but different impact parameters. These \arepo\ simulations of stellar collisions are characterized by $b=0.1$, $0.25$, and $0.5$ (from top to bottom) and shown here at $t=0.0$, $0.01$, $0.05$, and 0.5\,d (from left to right).}
    \label{fig:modelset1}
\end{figure*}

\begin{figure*}
    \centering
    \includegraphics[width=0.85\linewidth,height=12cm]{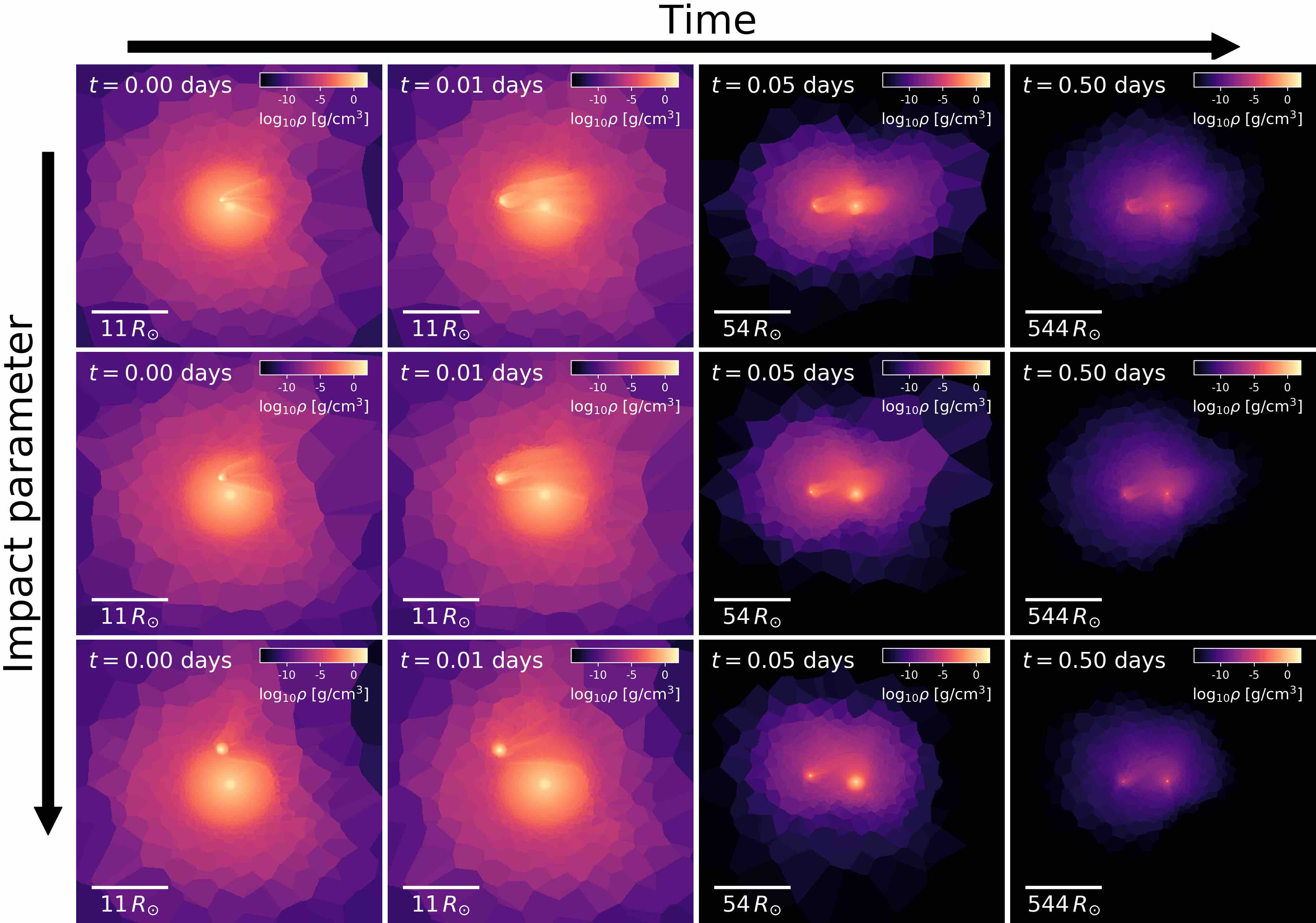}
    \caption{Same as Fig. \ref{fig:modelset1}, but for collisions with $M_{1}=10\Msol$ and $M_{2}=1\Msol$, $V_{\rm rel}$=5\,000\,\kms. In all cases, the collision unbinds only a fraction of the envelopes of both stars, leaving behind remnants. }
    \label{fig:modelset5}
\end{figure*}

\begin{figure*}
    \centering
    \includegraphics[width=18cm]{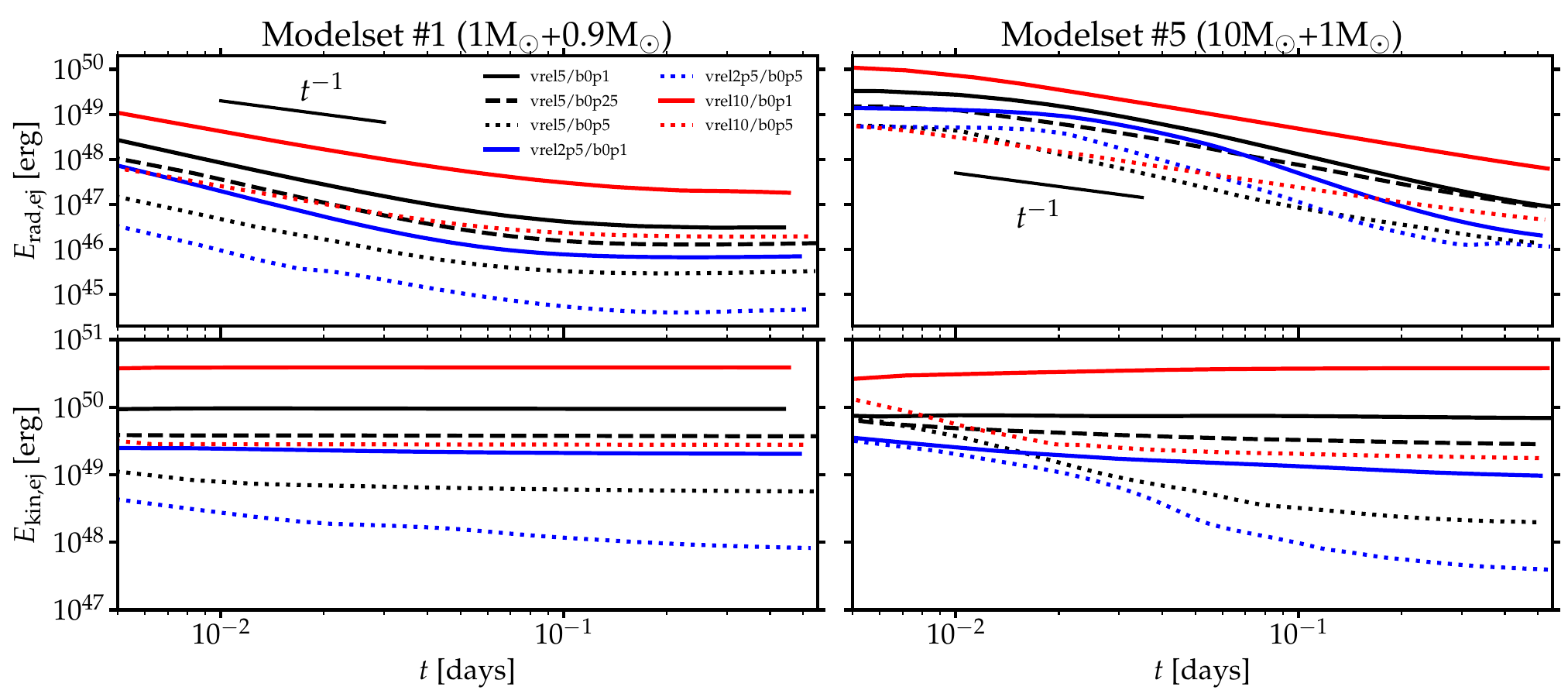}
    \caption{Evolution of the ejecta kinetic and radiative energies in a sample of stellar-collision simulations. We show radiation energy $E_{\rm rad, ej}$ (top) and kinetic energy $E_{\rm kin,ej}$ (bottom) for modelsets \#1 (left, $1\Msol+0.9\Msol$) and \#5 (right, $10\Msol+1\Msol$). The legend indicates the initial relative velocity (e.g., vrel5 means $V_{\rm rel}=5\,000$ km s$^{-1}$) and impact parameter (e.g., b0p5 means $b=0.5$). The diagonal black lines in the top panels indicate the power law of $t^{-1}$, expected for an adiabatic expansion of a radiation-dominated plasma. }
    \label{fig:erad15}
\end{figure*}

\subsection{Dynamical overview}\label{subsec:overview}

We show the evolution of the density for collisions with different impact parameter $b$, first for three models from the set \# 1 ($M_{1}=1\Msol$ and $M_{2}=0.9\Msol$) in Fig.~\ref{fig:modelset1} and  and then for three models from the set \# 5 ($M_{1}=10\Msol$ and $M_{2}=1\Msol$) in Fig.~\ref{fig:modelset5}. The overall evolution is qualitatively similar to collisions between giants studied in \citet{Ryu+2024}. At the onset of the collision, strong shocks form along the contact surface, as shown in the first columns of both figures, converting $0.1\%-50\%$ of the total kinetic energy into thermal energy, depending on the collision parameters. We note that the conversion efficiency for collisions with $b=0.5$ is relatively low ($0.1-4\%$) and independent of the stellar masses and $V_{\rm rel}$. Although the shocks are weak in some cases, they lead to the formation of homologously, supersonically expanding debris after a few hours. While small structures exist within such ejecta, the overall shape is approximately spherical, which is consistent with the assumption for the debris made in \citet{Balberg+2013} and \citet{AmaroSeoane2023}.

The shocks convert kinetic energy into radiation energy. 
Figure~\ref{fig:erad15} shows the radiation energy $E_{\rm rad,ej}$ and kinetic energy $E_{\rm kin,ej}$ within the ejecta in the two models considered in Fig.~\ref{fig:modelset1} (left) and ~\ref{fig:modelset5} (right), as a function of time. At collision ($t=0$), $E_{\rm rad,ej}$ peaks, with values ranging from $10^{47}-4\times10^{49}$ erg for modelset \#1 and from $6\times10^{48}-10^{50}$ erg for modelset \#5, depending on the collision parameters. It then drops by roughly two to three orders of magnitude in 0.5\,d, which is the start time of our radiative-transfer calculations (see \S~\ref{sec:RT}). Our hydrodynamics simulations do not account for radiative energy loss-- they assume adiabatic conditions. While this approximation should hold well for the optically thick regions of the debris, the radiation energy in the outer regions -- where the optical depth approaches unity -- may be overestimated. However, this should not introduce significant errors, as the excess heat is radiated away in {the first time step of the radiative transfer calculations. $E_{\rm kin,ej}$ ranges from $10^{48}-4\times10^{50}$ erg for both modelsets. In highly energetic collisions with small impact parameters, $E_{\rm kin,ej}$ remains constant, whereas in collisions with relatively low $V_{\rm rel}$ and large $b$, $E_{\rm kin,ej}$ initially tends to decrease gradually and then levels off within 0.5\,d after the collision.

Even if the collision kinetic energy exceeds the binding energy of both stars, the collision does not generally destroy them, but instead leaves behind stellar remnants, composed of the more compact, denser He-rich core of the star prior to collision. The only cases where both stars are completely destroyed are when $M_{1}=1\Msol$, $M_{2}=0.9\Msol$, $V_{\rm rel}\gtrsim 5\,000$ km s$^{-1}$, and $b=0.1$. 
 
The results for the full sample of simulations is stored in Table~\ref{tab:models} and indicate debris or ejecta masses $M_{\rm ej}$ in the range from a few 0.01$\Msol$ to slightly over 10$\Msol$, ejecta kinetic energies in the range from a few 10$^{47}$\,erg to several 10$^{51}$\,erg, and stored radiative energies at 0.5\,d in the range from a few 10$^{47}$\,erg to about $6 \times 10^{48}$\,erg.

\section{Radiative-transfer simulations}\label{sec:RT}

\begin{table}
\caption{Radiative properties of our sample of high velocity collisions at 1\,d. }
\label{tab:res}
\small
\centering
\begin{tabular}{l c c c c}
\hline
\hline
Model &   $L_{\rm bol, 1\,d}$  & $M_{UVW1, {\rm 1\,d}}$ &   $M_{V,{\rm 1\,d}}$  & $V_{{\rm H}\alpha, {\rm 1\,d}}$ \\
      &       [erg]          &    [mag]    & [mag]   & [\kms] \\
\hline
    ms1/ms0p9/vrel5/b0p1 &   2.17(41) &   -15.3 &   -13.4 &  12965 \\
   ms1/ms0p9/vrel5/b0p25 &   5.14(40) &   -13.5 &   -12.4 &  10537 \\
    ms1/ms0p9/vrel2p5/b0p1 &   3.82(40) &   -13.1 &   -12.1 &   9482 \\
   ms1/ms0p9/vrel10/b0p1 &   5.10(41) &   -16.2 &   -14.3 &  20267 \\
   ms1/ms0p9/vrel10/b0p5 &   9.73(40) &   -14.4 &   -12.8 &  11231 \\
\hline
    ms3/ms2p7/vrel5/b0p1 &   2.35(41) &   -15.4 &   -13.4 &  13235 \\
   ms3/ms2p7/vrel5/b0p25 &   7.25(40) &   -14.1 &   -12.5 &  10510 \\
    ms3/ms2p7/vrel2p5/b0p1 &   6.55(40) &   -14.0 &   -12.4 &   9336 \\
   ms3/ms2p7/vrel10/b0p1 &   1.06(42) &   -16.7 &   -14.6 &  20692 \\
   ms3/ms2p7/vrel10/b0p5 &   1.62(41) &   -15.0 &   -12.9 &  11184 \\
\hline
     ms10/ms9/vrel5/b0p1 &   7.04(41) &   -16.2 &   -14.0 &  15465 \\
    ms10/ms9/vrel5/b0p25 &   2.20(41) &   -15.2 &   -13.1 &  10970 \\
     ms10/ms9/vrel5/b0p5 &   7.27(40) &   -14.1 &   -12.2 &   7637 \\
     ms10/ms9/vrel2p5/b0p1 &   1.93(41) &   -15.1 &   -13.1 &  10794 \\
    ms10/ms9/vrel10/b0p1 &   3.33(42) &   -17.6 &   -15.3 &   9796 \\
    ms10/ms9/vrel10/b0p5 &   5.54(41) &   -15.9 &   -13.6 &  13353 \\
\hline
      ms3/ms1/vrel5/b0p1 &   1.73(41) &   -15.1 &   -13.2 &  12395 \\
     ms3/ms1/vrel5/b0p25 &   1.27(41) &   -14.7 &   -12.9 &  10868 \\
      ms3/ms1/vrel5/b0p5 &   1.71(40) &   -11.8 &   -11.5 &   9181 \\
      ms3/ms1/vrel2p5/b0p1 &   3.72(40) &   -13.2 &   -12.0 &   8240 \\
      ms3/ms1/vrel2p5/b0p5 &   4.97(39) &    -9.7 &   -10.5 &   6096 \\
     ms3/ms1/vrel10/b0p1 &   8.08(41) &   -16.5 &   -14.4 &  19141 \\
     ms3/ms1/vrel10/b0p5 &   7.71(40) &   -13.9 &   -12.6 &  12740 \\
\hline
     ms10/ms1/vrel5/b0p1 &   9.00(40) &   -14.4 &   -12.6 &   9732 \\
    ms10/ms1/vrel5/b0p25 &   9.86(40) &   -14.5 &   -12.6 &   9805 \\
     ms10/ms1/vrel5/b0p5 &   3.01(40) &   -12.7 &   -11.9 &   9439 \\
     ms10/ms1/relv2p5/b0p1 &   3.11(40) &   -13.1 &   -11.7 &   6138 \\
     ms10/ms1/vrel2p5/b0p5 &   1.29(40) &   -11.8 &   -11.1 &   6212 \\
    ms10/ms1/vrel10/b0p1 &   2.94(41) &   -15.6 &   -13.6 &  14728 \\
    ms10/ms1/vrel10/b0p5 &   1.06(41) &   -14.3 &   -12.9 &  14374 \\
\hline
     ms10/ms3/vrel5/b0p1 &   5.60(41) &   -16.0 &   -13.8 &  14134 \\
    ms10/ms3/vrel5/b0p25 &   4.29(41) &   -15.7 &   -13.5 &  12579 \\
     ms10/ms3/vrel2p5/b0p1 &   1.24(41) &   -14.7 &   -12.7 &   8995 \\
    ms10/ms3/vrel10/b0p1 &   2.64(42) &   -17.4 &   -15.1 &  23420 \\
    ms10/ms3/vrel10/b0p5 &   5.34(41) &   -15.9 &   -13.7 &  13964 \\
\hline
\end{tabular}
\tablefoot{The table gives results from the radiative transfer calculations with \cmfgen. The columns give the model name, the bolometric luminosity, the $UVW1$-band and $V$-band magnitudes and the Doppler velocity at maximum absorption measured in the H$\alpha$ line profile, all given at one day after the onset of the collision.
The model name in the first column consists of the masses of the two colliding stars, their relative velocity, and the impact parameter at
collision. Numbers appearing within parentheses correspond to powers of ten.}
\end{table}

\subsection{Numerical setup}\label{subsec:RT_setup}
Our main goal is to obtain accurate light curves and spectra of high-velocity collisions. As in \citet{Ryu+2024}, the bolometric luminosities from the debris can be estimated based on the radiation energy and photon diffusion time derived from the 3D hydrodynamic simulations. However, such an estimate is crude. Moreover, spectral calculations are not possible solely using the hydrodynamic properties of the debris. Therefore, once the debris begin to expand in a quasi-spherical and homologous fashion, which occurs at $\simeq0.5$ d, we stopped the 3D hydrodynamic simulations and performed radiative transfer calculations using the properties of the debris obtained from the 3D simulations as initial conditions to calculate light curves and spectra. The radiative-transfer calculations performed here on main-sequence star collisions follow the same procedure as described in \citet{Dessart+2024}, where collisions between giant stars were studied. Hence, we refer the reader to that work and in particular their Section~2 for a detailed presentation of the numerical approach. Here, we merely provide a concise summary. 

We performed 1D nonlocal thermodynamic equilibrium (NLTE) time-dependent radiative-transfer calculations with the code \cmfgen\ \citep{HD12} using the \arepo\ simulation results for a whole range of stellar collisions. Since this radiative-transfer code does not treat hydrodynamics and assumes spherical symmetry, we started the simulations at a sufficiently late time after the onset of the 
collision, that is once the debris of the collision were coasting. Because of the relatively compact structure of the main-sequence stars used (see Fig.~\ref{fig:initialprofile}), this occurs after a few hours. Furthermore, we built a spherical average of the fluid quantities obtained in the \arepo\ simulation. The spherical average was calculated so that the ejecta mass in each radius bin was the same in 1D and 3D profiles. However, the kinetic energy and radiation energies (or radial velocity and temperature) were not strictly conserved. However, as shown in Table~\ref{tab:models}, the difference between the two energy estimates is not significant, less than a factor of two in most cases. It becomes significant only for $b=0.5$ and $V_{\rm rel}=2\,500$ km s$^{-1}$ (a factor of $5-10$), indicating our assumption of spherical symmetry for expanding ejecta is not valid in such cases. This also implies that viewing angle effects are important for these collisions. 

To cover the early evolution of the radiative properties of such collisions when they are the most luminous, we started our simulations at a post-collision time of 0.5\,d. Unlike \citet{Dessart+2024} in which the evolution was followed until the debris became optically-thin (i.e., when the total radial electron-scattering optical depth drops below one), we evolved most of our simulations until the debris entered the H recombination phase, at a time when the luminosity from the debris has dropped by a factor of ten to a hundred. By that time, the associated transient would likely be undetectable at representative distances of 10--100\,Mpc. The other reason for starting early is that we are interested in documenting the properties of our simulations when they are UV bright, which is also when they are the most luminous, and thus exhibit favorable properties for UV observations with future satellites like UVEX \citep{uvex} or ULTRASAT \citep{ultrasat} -- these space observatories will routinely detect and monitor transients merely hours after explosion, collision, or disruption, depending on the type of transient. With this aim, we could make another adjustment in setting up the initial models by truncating the inner, dense, and slow moving regions of the debris. This adjustment should not affect our main results for light curves and spectra over the duration considered ($\lesssim 10$ d). Indeed, at 0.5\,d, these regions tend to be exceptionally dense and slow, where an optical depth of $\gtrsim10^5$, and containing merely a few percent of the total stored radiative energy at that time (see penultimate column in Table~\ref{tab:models}). This means that because of relatively long photon diffusion times, these regions would influence escaping radiation at later times -- days and weeks -- only marginally. A further complication is that at such later times, accretion of debris material into the supermassive black hole may lead to power injection into the expanding debris \citep[see][]{Ryu+2024,Brutman+2024} and alter its energy budget -- this influence should be small or even negligible at earlier times when the debris are totally optically-thick but would completely dominate at late times. The longer-time evolution on week-timescales of our models would require a more global approach to account for both the debris and the supermassive black hole, which we delay to a forthcoming study.

Collisions with small impact parameters and high stellar masses reached debris masses and kinetic energy comparable to what is found in Type II supernovae (i.e., about 10$\Msol$ and 10$^{51}$\,erg) but with a very small radiative energy content on the order of 10$^{48}$\,erg more typical of (compact) blue-supergiant \citep{DH_2pec_19} rather than (extended) red-supergiant star explosions (see, for example, the simulations of \citealt{d13_sn2p} and \citealt{DH_2pec_19}). These quantities are stored in Table~\ref{tab:models}. In this work, we performed radiative-transfer calculations for 35 collisions but excluded the weakest collisions (some in the sample having small relative velocities and large impact parameters) with the lowest radiative-energy content since their transients would be too faint to be observed. In a few cases (e.g., model ms10/ms9/vrel2p5/b0p5), the model was not included because of numerical difficulties when evolving the ejecta with \cmfgen, in particular when the initial temperature structure was highly nonmonotonic (e.g., a strong temperature peak at intermediate velocities in the debris but low temperatures in the inner, dense debris regions). The suite of radiation transfer simulations is sufficiently large to establish correlations between collision parameters and observables.

\begin{figure}
   \centering
    \begin{subfigure}[b]{0.49\textwidth}
       \centering
       \includegraphics[width=\textwidth]{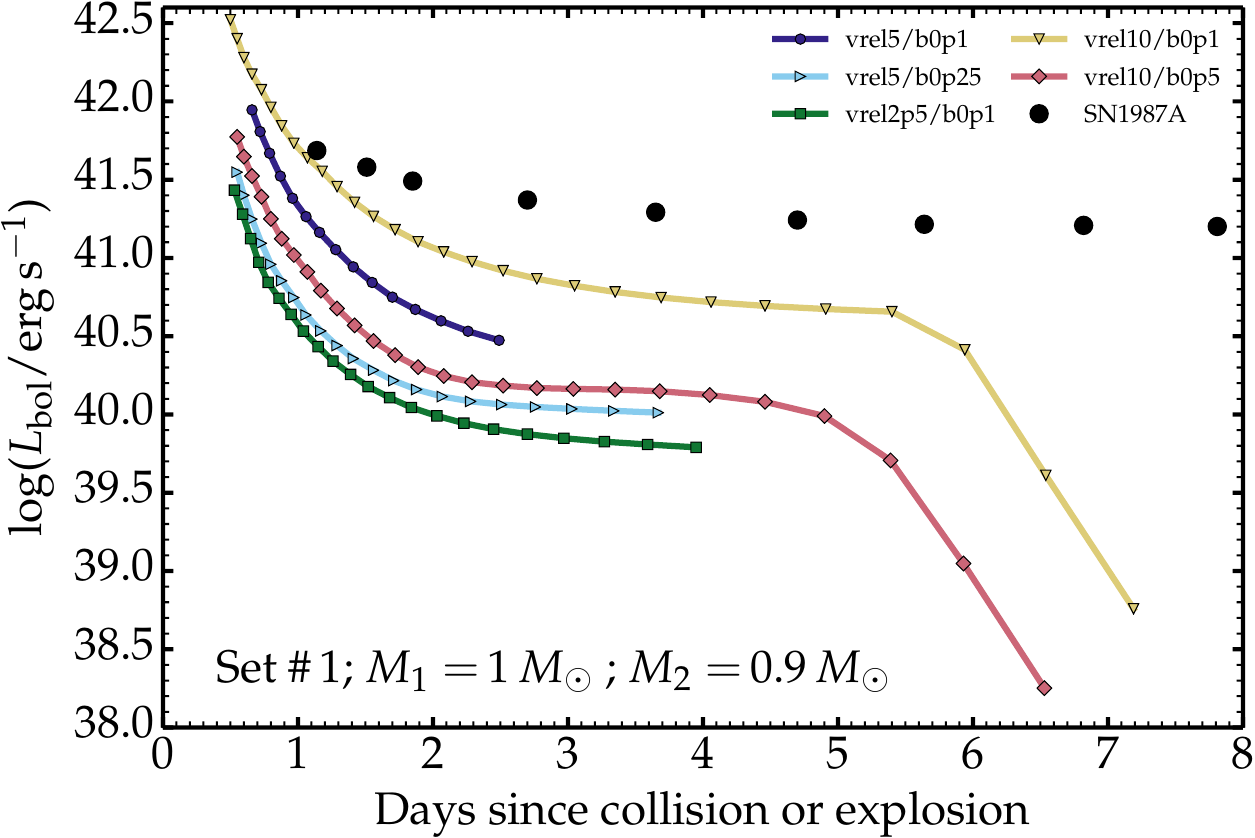}
    \end{subfigure}
    \hfill
    \begin{subfigure}[b]{0.49\textwidth}
       \centering
       \includegraphics[width=\textwidth]{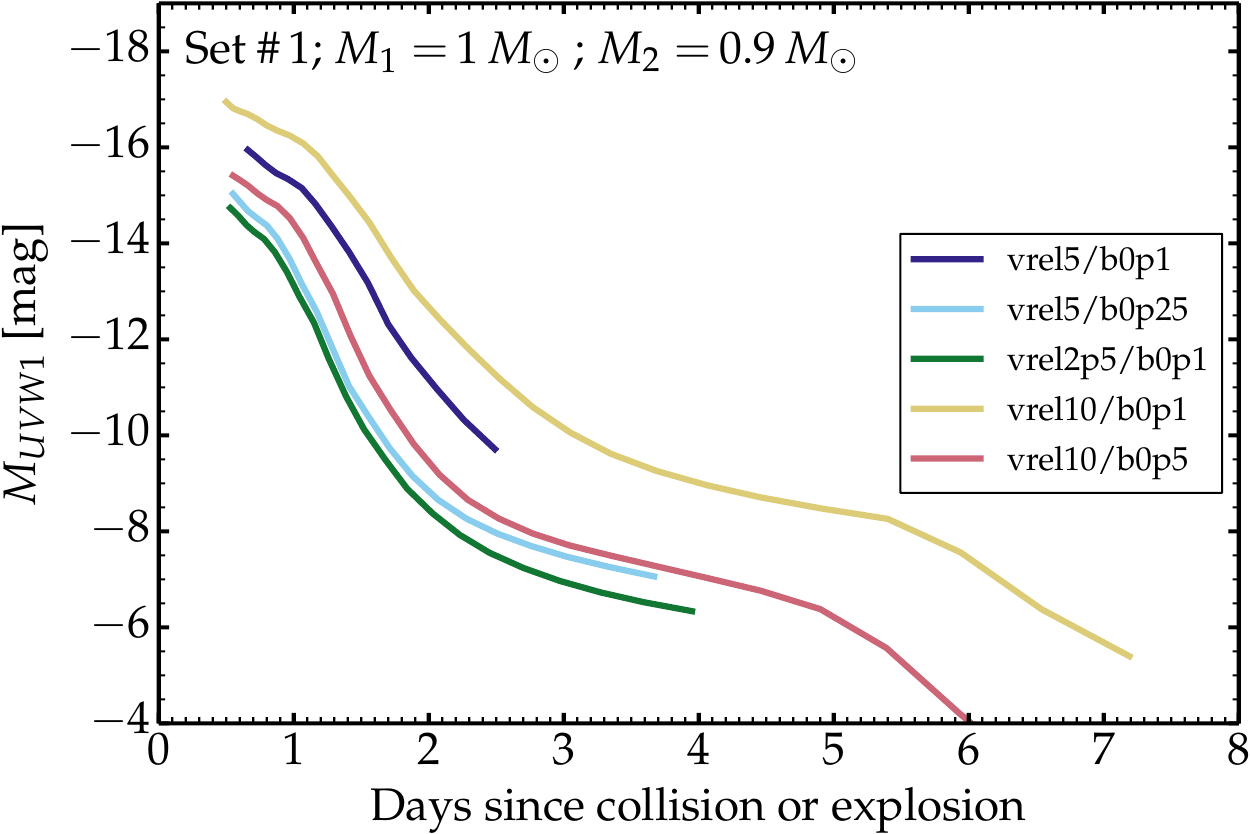}
    \end{subfigure}
     \hfill
    \begin{subfigure}[b]{0.49\textwidth}
       \centering
       \includegraphics[width=\textwidth]{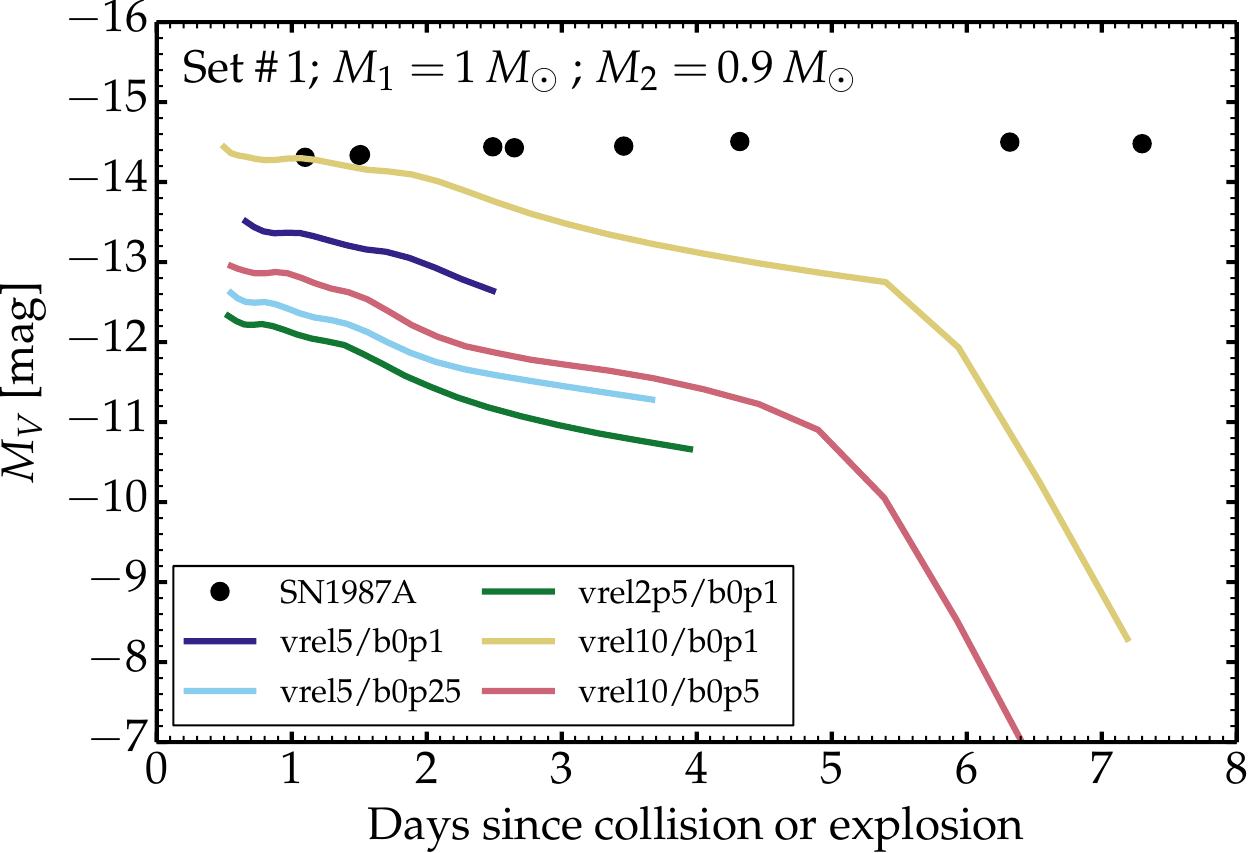}
    \end{subfigure}
     \hfill
\caption{Photometric properties for main-sequence star collisions from the modelset \#\,1. From top to bottom, we show the bolometric, $UVW1$-band, and $V$-band light curves. Filled-dot indicate the corresponding quantities inferred for SN\,1987A (after correction for reddening and distance). Corresponding results for all sets are shown in Appendix \S~\ref{sec:appendix}.
\label{fig:phot_modelset1}
}
\end{figure}

\begin{figure}
    \centering
    \includegraphics[width=1\linewidth]{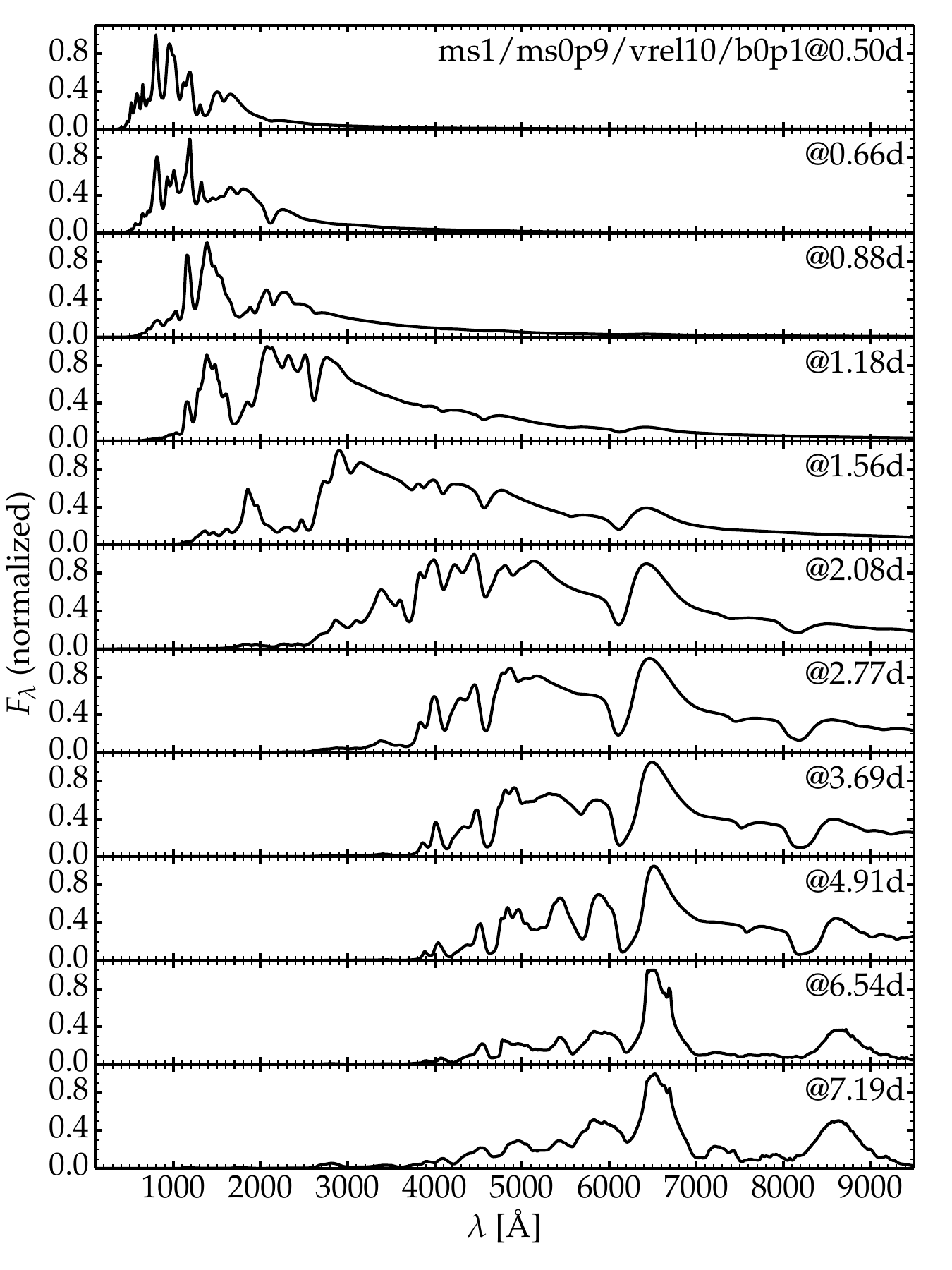}
    \caption{Spectral evolution from the UV to the optical and from $\sim$\,0.5 until $\sim$\,7.2\,d for
model ms1/ms0p9/vrel10/b0p1 (modelset \#\,1). $F_{\lambda}$ represents the spectral flux density or the flux per unit wavelength interval. Corresponding results for all models from the six sets are shown in Appendix \S~\ref{sec:appendix}.}
\label{fig:montage_spec_ex}
\end{figure}

For the \cmfgen\ calculations, we assume a solar-metallicity composition for all metals above O but take the composition profile from the \arepo\ simulations for species H, He, C, N, and O (the N/C abundance ratio is increased by a factor of a few in the H-depleted He-rich core of the original stellar models; see also Fig.~\ref{fig:initialprofile}). Specifically, we treat H, He, C, N, O, Ne, Na, Mg, Al, Si, S, Ar, Ca, Sc, Ti, Cr, Fe, Co, Ni, and Ba. Following \citet{Dessart+2024}, we include the following ions for the above species: H\one, He\one-\two, C\one--\four, N\one--\five, O\one--\five, Ne\two--\four, Na\one, Mg\two--\four, Al\two--\three, Si\one--\four, S\two--\four, Ar\one--\three, Ca\two--\four, Sc\two--\three, Ti\two--\three, Cr\two--\five, Fe\one--\five, Co\two--\five, Ni\one--\five, and Ba\two. All associated isotopes being stable, there is no contribution from radioactive decay and no non-thermal effects associated with such decays. All spherical-averaged debris inputs are remapped onto a grid with 80 points uniformly spaced on a logarithmic optical-depth scale. Remapping the grid is also necessary at the start of every time step as well as during the calculation in order to track the H\one\ and He\one-He\two\ recombination fronts that form as the debris cool through expansion and radiation escape. To cover the full evolution, a total of about 20--30 timesteps (depending on the timespan covered) were computed for most models. In some models, the calculation was stopped when the debris were becoming very faint (e.g., luminosity below 10$^6$\,\lsun) during the recombination phase. The main \cmfgen\ output of interest for this study is the emergent flux, which is calculated at each time step in the observer's frame from the far-UV to the far-IR. This spectrum can be shown directly or used to compute the bolometric luminosity $L_{\rm bol}$ or the absolute magnitude in various filters. We selected the Swift $UVW1$-band and the optical $V$-band filters. The $UVW1$ filter is a close analogue of the NUV filter that will be on board ULTRASAT \citep{ultrasat}.

\begin{figure*}
   \centering
    \begin{subfigure}[b]{0.33\textwidth}
       \centering
       \includegraphics[width=\textwidth]{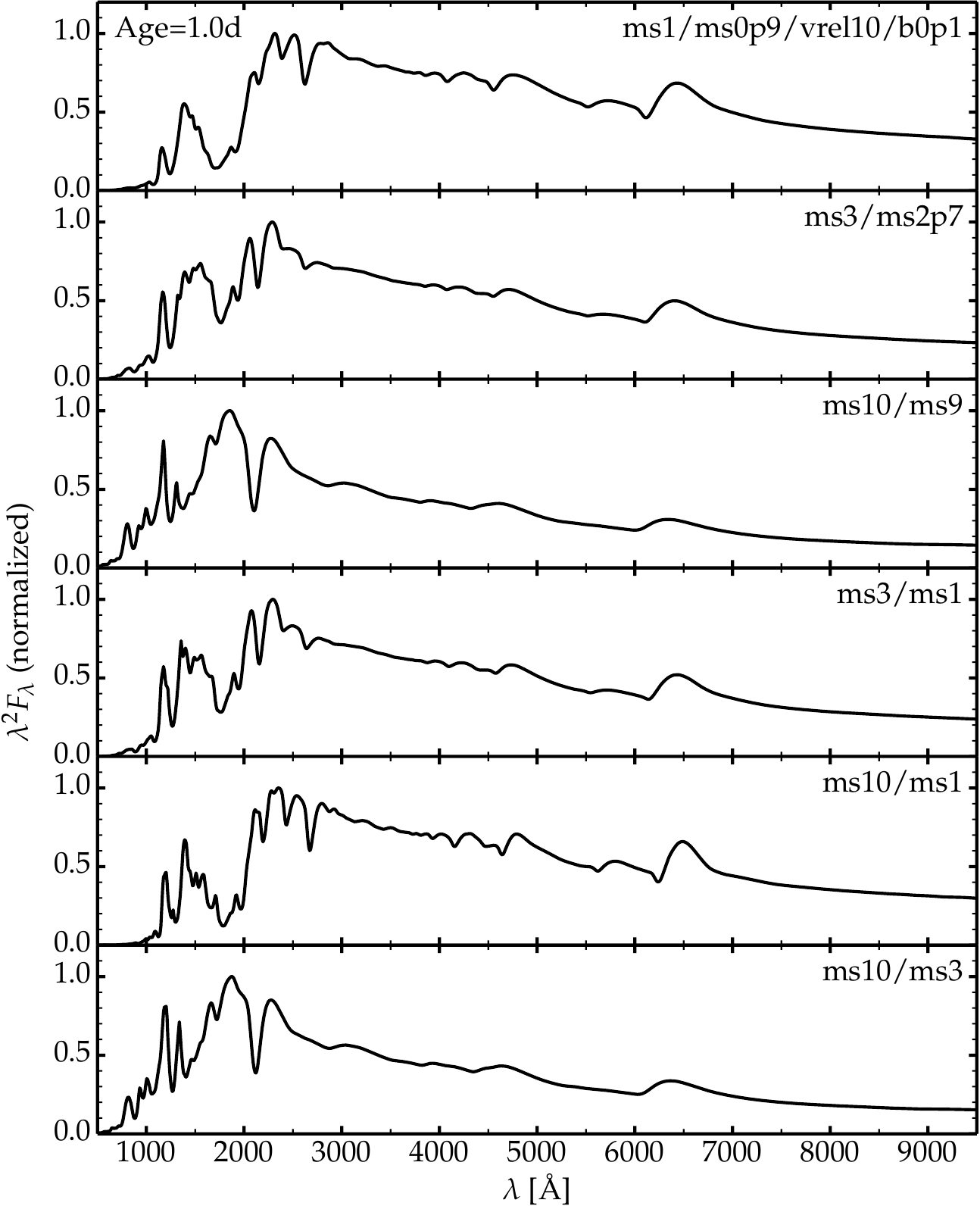}
    \end{subfigure}
    \hfill
    \begin{subfigure}[b]{0.33\textwidth}
       \centering
       \includegraphics[width=\textwidth]{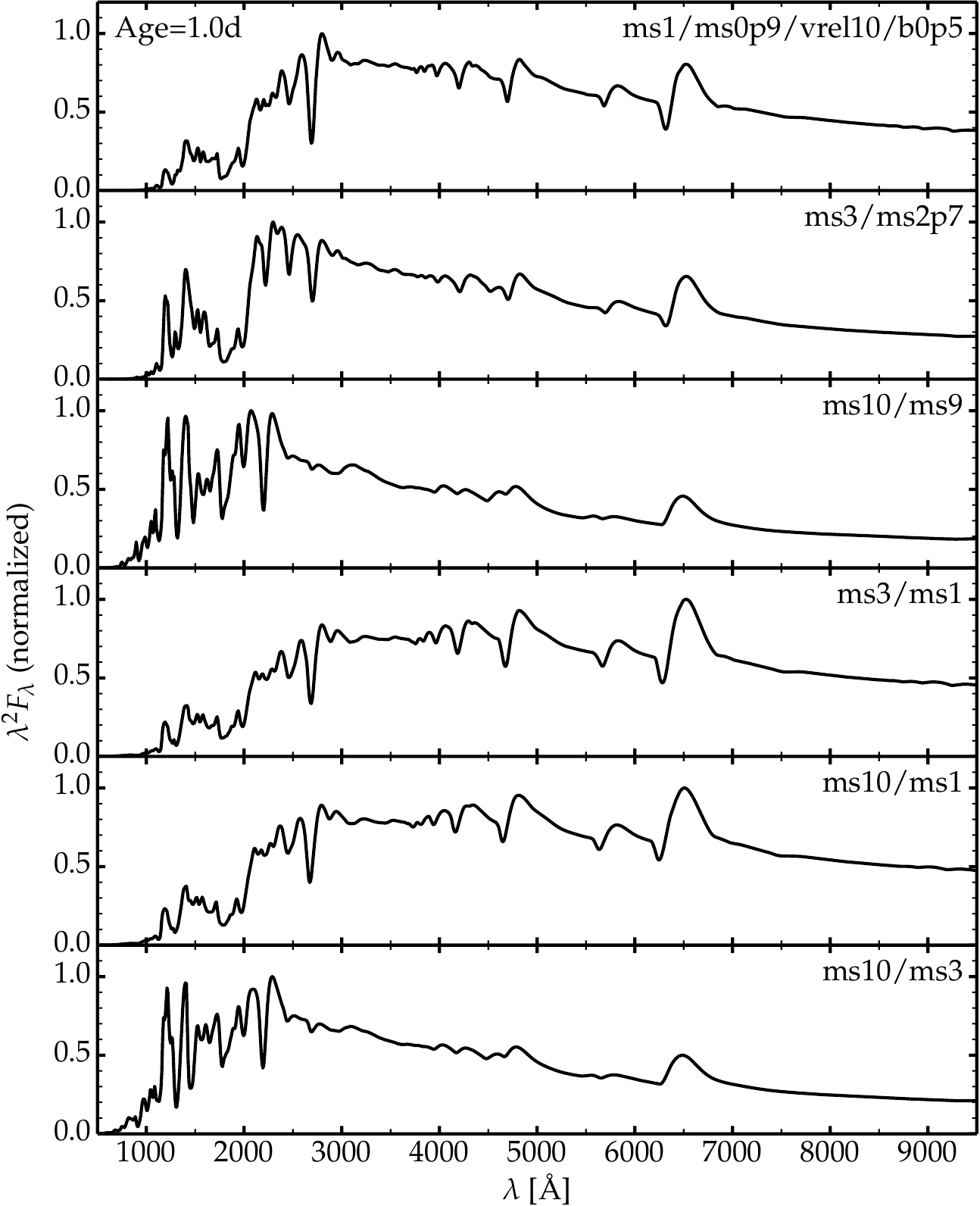}
    \end{subfigure}
     \hfill
    \begin{subfigure}[b]{0.33\textwidth}
       \centering
       \includegraphics[width=\textwidth]{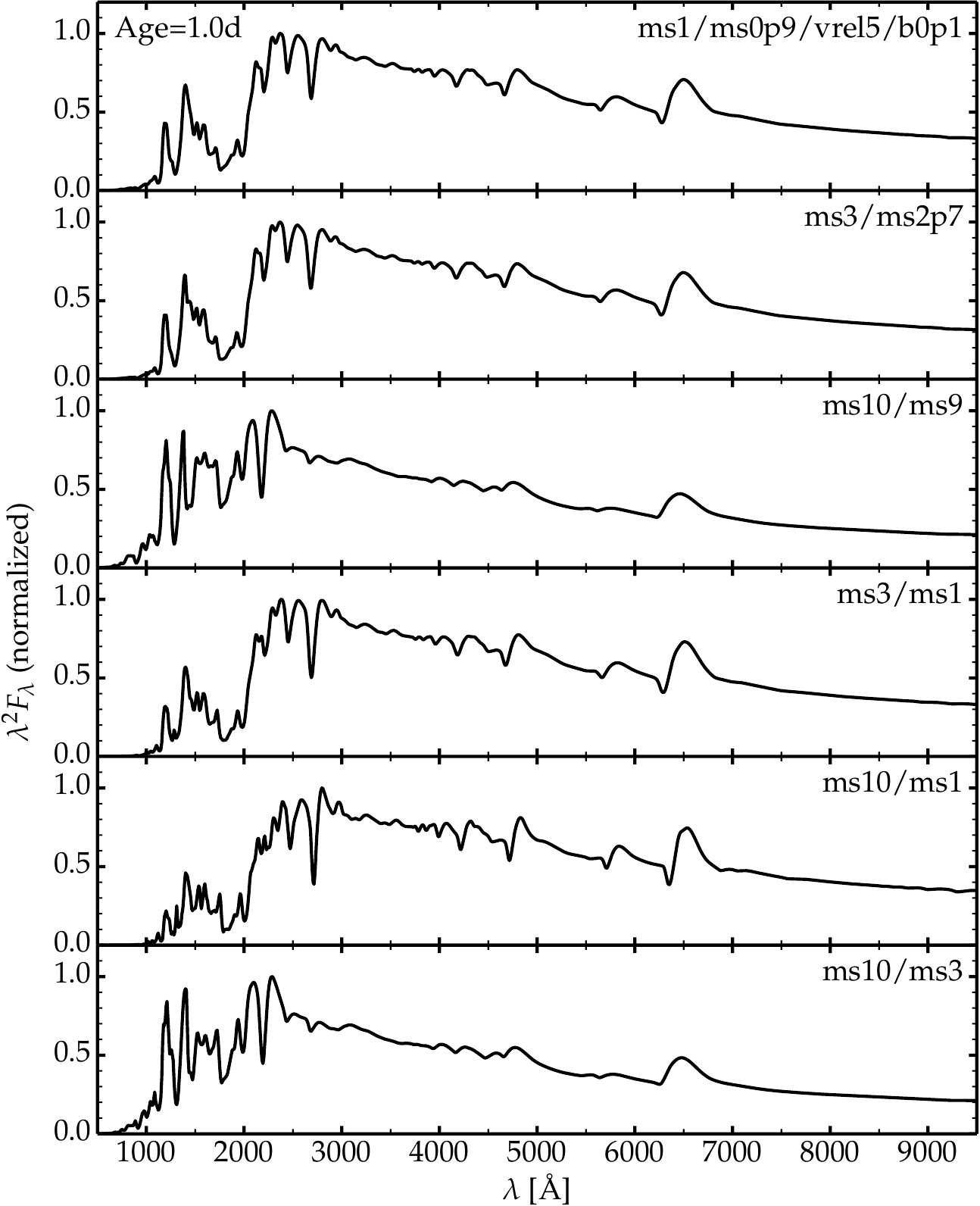}
    \end{subfigure}
     \hfill
    \begin{subfigure}[b]{0.33\textwidth}
       \centering
       \includegraphics[width=\textwidth]{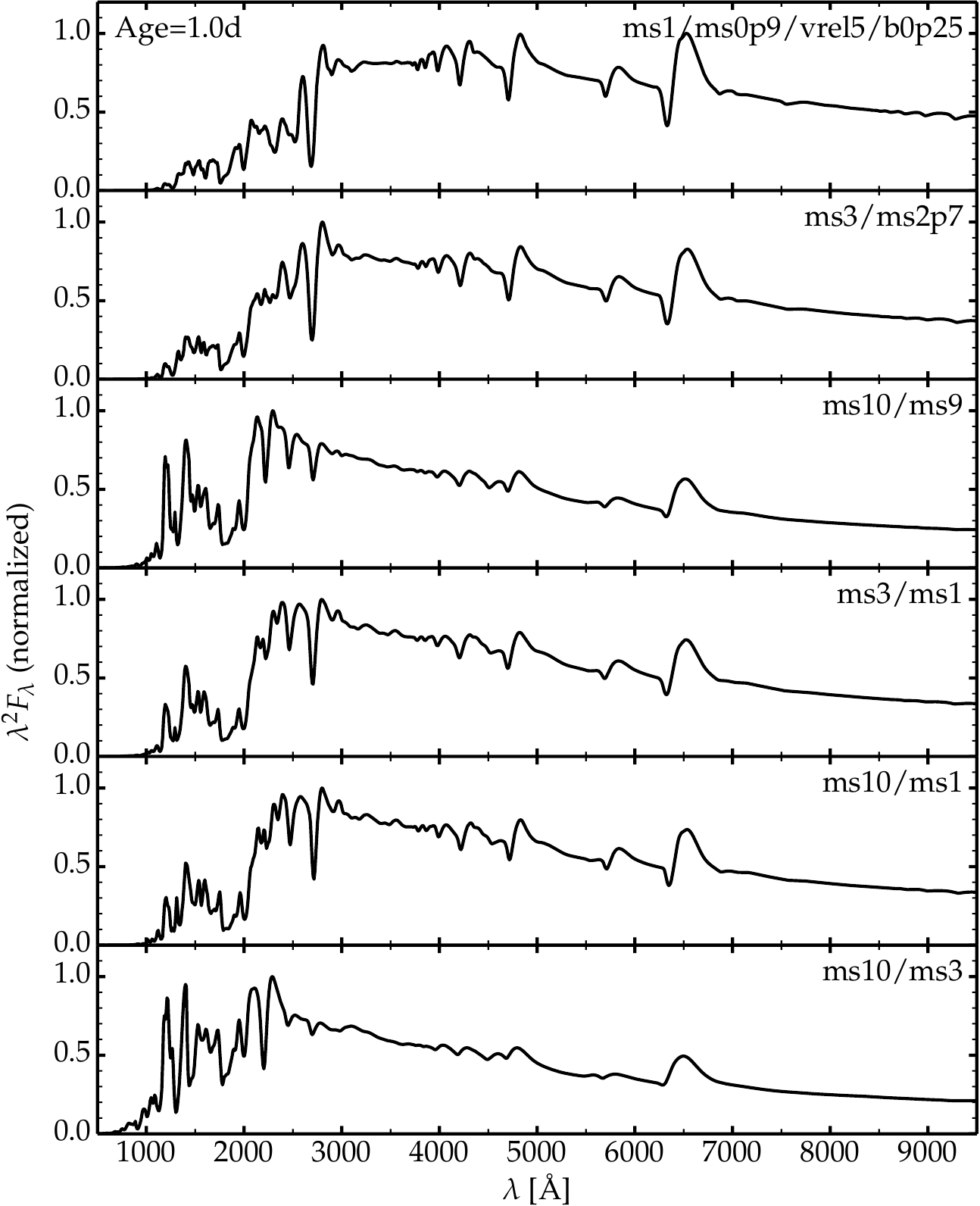}
    \end{subfigure}
     \hfill
     \begin{subfigure}[b]{0.33\textwidth}
       \centering
        \includegraphics[width=\textwidth]{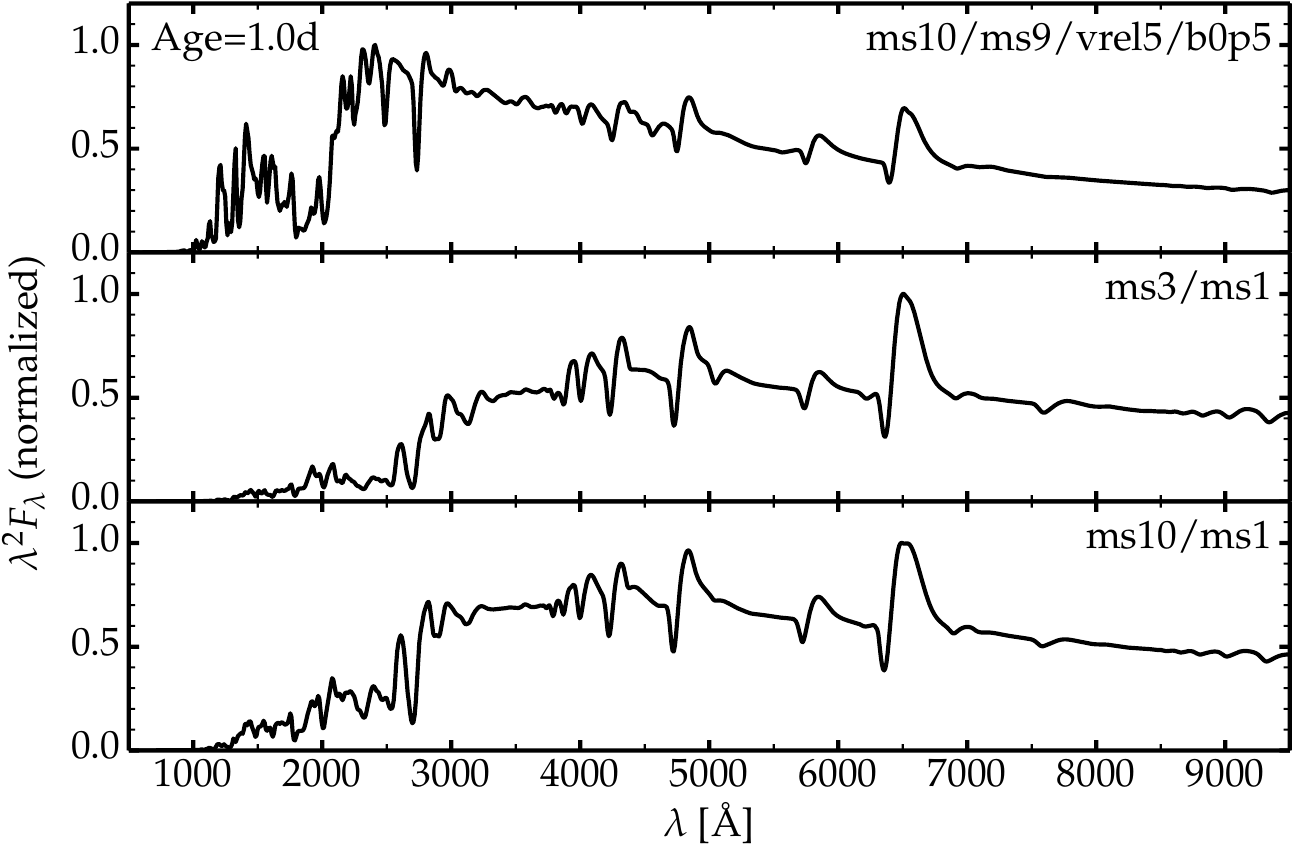}
        \includegraphics[width=\textwidth]{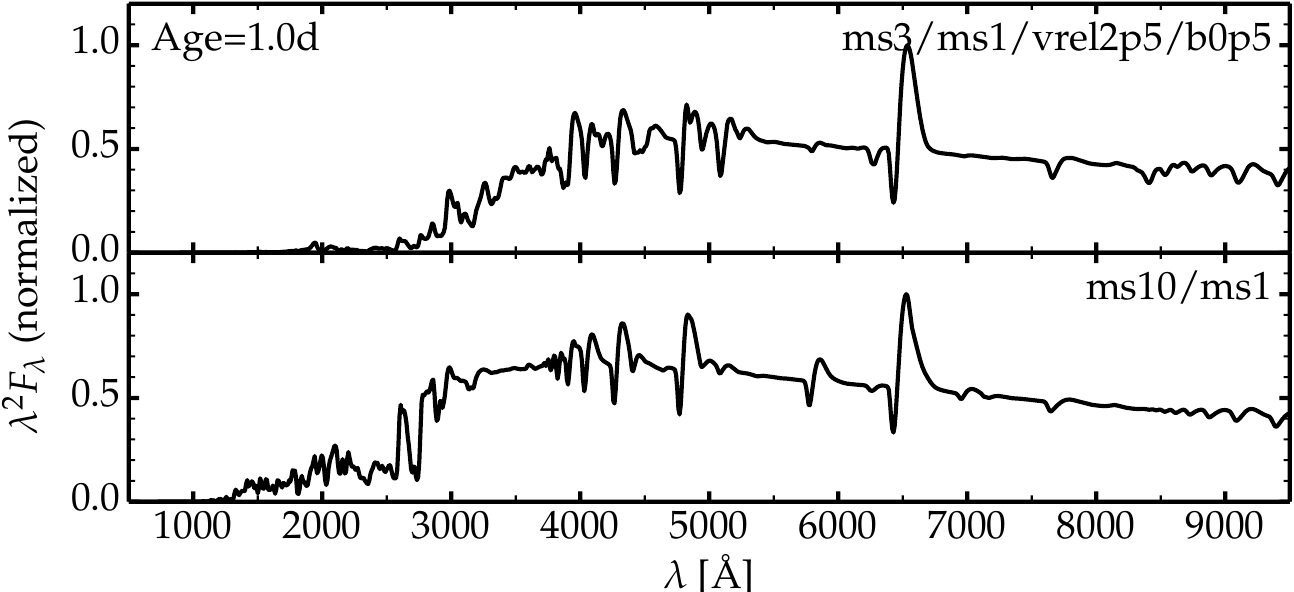}
    \end{subfigure}
     \hfill
     \begin{subfigure}[b]{0.33\textwidth}
       \centering
        \includegraphics[width=\textwidth]{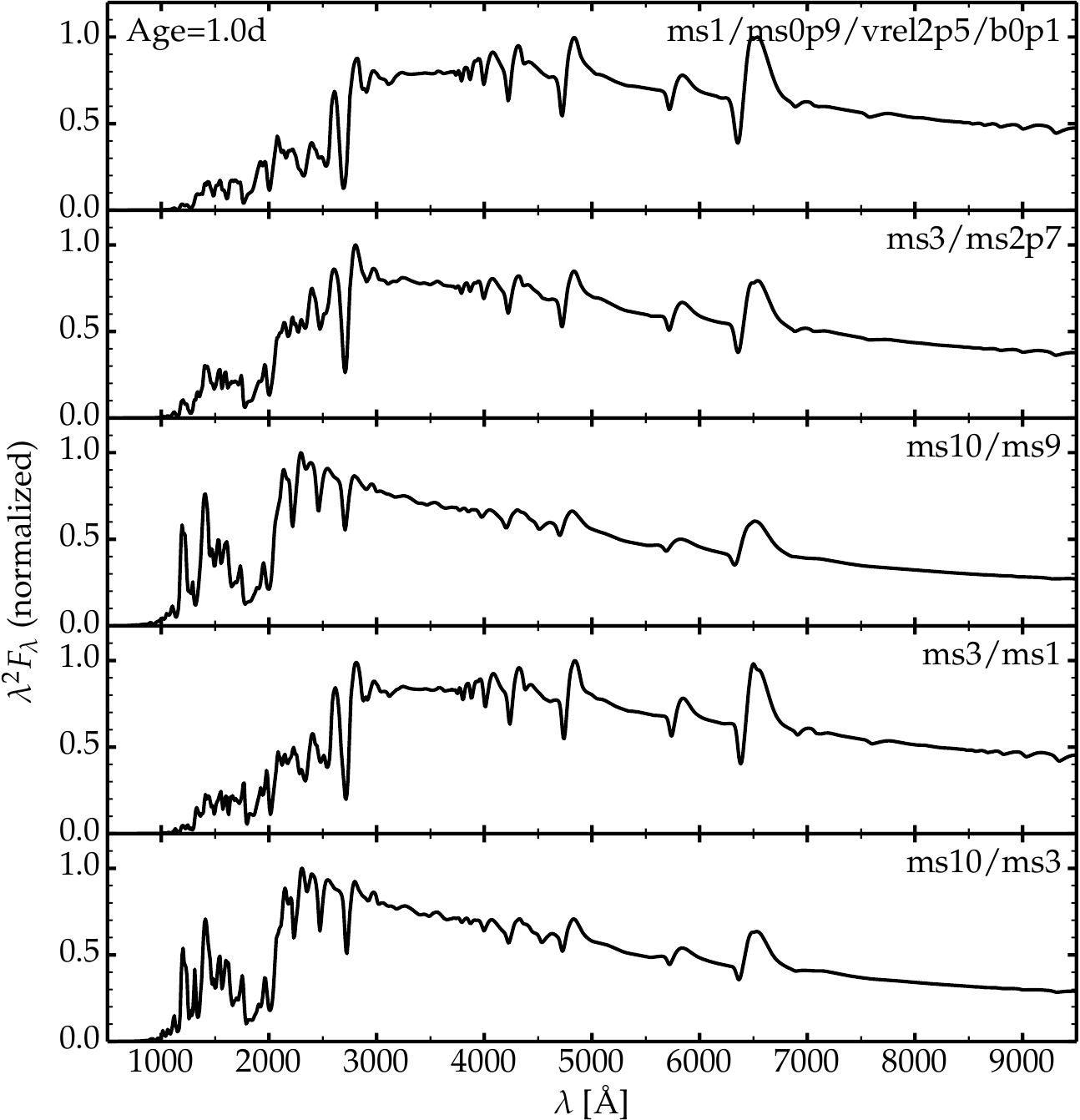}
    \end{subfigure}
\caption{Spectral comparisons of models grouped by the same values of $V_{\rm rel}$ and $b$ but differing in the stellar masses of the colliding pair. The time is one day after the onset of the collision. To better reveal the weaker flux in the optical, we show the quantity $\lambda^2 F_\lambda$, which is analogous to an $F_\nu$.
\label{fig:montage_spec_at_1d}
}
\end{figure*}

\begin{figure}
    \centering
    \includegraphics[width=8.9cm]{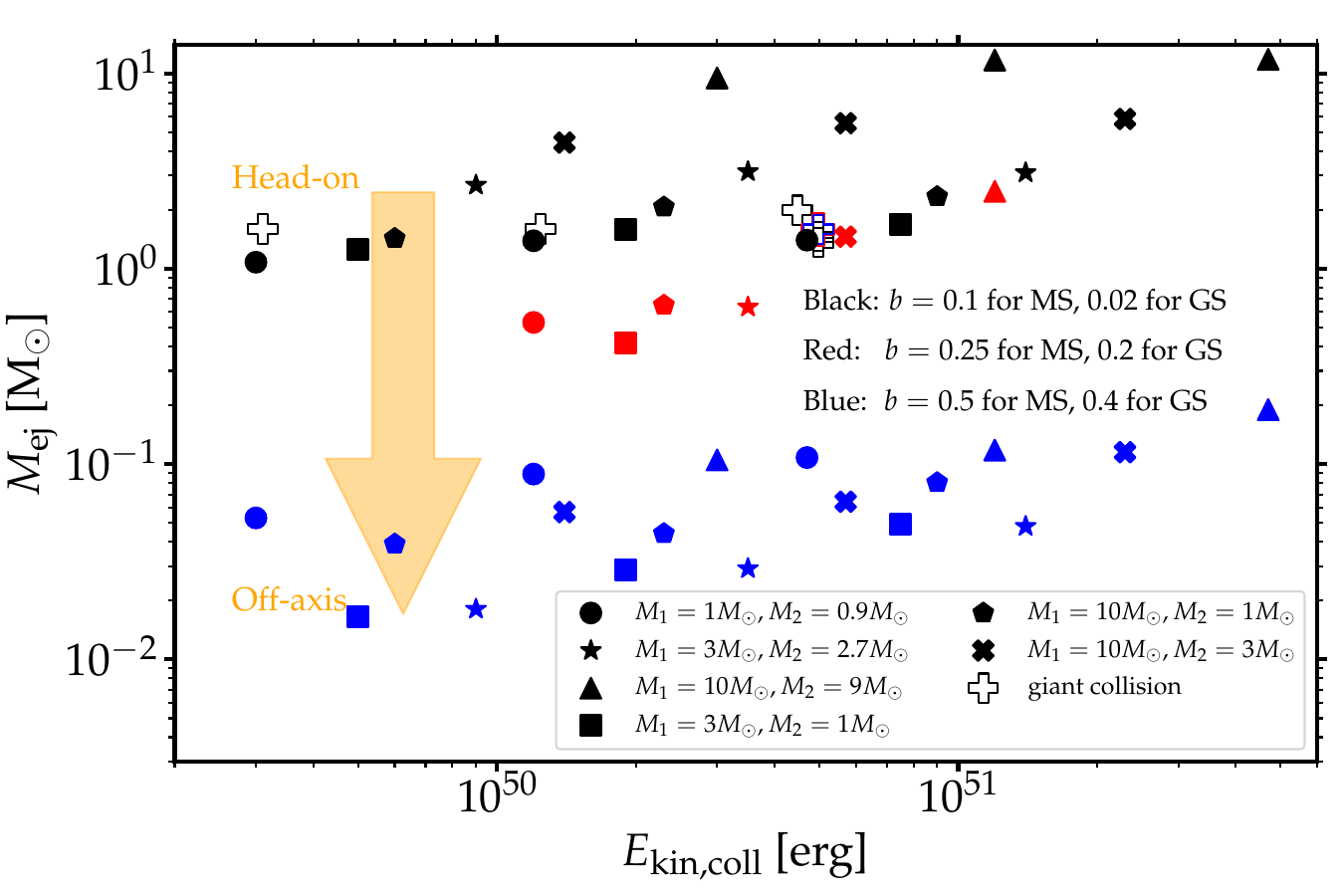}
    \includegraphics[width=8.9cm]{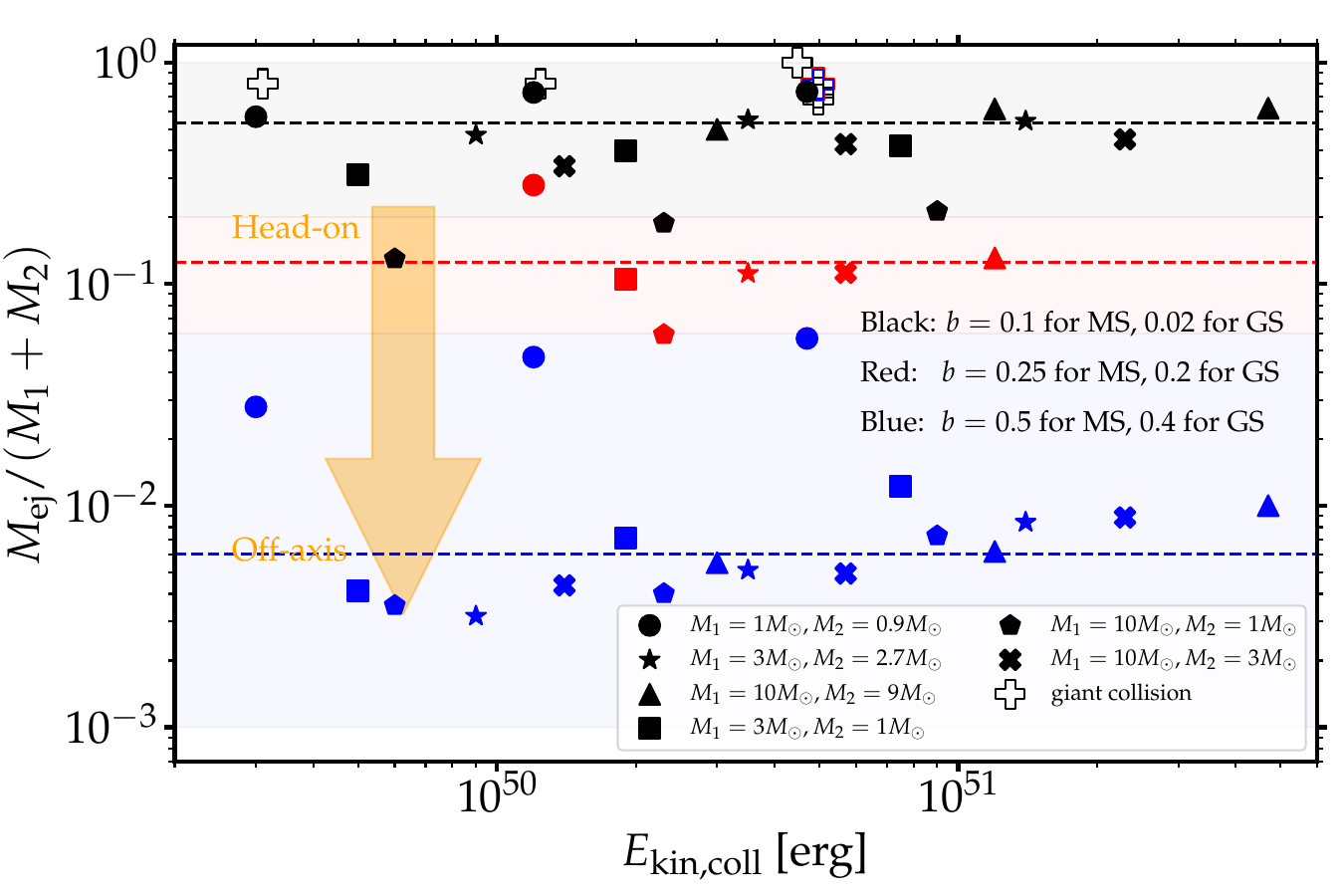}
    \caption{ Ejecta mass $M_{\rm ej}$ (top) and its fraction of the total mass $M_{\rm ej}/(M_{1}+M_{2})$ (bottom), at 0.5\,d after the collision as a function of initial kinetic energy $E_{\rm kin, coll}$. We use different markers to distinguish collisions of different masses, while different colors represent different impact parameters. The horizontal dashed lines in the bottom panel correspond to estimates using the fitting formula in Eq.~\ref{eq:mej}. The empty crosses indicate $M_{\rm ej}$ for giant collisions considered in \citet{Ryu+2024}.}
    \label{fig:correlation_Mej}
\end{figure}

\begin{figure}
    \centering
    \includegraphics[width=8.9cm]{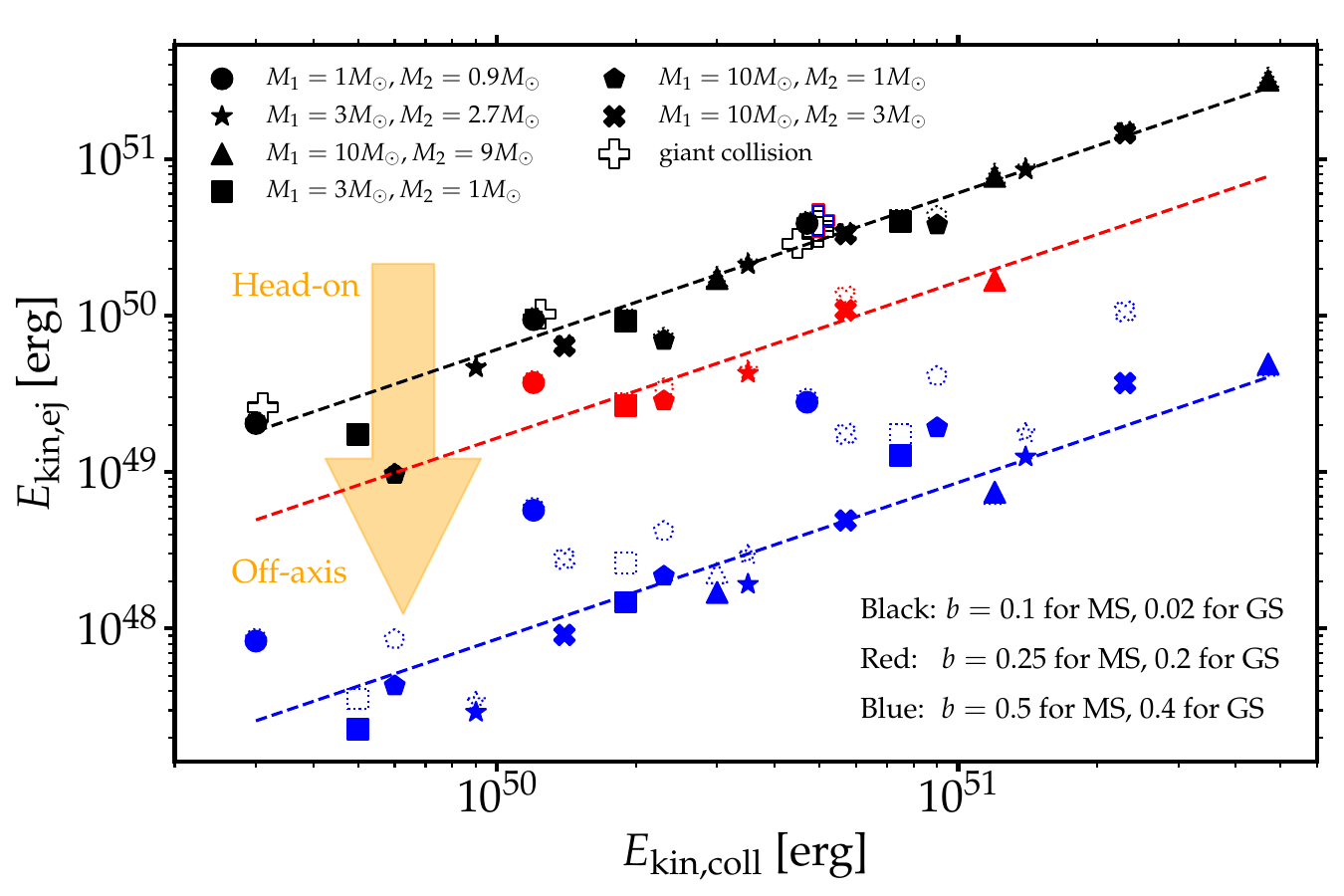}\\
    \includegraphics[width=8.9cm]{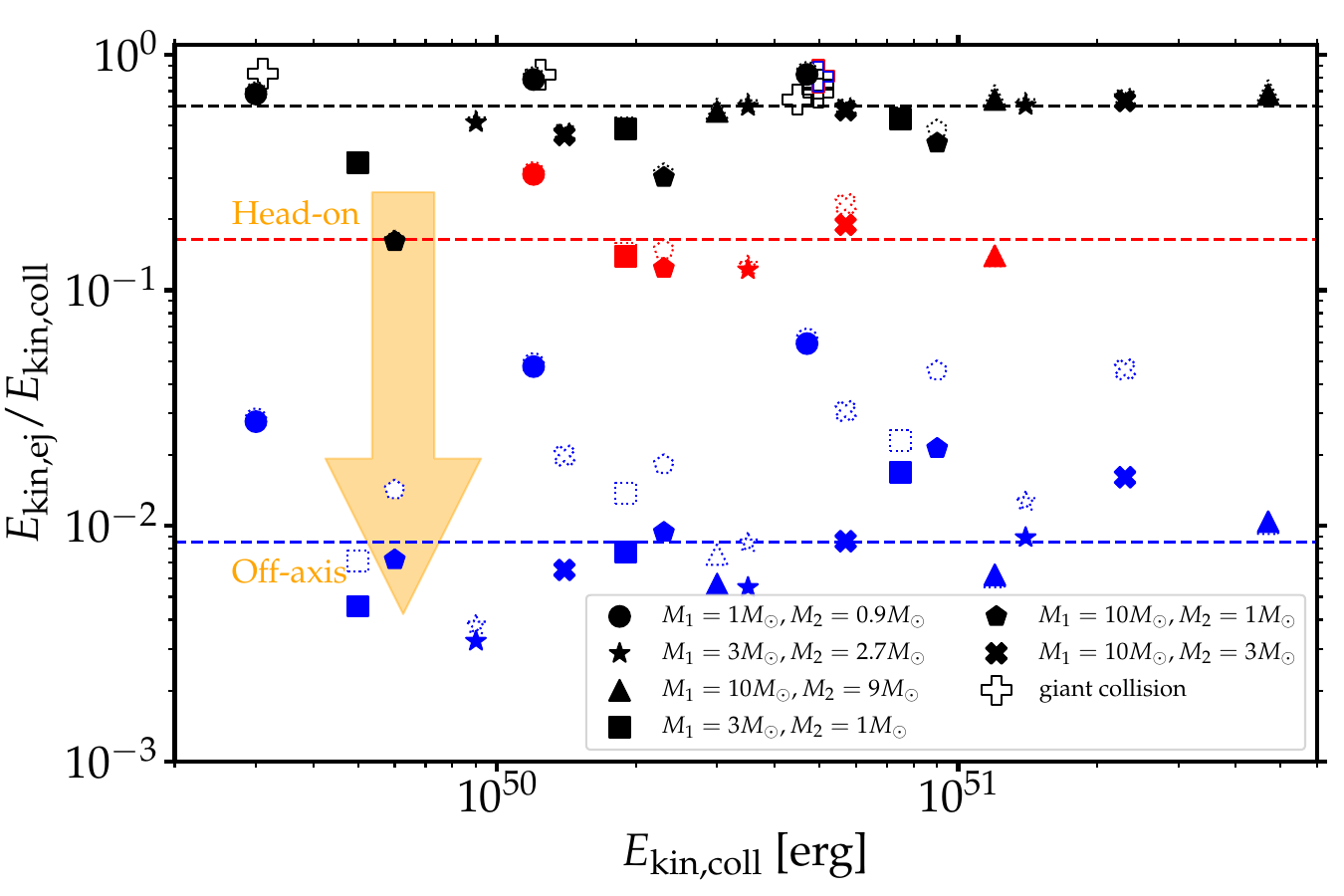}\\
    \includegraphics[width=8.9cm]{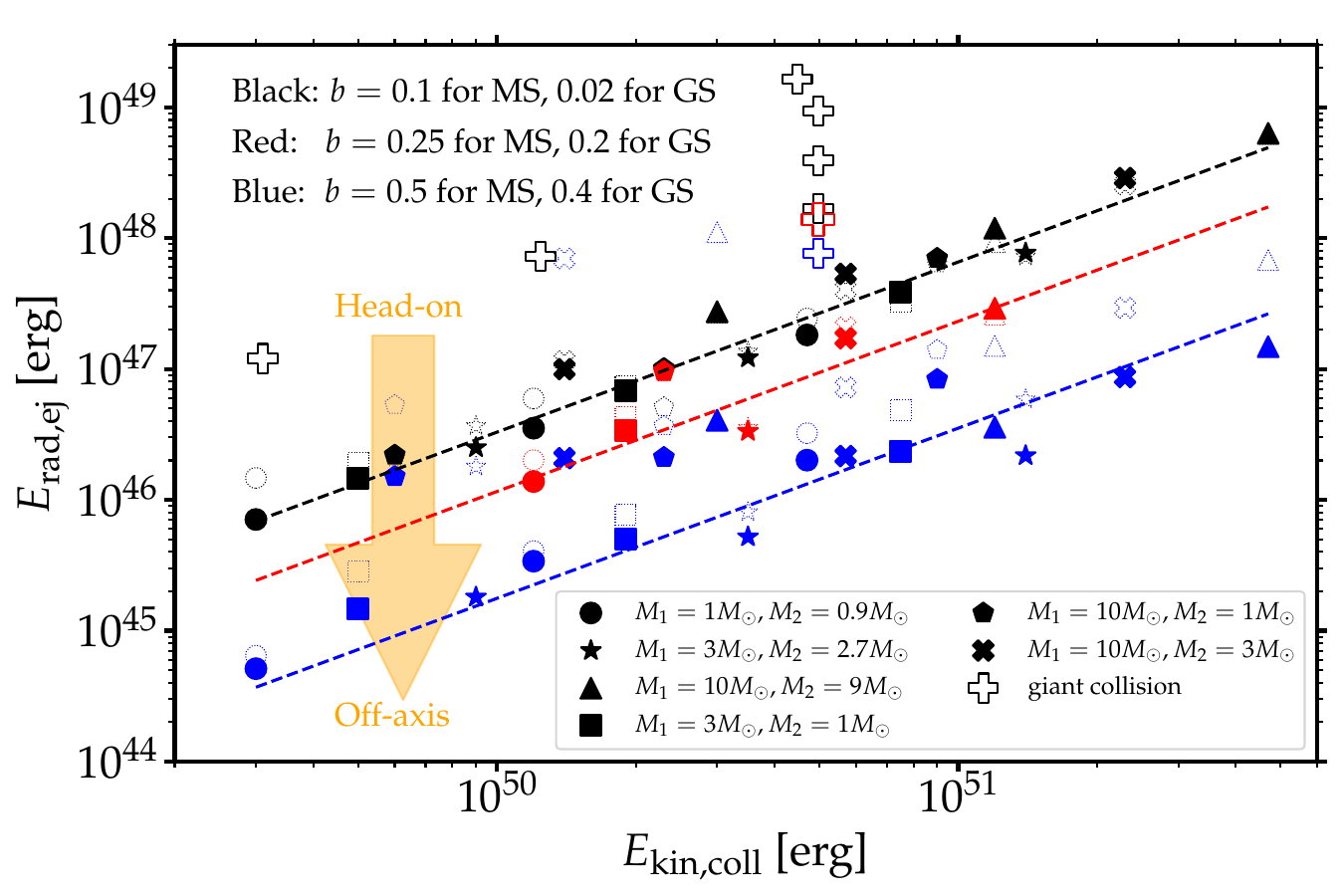}
    \caption{ Same as Fig.~\ref{fig:correlation_Mej}, but for the kinetic energy $E_{\rm kin, ej}$ (top and middle) and radiation energy $E_{\rm rad,ej}$ of the ejecta. The dashed lines indicate estimates by Eq.~\ref{eq:Ekin_Ekinej} for $E_{\rm kin, ej}$ and Eq.~\ref{eq:Ekin_Eradej} for $E_{\rm rad,ej}$. Solid markers represent energy values directly obtained from 3D simulations, while open markers (positioned above or below the solid markers) indicate those from 1D profiles used for radiative-transfer simulations. The empty cross markers depict values for giant collisions considered in \citet{Ryu+2024}. }
    \label{fig:correlation_Ekinej}
\end{figure}

\subsection{Overview of photometric and spectroscopic evolution}\label{subsec:RT_overview}

The radiative properties of main-sequence star collisions are found to be qualitatively similar to those of giant-star collisions and discussed in detail in \citet{Dessart+2024}. We thus summarize the properties obtained for modelset \#\,1 as a representative case and compare the results obtained for different collision configurations. We summarize the results of our radiative transfer calculations in Table~\ref{tab:res} and provide some additional results and illustrations for the full sample in Appendix \S~\ref{sec:appendix}. The main novel aspect of this work is the extent of the model grid and the correlations drawn from it (see \S~\ref{sec:correlation}). 

Figure~\ref{fig:phot_modelset1} illustrates the photometric properties of models in the set \#\,1 characterized by $M_{1}=1\Msol$ and $M_{2}=0.9\Msol$. All models exhibit the same bolometric light curve morphology, with values in the range 10$^{41.5}$ to 10$^{42.5}$\,\ergs\ at 0.5\,d and dropping rapidly ($L_{\rm bol}\propto t^{-2}- t^{-4}$ at 0.5--1\,d) to a plateau at 10$^{40}$ to 10$^{41}$\,\ergs\ at about 1.5\,d. Given the fading rate at 0.5\,d, the peak luminosities probably reach up to 10$^{42}-10^{43}$\ergs. The small radiative energy content (roughly 10$^{46}$ to 10$^{47}$\,erg) and mass ($\sim$\,0.05 to $\sim1\Msol$) limit the luminous, optically-thick phase to about a week (this was not calculated for all models in the set). All these models are less luminous than SN\,1987A at the same epochs. The UV brightness (shown here with the Swift $UVW1$ filter) follows a similar evolution as the bolometric luminosity. The models are brightest in the UV at the earliest times. As a result of expansion and radiation cooling, they rapidly cool and redden and fade in the UV. For all models, the fading rate in the UV is large and roughly in the range 4 to 6\,mag per day at one day. In contrast, the concomitant drop in luminosity and ejecta temperature leads to a quasi-plateau $V$-band light curve, with values in the range of -11 to -14\,mag a few days after the onset of the collision. Here again, all these models are optically fainter than SN\,1987A at the same epochs. Overall, the contrast between models follows directly from the initial conditions in the sense that the models with the greater stored radiative energy at 0.5\,d are brighter (in any filter) or bolometrically more luminous (all models have the same color evolution). Lacking any long-lived power source, the present models become exceedingly faint at late times. Accretion onto the supermassive black hole would dramatically change this picture.

Figure~\ref{fig:montage_spec_ex} illustrates the spectral evolution from UV to optical for model ms1/ms0p9/vrel10/b0p1 (also from set \#\,1) -- the evolution for other models in any modelset is qualitatively similar and also analogous to results obtained previously by \citet{Dessart+2024}. It reveals in more details what is conveyed by the photometric evolution, with the rapid shift of the spectral energy distribution from the UV (brighter than the optical until about 2\,d) to the optical. Initially strong and isolated lines from low-lying levels (e.g., \heiiuv, \ciiiuv, \civuv, or \mgiidoub) co-exist with extended spectral regions blanketed by a multitude of metal lines (in particular those due to Fe\,\two-\four). By two days after the onset of the collision, most of the flux falls in the optical and exhibits the whole series of H\,\one\ Balmer lines, with H$\alpha$ the strongest and broadest of all lines. Strong Fe\,\two\ lines at and below 5000\,\AA\ are present, together with \nad\ or the Ca\,\two\ NIR triplet near 8500\,\AA. At $\sim$\,6\,d, the rapid drop in luminosity shown in Fig.~\ref{fig:phot_modelset1} corresponds spectroscopically to the weakening of the continuum flux and the transition to the nebular phase. As the debris release the small residue of stored radiative energy, the spectrum exhibits a weakening continuum flux and strong, broad, and quasi-symmetric emission lines. This occurs because of the low optical depth of the debris. This apparent strength of lines is only relative since all fluxes are dropping (i.e., the bolometric flux drops but the flux in the lines drops slower than that in the continuum during that transition).

Figure~\ref{fig:montage_spec_at_1d} illustrates the full diversity of spectra from the UV to the optical at one day, including all models from all sets (few models were not computed -- see list in Table~\ref{tab:models}) but grouped according to the value of the collision velocity $V_{\rm rel}$ and impact parameter $b$. To allow the simultaneous assessment of the UV and the optical, we show the quantity $\lambda^2 F_\lambda$. In essentially all models, the spectral energy distribution peaks around 2000\,\AA, although models with a greater $V_{\rm rel}$ and smaller $b$ tend to be brighter in the UV. Most models exhibit broad lines at one day, but these Doppler-broadened lines tend to be narrower in configurations involving smaller collision velocities or higher stellar masses. The spectral properties obtained here are very similar to those of SN\,1987A or Type II supernovae in general (see discussion in \citealt{Dessart+2024}). Therefore, detailed studies of SNe II can be used as a source of information on what is displayed here \citep[e.g.,][]{DH11_pol, d13_sn2p}.

\section{Correlations}\label{sec:correlation}

In this section, we examine possible correlations between ejecta properties and collision parameters, which can be used for building semi-analytic models for destructive collisions, as well as those between observables and collision parameters, which can be used to infer collision parameters from observations. We used the ejecta properties at 0.5\,d and stored in Table~\ref{tab:models} (see \S~\ref{subsec:mje} and ~\ref{subsec:eej}), and the radiative-transfer results at 1\,d and stored in Table~\ref{tab:res} (see \S~\ref{subsec:observables}).
 We also confronted these to their counterparts from simulations of collisions between giant stars (GS) (measured at 1\,d) in \citet{Ryu+2024} and \citet{Dessart+2024} -- these works considered collisions between two identical $1\Msol$ giants with different radii and impact parameters of $b = 0.02-0.4$.\footnote{In those simulations of giant-star collisions, the impact parameter $b$ was defined as the pericenter distance in units of the radius of one star rather than the sum of the radii of the two colliding stars.}

\subsection{Ejecta mass}\label{subsec:mje}

The ejecta mass $M_{\rm ej}$ (see Table~\ref{tab:models}) primarily depends on the impact parameter $b$, with larger $M_{\rm ej}$ resulting from more head-on collisions. It also exhibits a weak positive correlation with the initial kinetic energy $E_{\rm kin}$. This trend is illustrated in Fig.~\ref{fig:correlation_Mej}. As shown in the top panel, more head-on collisions produce larger $M_{\rm ej}$. $M_{\rm ej}$ falls into three distinct ranges depending on $b$: $M_{\rm ej}\gtrsim1\Msol$ for $b=0.1$, $0.1\Msol\lesssim M_{\rm ej}\lesssim 1\Msol$ for $b=0.25$, and $M_{\rm ej}\lesssim 0.1\Msol$ for $b=0.5$. However, the distribution of $M_{\rm ej}$ is rather horizontal, indicating its weak dependence on $E_{\rm kin}$.

\begin{figure*}
    \centering
    \includegraphics[width=8.9cm]{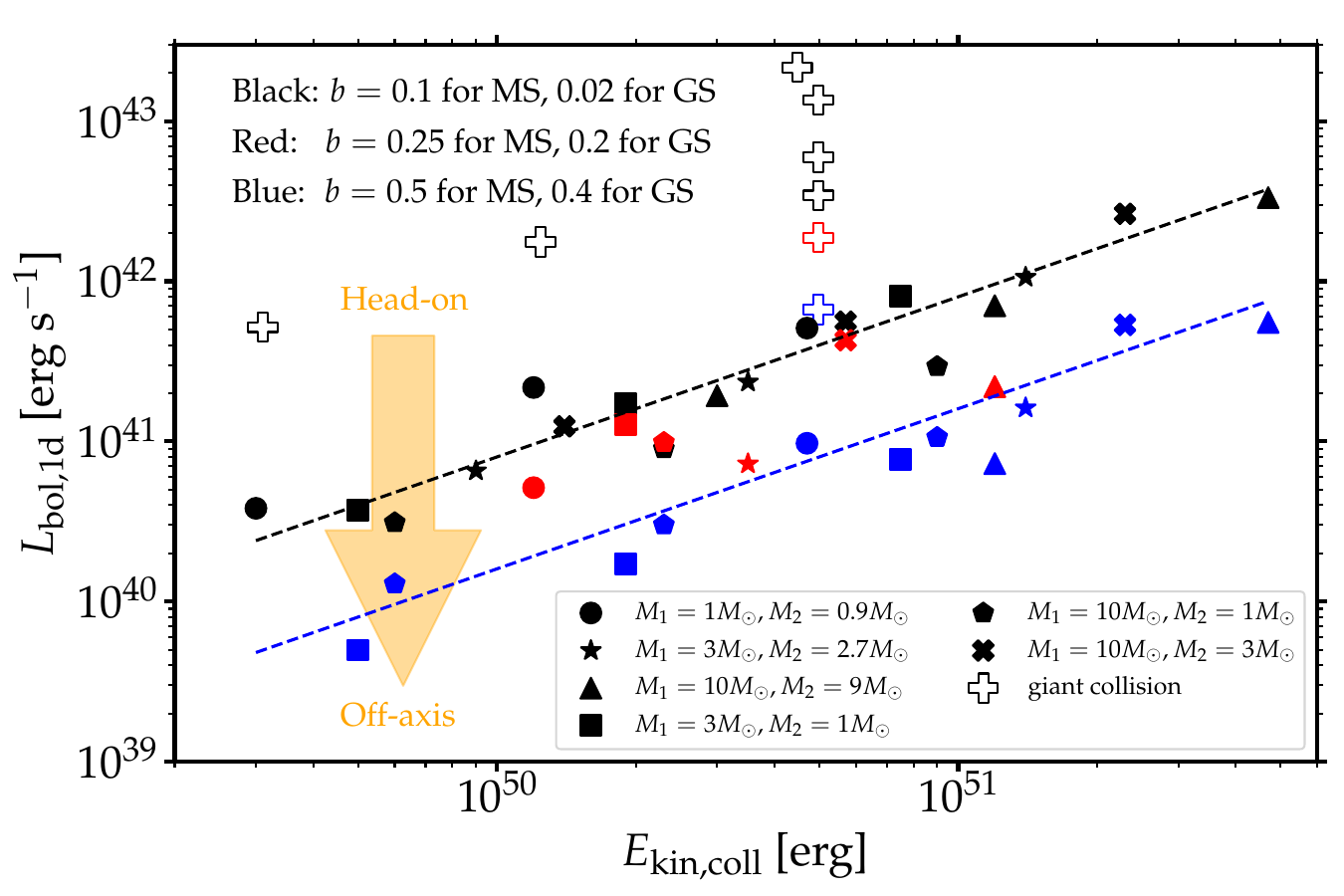}
    \includegraphics[width=8.9cm]{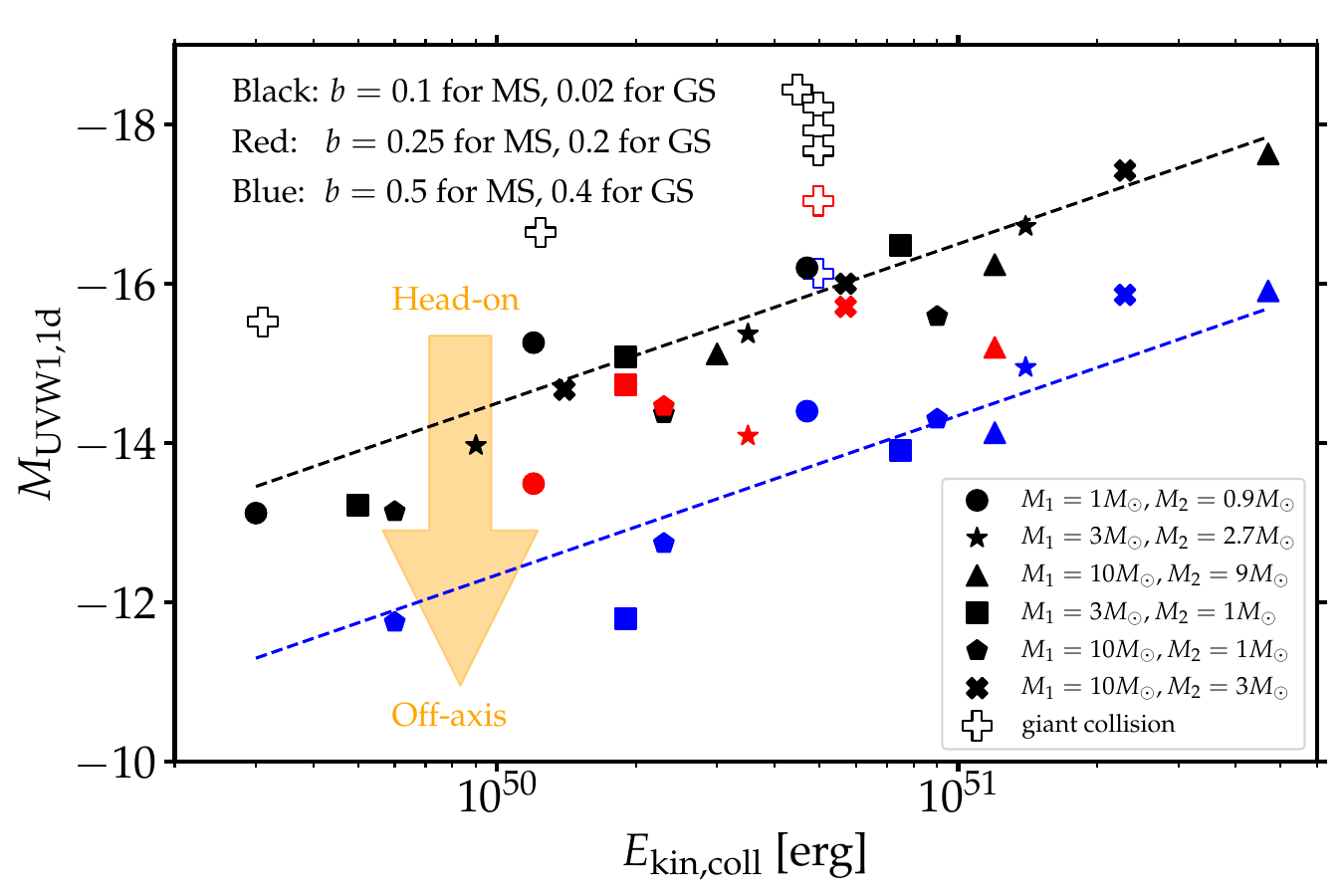}\\
    \includegraphics[width=8.9cm]{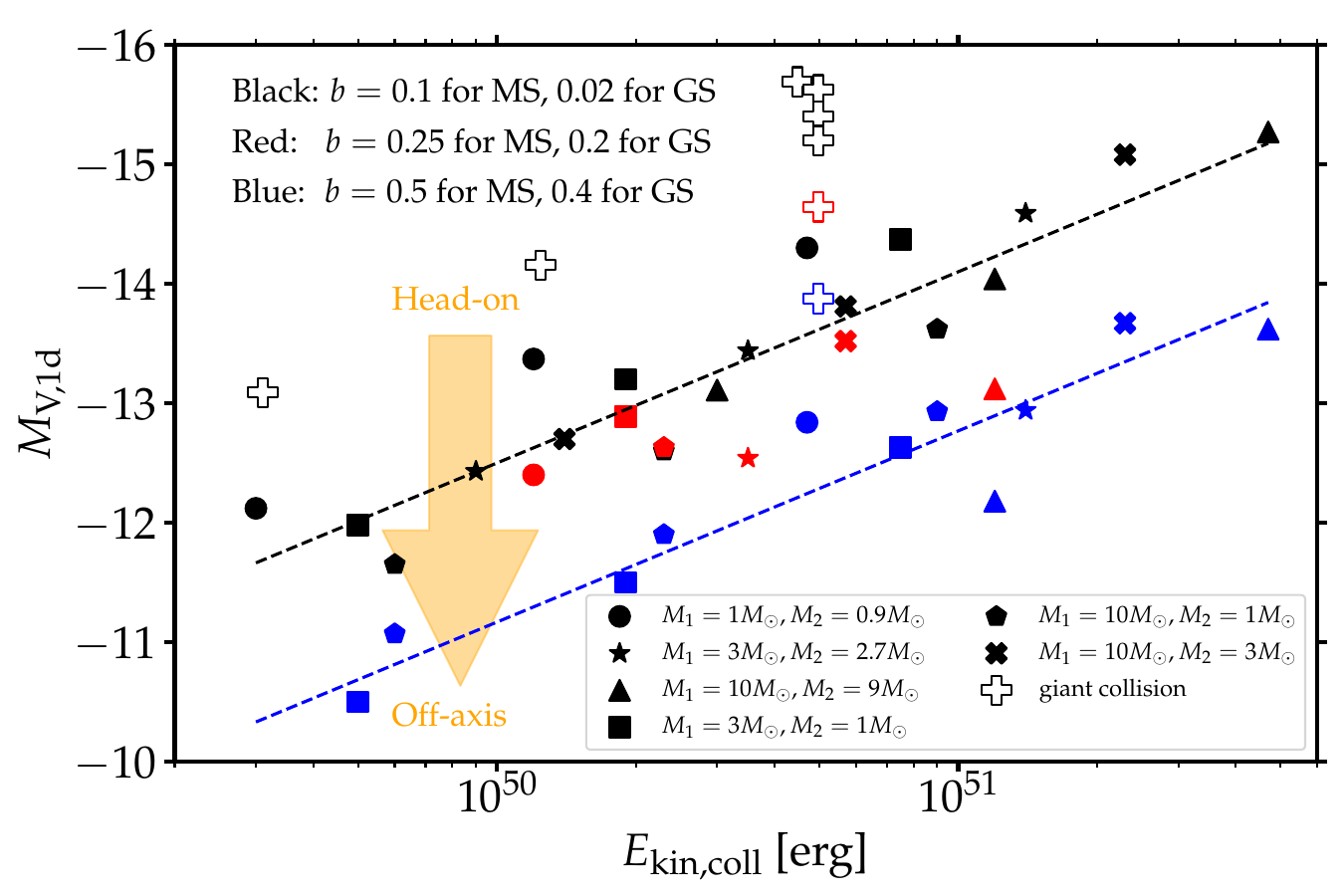}
    \includegraphics[width=8.9cm]{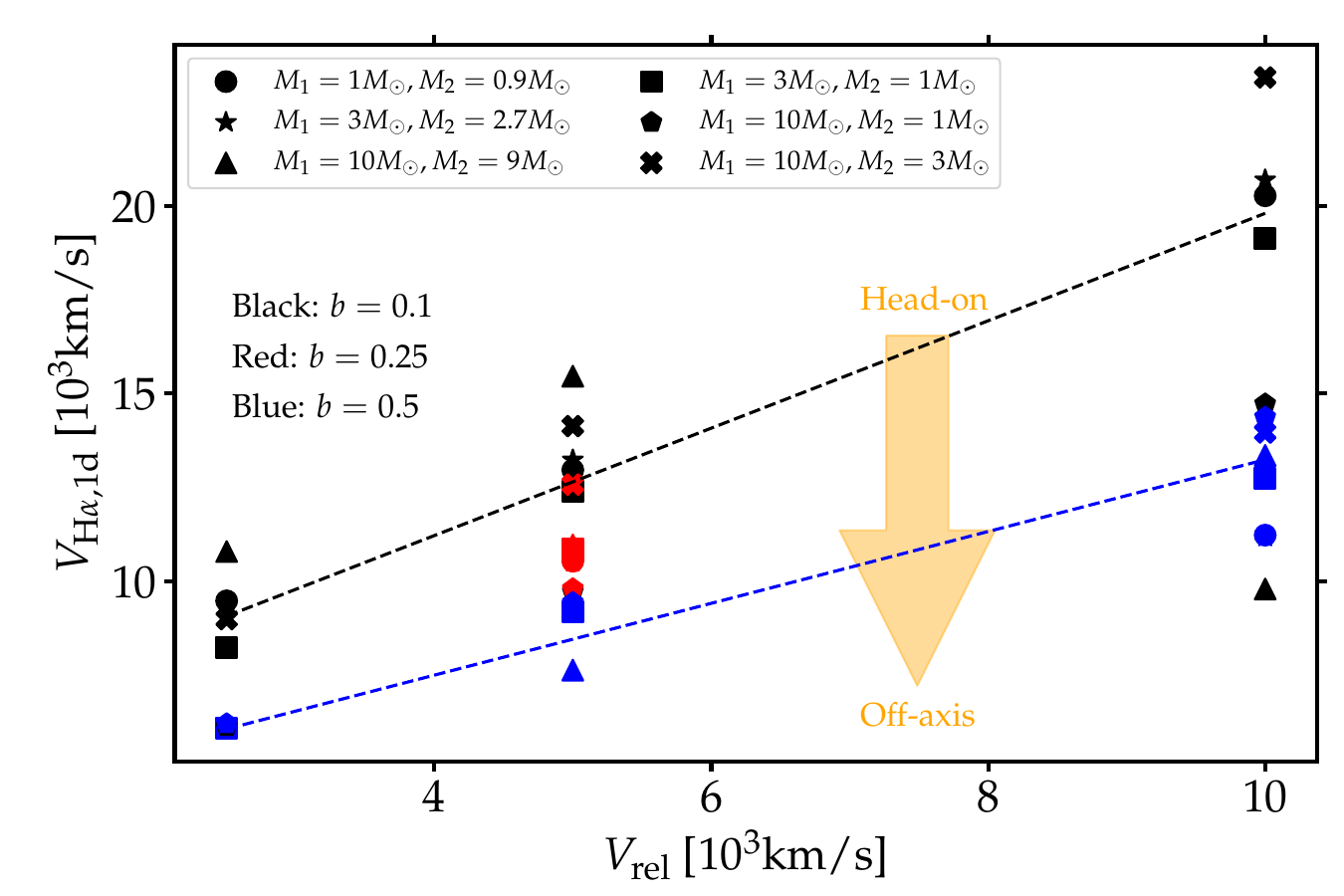}
    \caption{ Bolometric luminosity $L_{\rm bol,1d}$ (top-left) and $UVW1$-band magnitudes $M_{\rm UVW1,1d}$ (top-right), $V$-band magnitudes $M_{\rm V,1d}$ (bottom-left) as a function of initial kinetic energy $E_{\rm kin,coll}$ and the Doppler velocity $V_{\rm H\alpha,1d}$ (bottom-right) at maximum absorption measured in the H$\alpha$ line profile as a function of $V_{\rm rel}$. All these values are measured at 1\,d since collision. We distinguish collisions with different masses using different markers and those with different impact parameters using different colors. The black and blue diagonal lines show the relations given by the fitting formulae in Equations~\ref{eq:fitting1}$-$\ref{eq:fitting4} for $b=0.1$ and $0.5$, respectively. In the top panels and bottom-left panel, the empty crosses represent values for giant collisions considered in \citet{Ryu+2024}.}
    \label{fig:correlation}
\end{figure*}
These three mass ranges can also be expressed in terms of the fractional mass $M_{\rm ej}/(M_{1}+M_{2})$, as shown in the bottom panel of Fig.~\ref{fig:correlation_Mej}. Generally, $M_{\rm ej}/(M_{1}+M_{2})\gtrsim 0.1$ for $b=0.1$, $\simeq 0.1$ for $b=0.25$, and $\lesssim 0.05$ for $b=0.5$. The nearly constant mass fraction for each $b$ suggests that $M_{\rm ej}$ is not directly correlated with $M_{1}+M_{2}$, although $M_{1}+M_{2}$ sets an upper limit on $M_{\rm ej}$, i.e., $M_{\rm ej}<M_{1}+M_{2}$. This plot highlights the effect of the central concentration of the colliding stars. Notably, for the collisions between $1\Msol$ and $0.9\Msol$ stars with $b=0.5$ (the three blue circles at $0.03\lesssim M_{\rm ej}/(M_{1}+M_{2})\lesssim 0.05$), the mass fractions are particularly high compared to other $b=0.5$ cases. We attribute this to how much a He-rich core develops within the colliding stars. As a He-rich core forms, the star becomes increasingly resistant to complete destruction. In $1\Msol$ and $0.9\Msol$ stars, the He-rich core constitutes a noticeably small fraction compared to more massive stars (see Fig.~\ref{fig:initialprofile}). 

As a more extreme case than the relatively massive main-sequence stars with large He-rich cores, collisions between giants 
\citep{Ryu+2024,Dessart+2024} do not exhibit a correlation of $M_{\rm ej}$ with $b$ or $E_{\rm kin,coll}$, where $M_{\rm ej}/(M_{1}+M_{2})\gtrsim 80\%$ for $b=0.02-0.4$ (empty crosses in Fig.~\ref{fig:correlation_Mej}), roughly corresponding to the envelope mass of the giant stars. The different correlations are directly related to the structural differences between main-sequence and giant stars. Because almost entire main-sequence star is subject to being shocked and launched as ejecta upon collision, the ejecta mass generally increases as $b$ decreases.  While the central part of the star is more resistant to destruction than the envelope is, the density contrast is usually not large enough to make a significant difference. However, in giant stars, the He-core is highly resistance to destruction due to their compactness ($\simeq 10^{-2}\Rsol$ for $1\Msol$ giants; see Fig.~2 in \citealt{Ryu+2024}) and high binding energy ($\gtrsim 10^{49}$ erg for $1\Msol$ giants). For example, collisions that can result in even partial destruction of the cores in two $1\Msol$ giants occur only when $b\lesssim O(10^{-2})$ and $V_{\rm rel}\gtrsim 5000$ \kms. Consequently, most of the ejecta mass produced in giant collisions originally belonged to the envelope. Furthermore, because the envelope is relatively loosely bound, it can be easily stripped away by the collision, resulting in the weak dependence on $b$. The large size of the envelope in giants also leads to a more efficient conversion of kinetic energy to radiation energy, producing larger luminosities, which will be shown in the following section.

 The dependence of $M_{\rm ej}/(M_{1}+M_{2})$ on $b$ for collisions between main-sequence stars can be fitted with the formula
\begin{align}\label{eq:mej}
    \frac{M_{\rm ej}}{M_{1}+M_{2}} = \exp[-0.63\left(\frac{b}{0.1}\right)^{1.3}],
\end{align}
with a $1\sigma$ uncertainty of $\simeq 40-50\%$\footnote{All fitting coefficients presented in the paper were obtained using the curve\_fit function from the SciPy package \citep{2020SciPy-NMeth}.}.

\subsection{Ejecta energy}\label{subsec:eej}
The ejecta kinetic energy $E_{\rm kin,ej}$ exhibits a similar trend with $b$ as $M_{\rm ej}$, yielding a larger $E_{\rm kin,ej}$ for more head-on collisions. However, its correlation with $E_{\rm kin}$ is stronger, following $E_{\rm kin,ej}\propto E_{\rm kin}$, as depicted in the top panel of Fig.~\ref{fig:correlation_Ekinej}. As a consequence of the linearity between $E_{\rm kin, ej}$ and $E_{\rm kin}$, their ratio is nearly independent of $E_{\rm kin, ej}$, which is shown in the middle panel. In most of our models, $E_{\rm kin,ej}/E_{\rm kin}\simeq0.6$ for $b=0.1$, $\simeq 0.1$ for $b=0.25$, and $\simeq 0.01$ for $b=0.5$. Accounting for the dependence on $b$, the correlation between $E_{\rm kin, ej}$ and $E_{\rm kin}$ for $0.1\lesssim b \lesssim 0.5$ can be expressed as,
\begin{align}\label{eq:Ekin_Ekinej}
    E_{\rm kin, ej} = E_{\rm kin} \exp[-0.5\left(\frac{b}{0.1}\right)^{1.4}],
\end{align}
with a $1\sigma$ uncertainty of $\simeq 25\%$ (for $b\lesssim 0.25$) and $\simeq 60\%$  (for $b= 0.5$). Eq.~\ref{eq:Ekin_Ekinej} may not satisfactorily describe the correlation outside the considered range of $b$, in particular, for $b<0.1$. For $b>0.5$, we expect Eq.~\ref{eq:Ekin_Ekinej} to describe the continuous decay of $E_{\rm kin, ej}$ beyond $b=0.5$ down to zero. However, it also predicts that $E_{\rm kin, ej} \longrightarrow E_{\rm kin}$ as $b\longrightarrow0$. While this remains reasonable, our models cannot confirm it. 

The radiation energy $E_{\rm rad,ej}$ trapped in the unbound ejecta shares similar correlations with $E_{\rm kin}$ as $E_{\rm kin,ej}$ -- namely, larger $E_{\rm rad,ej}$ for larger $E_{\rm kin}$ and smaller $b$. However, $E_{\rm rad,ej}$ has a steeper correlation with $E_{\rm kin}$, i.e., $E_{\rm rad,ej}\propto E_{\rm kin}^{1.3}$. The overall dependence of $E_{\rm rad,ej}$ on $E_{\rm kin}$ is expressed as,
\begin{align}\label{eq:Ekin_Eradej}
    E_{\rm rad, ej} = 6\times 10^{46} ~{\rm erg} ~\left(\frac{E_{\rm kin}}{10^{50}{\rm erg}}\right)^{1.3} \exp[-0.6\left(\frac{b}{0.1}\right)^{1.1}]. 
\end{align}
with a $1\sigma$ uncertainty of $30-60\%$.

The fraction of $E_{\rm kin,coll}$ converted into $E_{\rm kin,ej}$ for giant collisions (empty cross markers in the top and middle panels of Fig.~\ref{fig:correlation_Ekinej}) is consistently high, independent of $b$, i.e., $E_{\rm kin,ej}/E_{\rm kin,coll}0.5-0.7$, which corresponds to the highest fraction for main-sequence collisions with $b=0.1$. Its dependence of $E_{\rm kin,ej}$ on $E_{\rm kin,coll}$ is reasonably described by Eq.~\ref{eq:Ekin_Eradej} for $b=0.1$. On the other hand, $E_{\rm rad,ej}$ for giant collisions is up to two orders of magnitude larger than that for main-sequence collisions at a given $E_{\rm kin,coll}$. This indicates a greater conversion of $E_{\rm kin,coll}$ to $E_{\rm rad,ej}$ and brighter events in giant collisions (see \citealt{Dessart+2024}).

\subsection{Ejecta observables}\label{subsec:observables}
Figure~\ref{fig:correlation} shows three strong correlations with observables at 1\,d. As illustrated in the top panels and bottom-left panel, the bolometric luminosity at 1d $L_{\rm bol,1d}$ ranges from $5\times 10^{39}$\,\ergs to $3\times 10^{42}$\,\ergs. More energetic collisions with smaller impact parameters (so more head-on) generate brighter events ($L_{\rm bol,1d}\propto E_{\rm kin}b^{-1}$). This is expected because the collision kinetic energy sets the energy budget available for conversion into the radiation energy of the ejecta, while the impact parameter determines the amount of mass that can be shocked at the collision, so the energy conversion efficiency. However, the brightness of the events appears to show no correlations with either the stellar masses or the relative velocity when considered individually. The correlations with the original kinetic energy suggest that it may be possible to infer the kinetic energy from the observables. However, the lack of correlations with the masses and the relative velocity indicates that it would be challenging to infer them separately based solely on the luminosity or magnitudes.

Note that the correlation between the event's brightness ($L_{\rm bol,1d}$, $M_{\rm UVW1,1d}$ and $M_{\rm V,1d}$) and $E_{\rm kin}$ for $b=0.25$ appears noticeably weaker than that for the other two cases, $b=0.1$ and $0.5$. In other words, the scatter for $b=0.25$ in the top panels and bottom-left panel is distributed more horizontally. We suspect that this weak correlation is largely due to the small sample size for $b=0.25$ (only six cases, compared to 17 cases for $b=0.1$ and 11 cases for $b=0.5$).

In comparison, giant collisions, with their high values of $E_{\rm kin,ej}/E_{\rm kin,coll}$, produce significantly brighter transients. As shown in the top panels and bottom-left panels of Fig.~\ref{fig:correlation} using empty cross markers, $L_{\rm bol,1d}$ for giant collisions is higher by up to two orders of magnitude and both $M_{\rm UVW1,1d}$ and $M_{\rm V, 1d}$ are brighter by a factor of 2--3\,mag.

However, the positive correlation between the Doppler velocity at maximum absorption measured in the H$\alpha$ line profile and the relative collisional velocity, shown in the bottom panel of Fig.~\ref{fig:correlation}, may help break the degeneracy between the masses and relative velocity. This correlation, $V_{\rm H\alpha,1d}\propto V_{\rm rel}$, can be understood by the fact that the debris expansion speed scales with the initial relative velocity, as has been also found in collisions involving giants \citep{Ryu+2024}. Consequently, accurate measurements of the H$\alpha$ velocity, along with either the luminosity or magnitudes, may place valuable constraints on collision parameters, particularly the initial relative velocity and the masses (e.g., the reduced mass if the center-of-mass of the two colliding stars was initially negligible compared to the relative velocity). This implies that  spectra are essential for parameter inference. 

The fitting formulae that describe the correlations are
\begin{align}\label{eq:fitting1}
    L_{\rm bol,1d} &= 8\times10^{40} {\rm erg} \left(\frac{b}{0.1}\right)^{-1} \left(\frac{E_{\rm kin}}{10^{50}{\rm erg}}\right),\\\label{eq:fitting2}
    M_{\rm UVW1,1d} &= -2 \log_{10}\left(\frac{E_{\rm kin}}{10^{50}{\rm erg}}\right) - 14.5\left(\frac{b}{0.1}\right)^{-0.1},\\\label{eq:fitting3}
    M_{\rm V,1d} &=-1.6 \log_{10}\left(\frac{E_{\rm kin}}{10^{50}{\rm erg}}\right) - 12.5\left(\frac{b}{0.1}\right)^{-0.07},\\\label{eq:fitting4}
    V_{\rm H\alpha,1d} &= 5500 \left(\frac{b}{0.1}\right)^{-0.25}\left[1.3\left(\frac{V_{\rm rel}}{5\,000{\rm km s}^{-1}}\right) + 1\right].
\end{align}
The $1\sigma$ uncertainty is $36\%$,  $2-3\%$, and $10\%$ for $L_{\rm bol,1d}$, both of the magnitudes, and $V_{\rm H\alpha,1d}$, respectively.

\section{Summary and conclusion}\label{sec:conclusion}
In this paper, we investigated the properties of ejecta produced in high-velocity collisions between terminal-age main-sequence stars in galactic nuclei
and their observables. We adopt the same methodology as in \citet{Dessart+2024}, combining 3D moving-mesh hydrodynamics and 1D radiation-transfer simulations. We consider a wide range of parameters for collisions, namely, stellar masses of $0.9-10\Msol$, mass ratios of $0.1-0.9$, impact parameters of $(0.1 - 0.5)(R_{1} + R_{2})$, and initial relative velocities of $2\,500-10\,000$ km s$^{-1}$. 

When two main-sequence stars collide at high velocities, strong shocks form, driving quasi-spherically expanding ejecta. In most of our models, the collisions leave behind single He-rich remnants. If we use stars are at an earlier evolutionary stage (e.g., bona fide main-sequence stars), they are more likely to be completely destroyed because their core is less gravitationally bound.

Because of the large stored radiation energy, the debris from high-velocity stellar collisions release copious amounts of electromagnetic radiation. Their bolometric luminosities range from $5\times 10^{39}$\ergs to $3\times 10^{42}$\ergs at 0.5\,d after the collision, during which the flux falls in the UV. If the rate of fading at 0.5--1\,d ($L_{\rm bol}\propto t^{-2}-t^{-4}$) holds at earlier times, the peak luminosity at the moment of collision might be significantly higher (potentially by a factor of $10$ or $2$ magnitude brighter), possibly reaching a brightness level comparable to tidal disruption events \citep{Gezari+2021, Wevers+2023}. However, the peak duration is typically short on the order of $O(0.1)$\,d, so transients surveys with cadence less than a day would be able to observe these events at their luminosity peak. Even after a photometric discovery, prompt spectroscopic follow-up is essential to confirm the nature of the events. Within a week, the luminosity drops by more than an order of magnitude and levels off on a plateau, at which point the spectrum has shifted to the optical.

The observables of these collisions are primarily determined by the mass and energy content of the debris at the onset of their expansion. The ejecta mass depends mostly on the impact parameter $b$, with the fractional mass $M_{\rm ej}/(M_{1}+M_{2})$ increasing as $b$ decreases ($\propto \exp[-b^{1.3}]$ in Eq.~\ref{eq:mej}). On the other hand, the energy content of ejecta and the transient brightness (e.g., luminosity and magnitude) generally correlate positively with $b$ and the initial kinetic energy $E_{\rm kin}$. For example, at one day, the radiation energy of the expanding ejecta scales as $E_{\rm rad,ej}\propto E_{\rm kin}^{1.3}\exp[-b^{1.1}]$ (Eq.~\ref{eq:Ekin_Eradej}) and the bolometric luminosity $L_{\rm d,1d}\propto E_{\rm kin}b^{-1}$ (Eq.~\ref{eq:fitting1}). The correlation with $E_{\rm kin}$, rather than the masses of the colliding stars and initial relative velocity individually, suggests a degeneracy between the stellar masses and the relative velocity when inferred from observations. This degeneracy can be broken with velocity measurements done on H\,\one\,Balmer lines like H$\alpha$, as those lines depend most strongly on the relative velocity and not on the stellar masses. 

In conclusion, our detailed hydrodynamics and radiative transfer calculations suggest that transients generated by high-velocity collisions in galactic nuclei are observable nuclear transients. This is expected for collisions between main-sequence stars (this work) as well as between giant stars \citep{Ryu+2024,Dessart+2024}. However, some of our assumptions require further investigation. For example, the quasi-spherical expansion found in our hydrodynamics simulations would be disrupted near the supermassive black hole, where tidal forces are strong enough to deform the expanding debris. In such cases, observable properties would depend on viewing angle. Additionally, the supermassive black hole would influence the debris on a longer times scale, extending the optically-thick and luminous phase, which must be thoroughly examined to develop a complete picture of transients associated with high-velocity collisions.

\section{Data availability}
     All radiative-transfer simulations presented in this work are available at  \url{https://zenodo.org/communities/snrt}.

\begin{acknowledgements}
We thank the anonymous referee for kindly reviewing the manuscript and providing constructive comments. This research project was conducted using computational resources (and/or scientific computing services) at the Max-Planck Computing \& Data Facility. The authors gratefully acknowledge the scientific support and HPC resources provided by the Erlangen National High Performance Computing Center (NHR@FAU) of the Friedrich-Alexander-Universität Erlangen-Nürnberg (FAU) under the NHR projects b166ea10 and b222dd10. NHR funding is provided by federal and Bavarian state authorities. NHR@FAU hardware is partially funded by the German Research Foundation (DFG) – 440719683. In addition, some of the simulations were performed on the national supercomputer Hawk at the High Performance Computing Center Stuttgart (HLRS) under the grant number 44232. This work was supported in France by the ``Programme National Hautes Energies'' of CNRS/INSU co-funded by CEA and CNES. This work was granted access to the HPC resources of TGCC under the allocation 2023 -- A0150410554 on Irene-Rome made by GENCI, France. 
\end{acknowledgements}

\bibliographystyle{aa} 
\bibliography{./biblio} 

\newpage
\begin{appendix}
\onecolumn
\section{Model parameters}
Tab.~\ref{tab:models} summarizes the model parameters for the collision configurations and corresponding ejecta characteristics. 

\begin{table*}[!htbp]
\caption{Parameters for the collision configurations and corresponding ejecta characteristics.}
\label{tab:models}
\centering
\begin{tabular}{c | c c c c c c c c c c c}
\hline
\hline
Set & $M_{1}$ & $M_{2}$  & $V_{\rm rel}$ & $b$ &  $E_{\rm kin,coll}$   & \mej\ & \multicolumn{2}{c}{$E_{\rm kin,ej}$ (3D | 1D)} & \multicolumn{2}{c}{$E_{\rm rad,ej}$  (3D | 1D)} & RT \\
   & $[\Msol]$  &  $[\Msol]$ & [$10^{3}$\,\kms]  & -  & [erg] & $[\Msol]$ & \multicolumn{2}{c}{[erg]} & \multicolumn{2}{c}{[erg]}  &     \\
\hline
\multirow{7}{*}{1} &   1          &     0.9      & 5       &   0.1   & 1.2(50)   &  1.39(0) & 9.40(49)  &  9.59(49) & 3.53(46) &  5.95(46)  &  Yes  \\
                   &   1         &     0.9        &   5     &   0.25    & 1.2(50)  &  5.31(-1) & 3.72(49)  &  3.79(49) & 1.38(46)   &  2.01(46)  &  Yes  \\
                   &   1         &     0.9        &  5      &   0.5     & 1.2(50)  &  8.89(-2) & 5.69(48)  &  5.86(48) &  3.40(45)   &  4.06(45)  &   No    \\
                   &   1         &     0.9        &  2.5    &   0.1     & 3.0(49)  &  1.08(0)  & 2.04(49)  &  2.07(49) &  7.08(45)   &  1.47(46)   &  Yes  \\
                   &   1         &     0.9        &  2.5    &   0.5     & 3.0(49)  &  5.30(-2) & 8.31(47)  &  8.54(47) &  5.13(44)   &  6.45(44)  &   No    \\
                   &   1         &     0.9        &  10     &   0.1     & 4.7(50)  &  1.40(0)  & 3.87(50)  &  3.95(50) &  1.82(47)   &  2.43(47)   &  Yes  \\
                   &   1         &     0.9        &  10     &   0.5     & 4.7(50)  &  1.08(-1) & 2.79(49)  &  2.92(49) &  2.01(46)   &  3.23(46)  &  Yes  \\
\hline                                                
\multirow{7}{*}{2} &   3         &     2.7      &  5      &   0.1     &  3.5(50)&   3.14(0) & 2.11(50) &   2.17(50) &  1.22(47) &  1.36(47)  &  Yes  \\
                   &   3         &     2.7      &  5      &   0.25    &  3.5(50)&   6.37(-1) &  4.28(49) &  4.47(49)  &  3.34(46) & 3.56(46) &  Yes  \\
                   &   3         &     2.7      &  5      &   0.5     &  3.5(50)&   2.92(-2) &  1.92(48) & 3.85(48)  &  5.22(45) &  8.08(45)  &   No    \\
                   &   3         &     2.7      &  2.5    &   0.1     &  9.0(49)&   2.68(0) &  4.63(49)  &  4.72(49) &  2.51(46) &  3.67(46)  &  Yes  \\
                   &   3         &     2.7      &  2.5    &   0.5     &  9.0(49)&   1.81(-2) &  2.92(47) & 3.42(47) &  1.81(45)  &  1.80(46) &   No    \\
                   &   3         &     2.7      &  10     &   0.1     &  1.4(51)&   3.10(0) &  8.52(50)  &  8.76(50) & 7.70(47) & 7.23(47)  &  Yes  \\
                   &   3         &     2.7      &  10     &   0.5     &  1.4(51)&   4.80(-2) & 1.25(49) &  1.78(49) &  2.18(46) & 5.87(46) &  Yes  \\
\hline                                                
\multirow{7}{*}{3} & 10  &     9       &  5      &   0.1     & 1.2(51)    &   1.17(1)  &  7.73(50) &  8.08(50) & 1.19(48) &    9.31(47)  &  Yes \\
       &   10         &     9       &  5      &   0.25    & 1.2(51)    &   2.49(0)  &   1.68(50) &  1.68(50)  &  2.91(47) &    2.61(47)  &  Yes \\
       &   10         &     9       &  5      &   0.5     & 1.2(51)    &   1.18(-1)  &  7.44(48) &  7.29(48)  &  3.59(46) &    1.50(47) &  Yes \\
       &   10         &     9       &  2.5    &   0.1     & 3.0(50)    &   9.47(0)  &    1.72(50) & 1.77(50)  & 2.74(47) &    2.70(47)  &  Yes \\
       &   10         &     9       &  2.5    &   0.5     & 3.0(50)    &   1.05(-1)  &   1.70(48) & 2.27(48)  &  4.06(46) &    1.11(48) &   No   \\
       &   10         &     9       &  10     &   0.1     & 4.7(51)    &   1.18(1)  &  3.18(51) &  3.32(51) &   6.32(48)  &    6.26(48)  &  Yes \\
       &   10         &     9       &  10     &   0.5     & 4.7(51)    &   1.90(-1)  &  4.90(49) &  4.81(49) & 1.48(47) &    6.79(47) &  Yes \\
\hline                                                
\multirow{7}{*}{4} & 3 &     1        &  5      &   0.1     &  1.9(50) &   1.59(0) &   9.20(49) &  9.29(49) & 6.81(46)  &    7.39(46)  & Yes  \\
       &   3         &     1        &  5      &   0.25    &  1.9(50)   &   4.19(-1) &  2.64(49) &    2.72(49) & 3.40(46) &    4.28(46) & Yes  \\
       &   3         &     1        &  5      &   0.5     &  1.9(50) &   2.86(-2) &   1.57(48) & 2.62(48) &   5.05(45) &  7.71(45) & Yes  \\
       &   3         &     1        &  2.5    &   0.1     &  5.0(49) &   1.25(0) &   1.73(49) & 1.74(49) & 1.46(46) &    1.91(46)  & Yes  \\
       &   3         &     1        &  2.5    &   0.5     &  5.0(49) &   1.65(-2) &    2.28(47) & 3.55(47) & 1.46(45) &    2.84(45) & Yes  \\
       &   3         &     1        &  10     &   0.1     &  7.5(50) &   1.68(0)  &   4.00(50)  &    4.05(50) & 3.88(47)  &    3.25(47)  & Yes   \\
       &   3         &     1        &  10     &   0.5     &  7.5(50) &   4.90(-2) &   1.27(49) & 1.73(49)  &  2.33(46) &    4.86(46) & Yes  \\
\hline                                                
\multirow{7}{*}{5}&   10   &     1       &  5      &   0.1     & 2.3(50)  &    2.07(0) &  6.93(49)  &  7.19(49) &  1.01(47) &  5.11(46)  & Yes  \\
       &   10         &     1       &  5      &   0.25    & 2.3(50)  &    6.51(-1) &  2.85(49) &   3.41(49) &  9.58(46) &   9.45(46) & Yes  \\
       &   10         &     1       &  5      &   0.5     & 2.3(50)  &    4.41(-2) &  2.16(48) &  4.19(48) &   2.11(46) &   3.67(46) & Yes  \\
       &   10         &     1       &  2.5    &   0.1     & 6.0(49)  &    1.43(0) &  9.66(48)  &  9.73(48) &    2.20(46) & 2.11(46)  & Yes  \\
       &   10         &     1       &  2.5    &   0.5     & 6.0(49)  &    3.90(-2) &  4.31(47) &   8.54(47) &   1.51(46) &   5.32(46) & Yes  \\
       &   10         &     1       &  10     &   0.1     & 9.0(50)  &    2.34(0)  & 3.79(50) &    4.32(50) &  7.01(47) &    6.42(47)  & Yes  \\
       &   10         &     1       &  10     &   0.5     & 9.0(50)  &    8.04(-2) &  1.92(49) &   4.10(49) &  8.34(46)  &    1.40(47) & Yes  \\
\hline                                                
\multirow{7}{*}{6} &   10  &     3       &  5      &   0.1     &  5.7(50) &  5.56(0) &  3.31(50) & 3.37(50) &   5.33(47) & 4.05(47)  & Yes \\
       &   10         &     3       &  5      &   0.25    &  5.7(50) &  1.46(0) &  1.08(50) & 1.32(50) &   1.72(47) & 2.08(47)  & Yes \\
       &   10         &     3       &  5      &   0.5     &  5.7(50) &  6.42(-2) &  4.91(48) & 1.75(49) &  2.16(46) & 7.23(46) &   No  \\
       &   10         &     3       &  2.5    &   0.1     &  1.4(50) &  4.42(0) &   6.38(49) & 6.46(49) &  9.97(46) &  1.15(47)  & Yes \\
       &   10         &     3       &  2.5    &   0.5     &  1.4(50) &  5.68(-2) &   9.11(47)  &  2.78(48) &  2.08(46) &  6.97(47) &  No   \\
       &   10         &     3       &  10     &   0.1     &  2.3(51) &  5.84(0) &    1.46(51) & 1.49(51) & 2.88(48) &  2.57(48)  & Yes \\
       &   10         &     3       &  10     &   0.5     &  2.3(51) &  1.15(-1) &  3.70(49) &  1.06(50) &   8.68(46) & 2.93(47) & Yes \\
\hline
\end{tabular}
\tablefoot{From left to right, the columns give the modelset number, the mass of the primary star $M_{1}$, the mass of the secondary star $M_{2}$, the impact parameter $b$, the initial relative velocity $V_{\rm rel}$, and the kinetic energy $E_{\rm kin,coll}$ of the collision, defined as $0.5 (M_1 M_2) / (M_1+M_2) V_{\rm rel}^2$. The last three columns give the total mass ejected \mej\ and the associated kinetic $E_{\rm kin,ej}$ and trapped radiative energy $E_{\rm rad,ej}$ at 0.5\,d, which corresponds to the time at which we started the radiative transfer calculations with \cmfgen. The twin columns for $E_{\rm kin,ej}$ and $E_{\rm rad,ej}$ correspond to values obtained from the 3D simulations and the corresponding 1D, spherical average. The last column with heading ``RT'' indicates whether we performed radiative-transfer modeling for the corresponding collision. Numbers appearing within parentheses correspond to powers of ten.}
\end{table*}

\newpage

\section{Light curves and spectra}\label{sec:appendix}
We provide the bolometric luminosity (Fig.~\ref{fig_lbol_set}), $UVW1-$band magnitude (Fig.~\ref{fig_muvw1_set}), $V-$band magnitude (Fig.~\ref{fig_mv_set}) for all our models as a function time. In addition, we also present their spectral evolution in Figs.~\ref{fig_montage_set_1}--\ref{fig_montage_set_6}.
\FloatBarrier
\begin{figure*}[!htbp]
   \centering
    \begin{subfigure}[b]{0.48\textwidth}
       \centering
       \includegraphics[width=\textwidth]{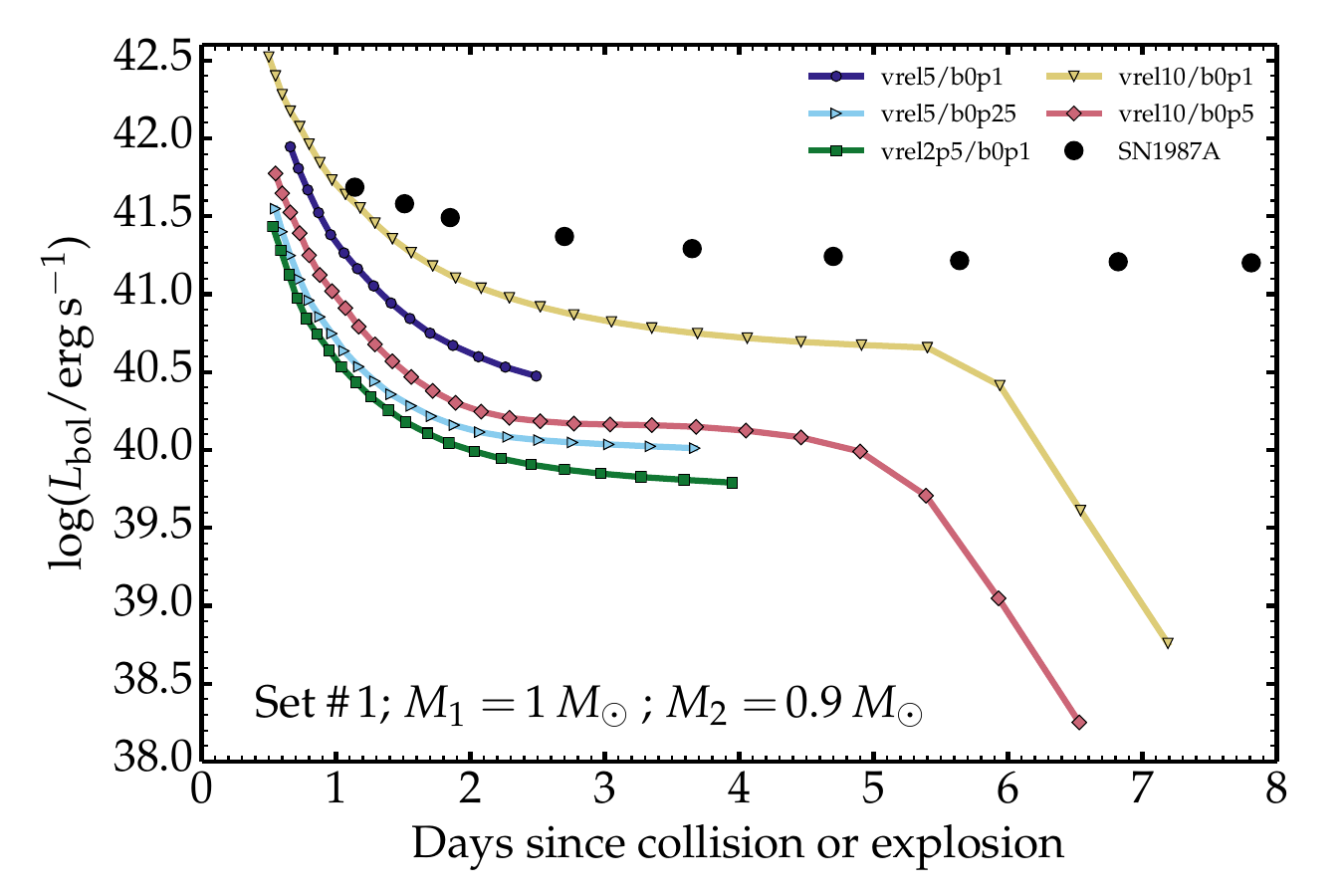}
    \end{subfigure}
    \begin{subfigure}[b]{0.48\textwidth}
       \centering
       \includegraphics[width=\textwidth]{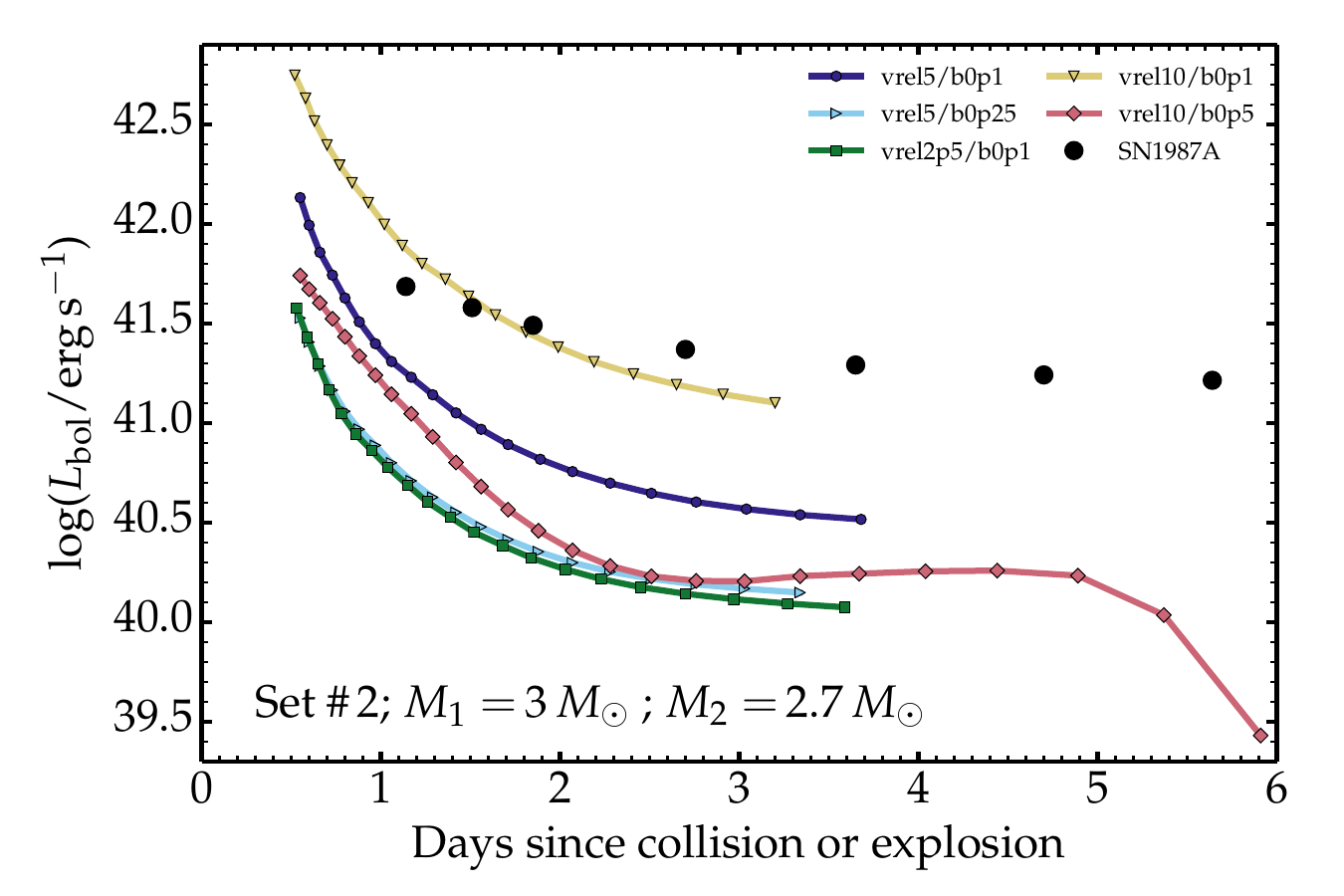}
    \end{subfigure}
    \begin{subfigure}[b]{0.48\textwidth}
       \centering
       \includegraphics[width=\textwidth]{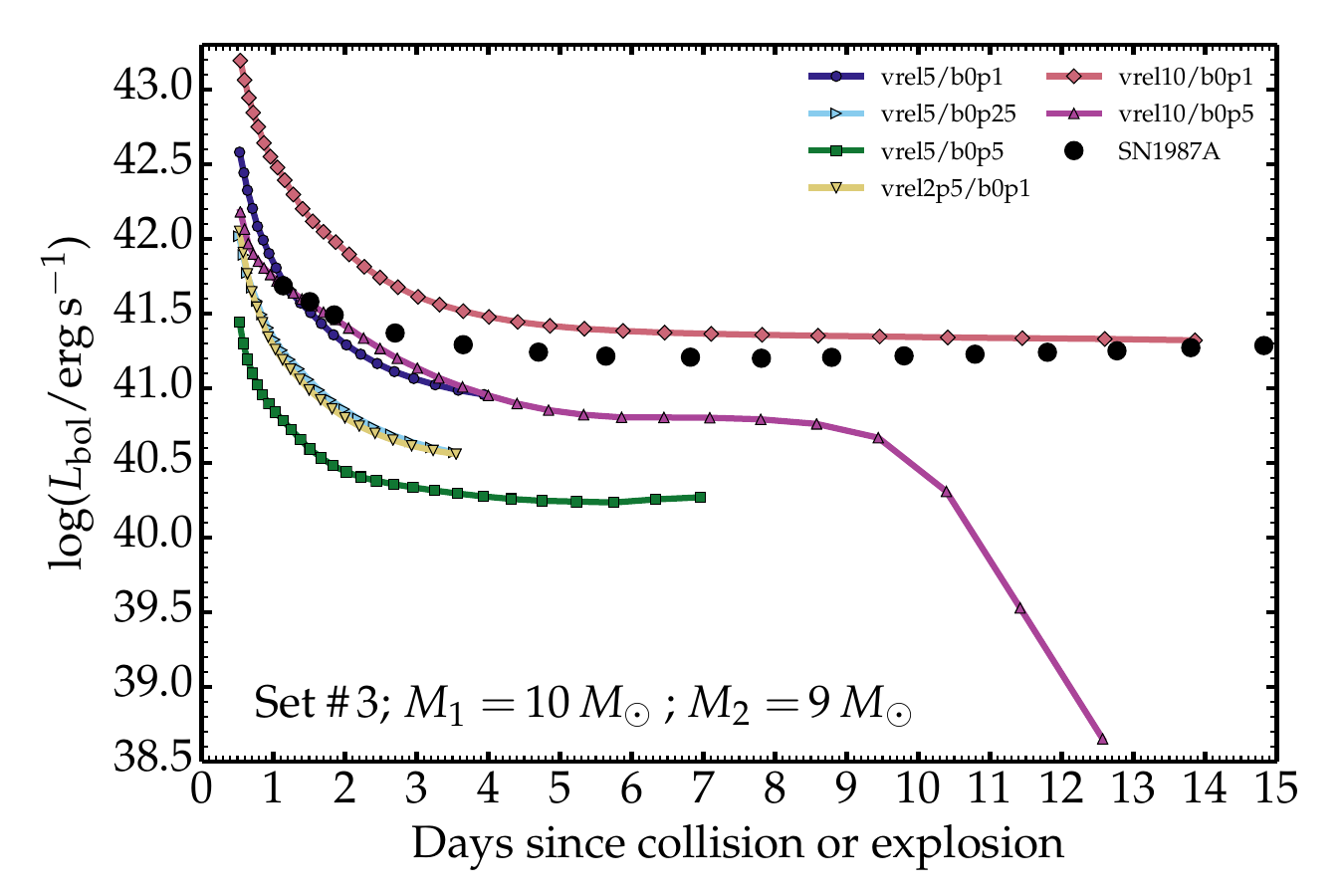}
    \end{subfigure}
    \begin{subfigure}[b]{0.48\textwidth}
       \centering
       \includegraphics[width=\textwidth]{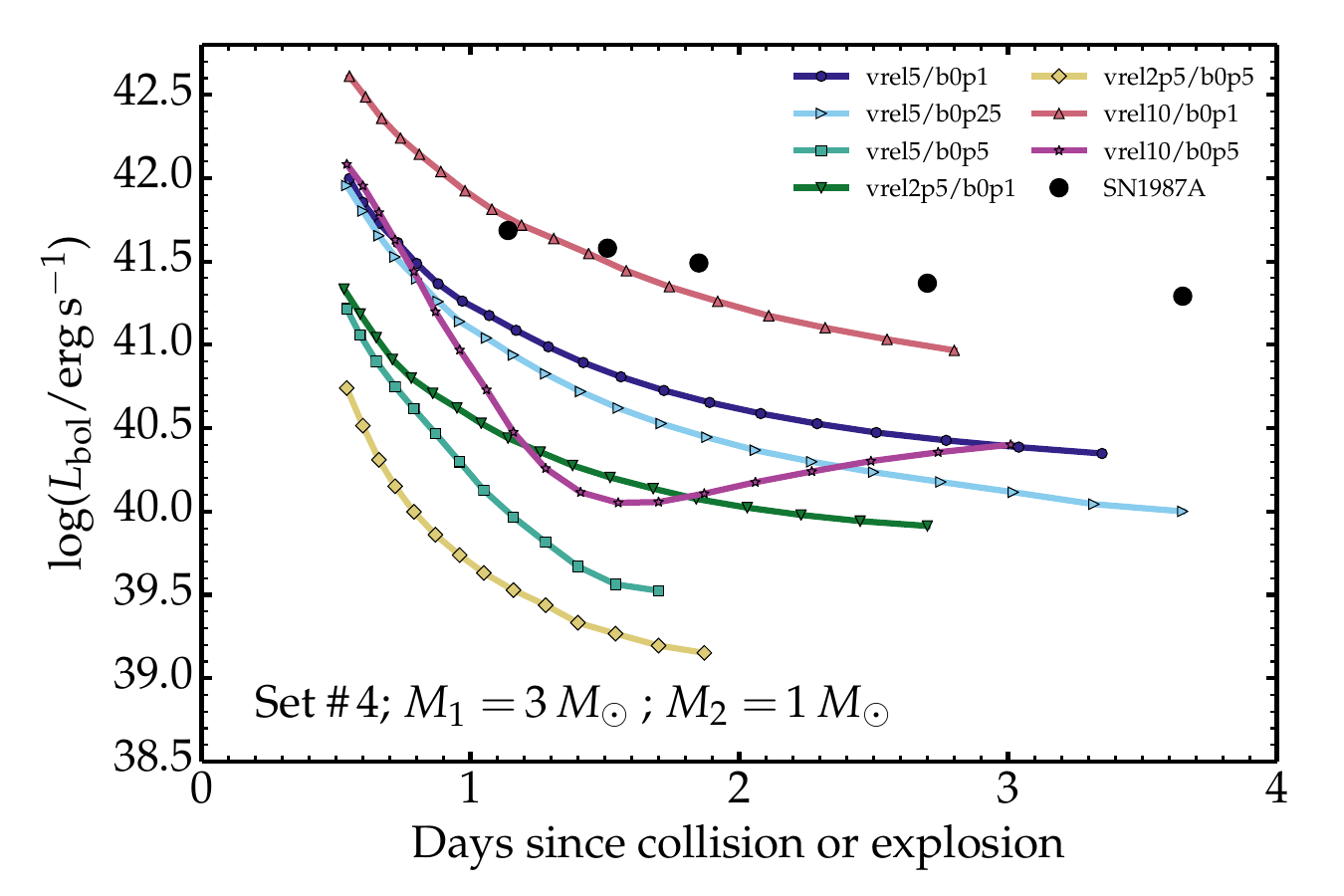}
    \end{subfigure}
     \begin{subfigure}[b]{0.48\textwidth}
       \centering
        \includegraphics[width=\textwidth]{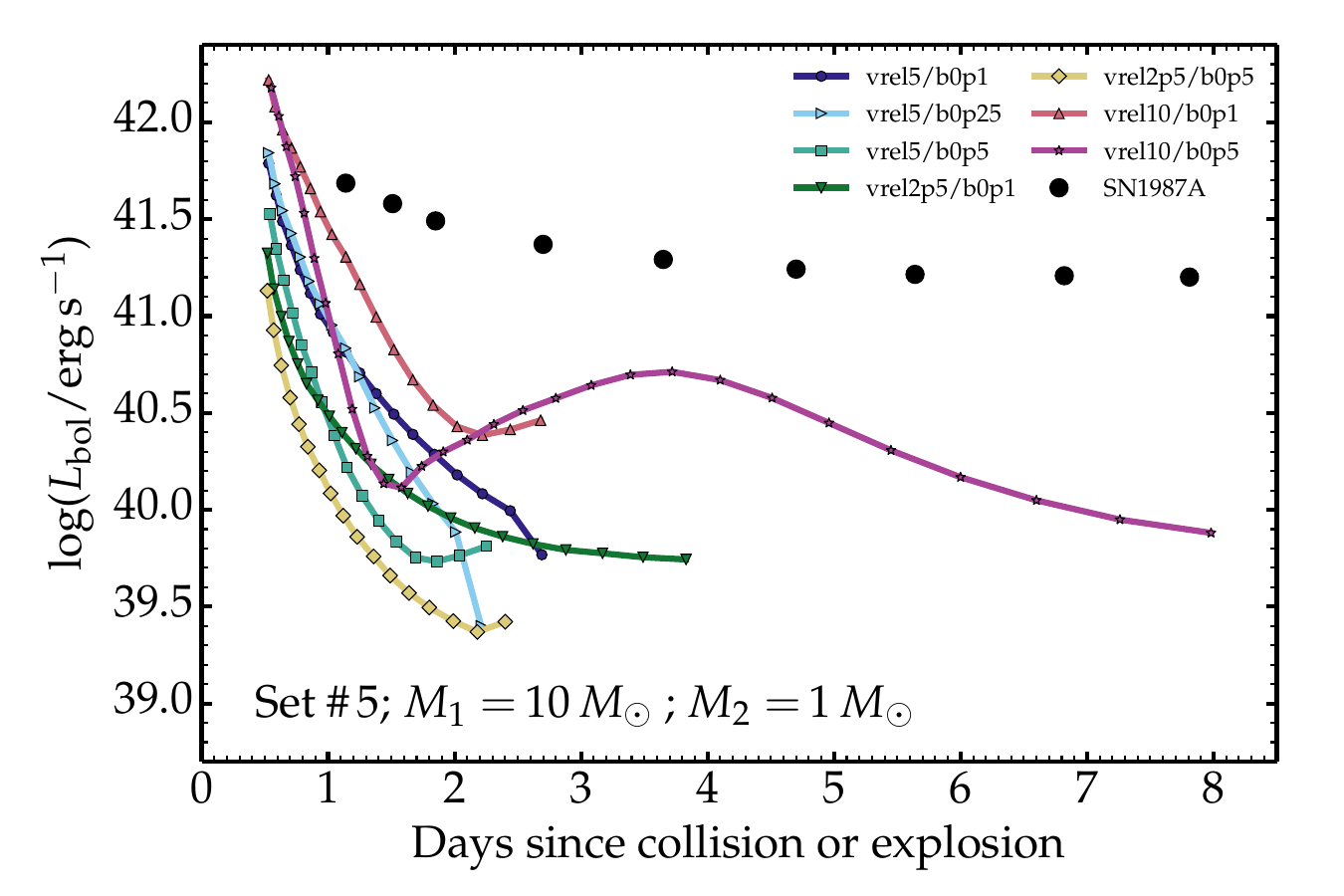}
    \end{subfigure}
     \begin{subfigure}[b]{0.48\textwidth}
       \centering
       \includegraphics[width=\textwidth]{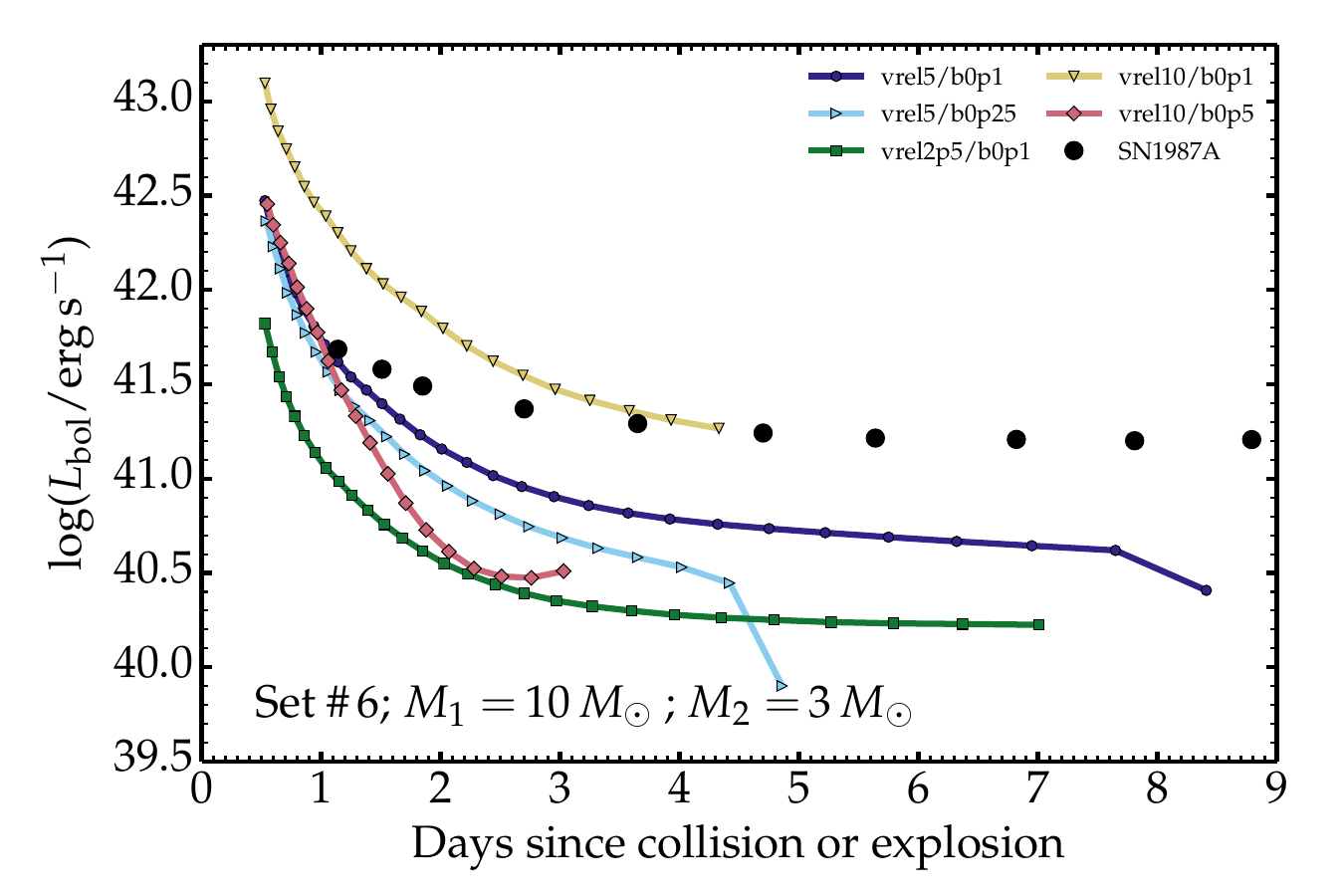}
    \end{subfigure}
\caption{Bolometric light curves for our six sets of models.
\label{fig_lbol_set}
}
\end{figure*}

\begin{figure*}
   \centering
    \begin{subfigure}[b]{0.48\textwidth}
       \centering
       \includegraphics[width=\textwidth]{Figures/RT/nlc_snobs_UVW1_modelset_1-crop.pdf}
    \end{subfigure}
    \begin{subfigure}[b]{0.48\textwidth}
       \centering
       \includegraphics[width=\textwidth]{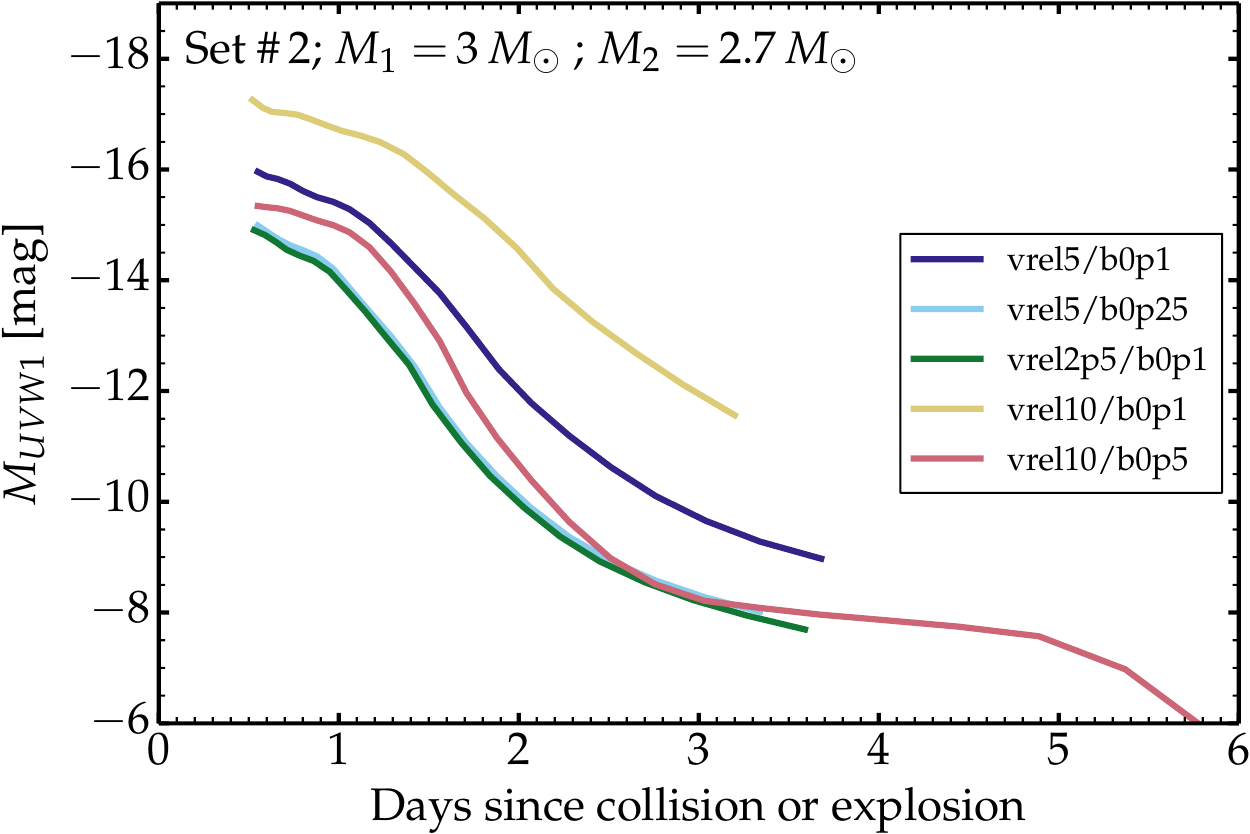}
    \end{subfigure}
    \begin{subfigure}[b]{0.48\textwidth}
       \centering
       \includegraphics[width=\textwidth]{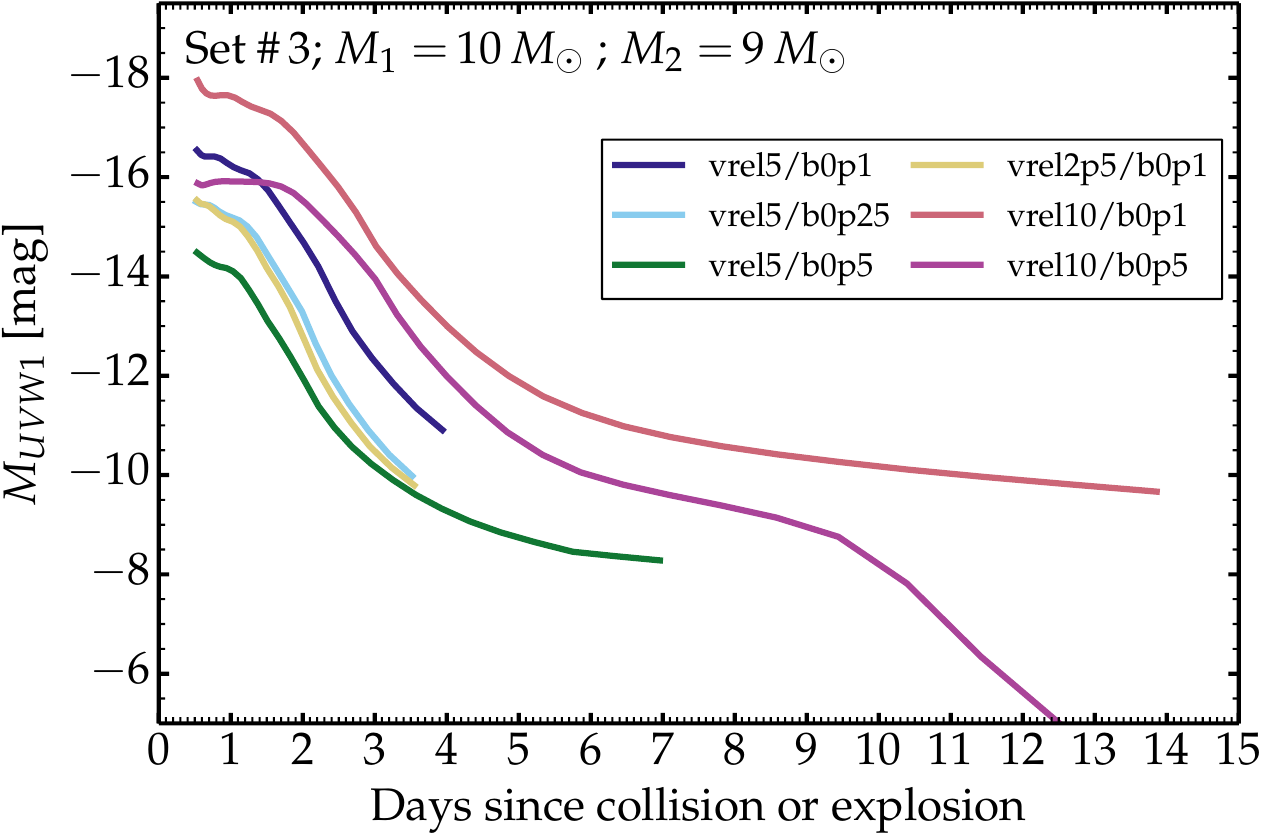}
    \end{subfigure}
    \begin{subfigure}[b]{0.48\textwidth}
       \centering
       \includegraphics[width=\textwidth]{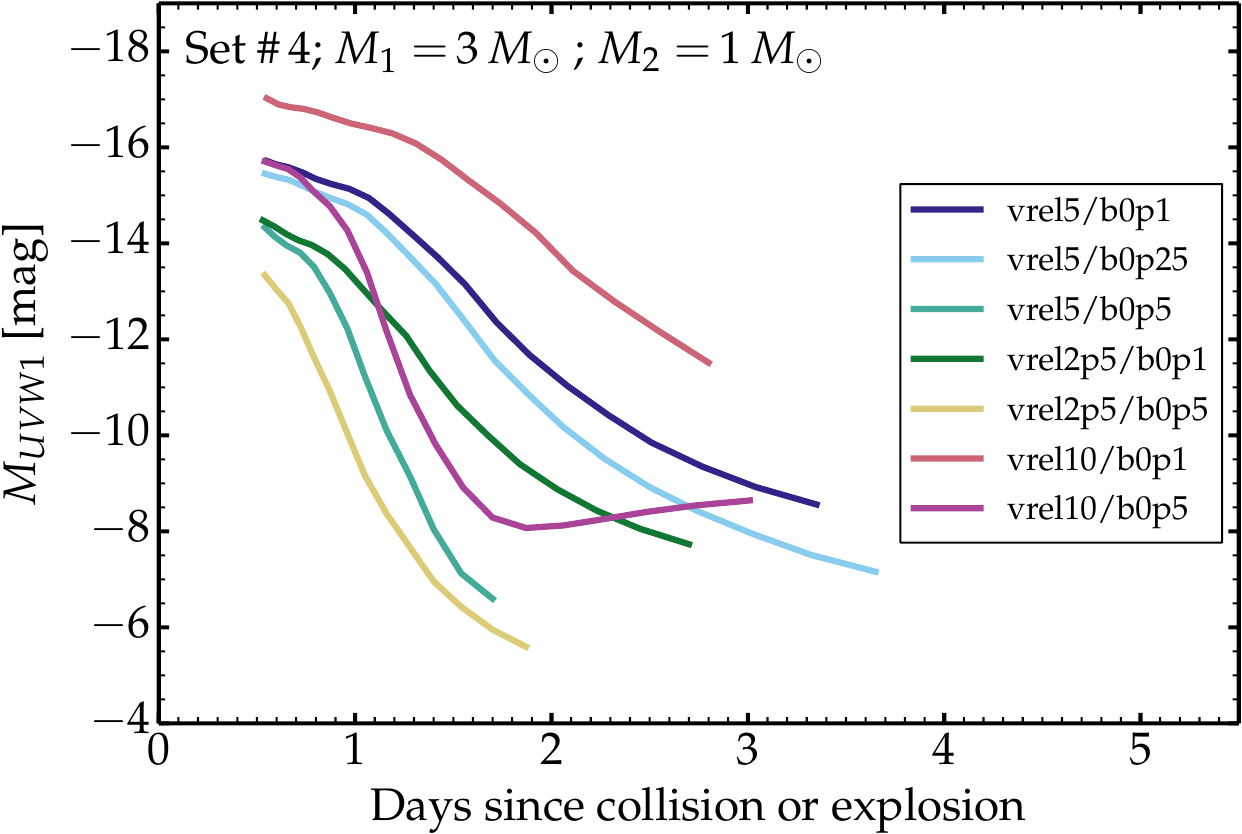}
    \end{subfigure}
     \begin{subfigure}[b]{0.48\textwidth}
       \centering
        \includegraphics[width=\textwidth]{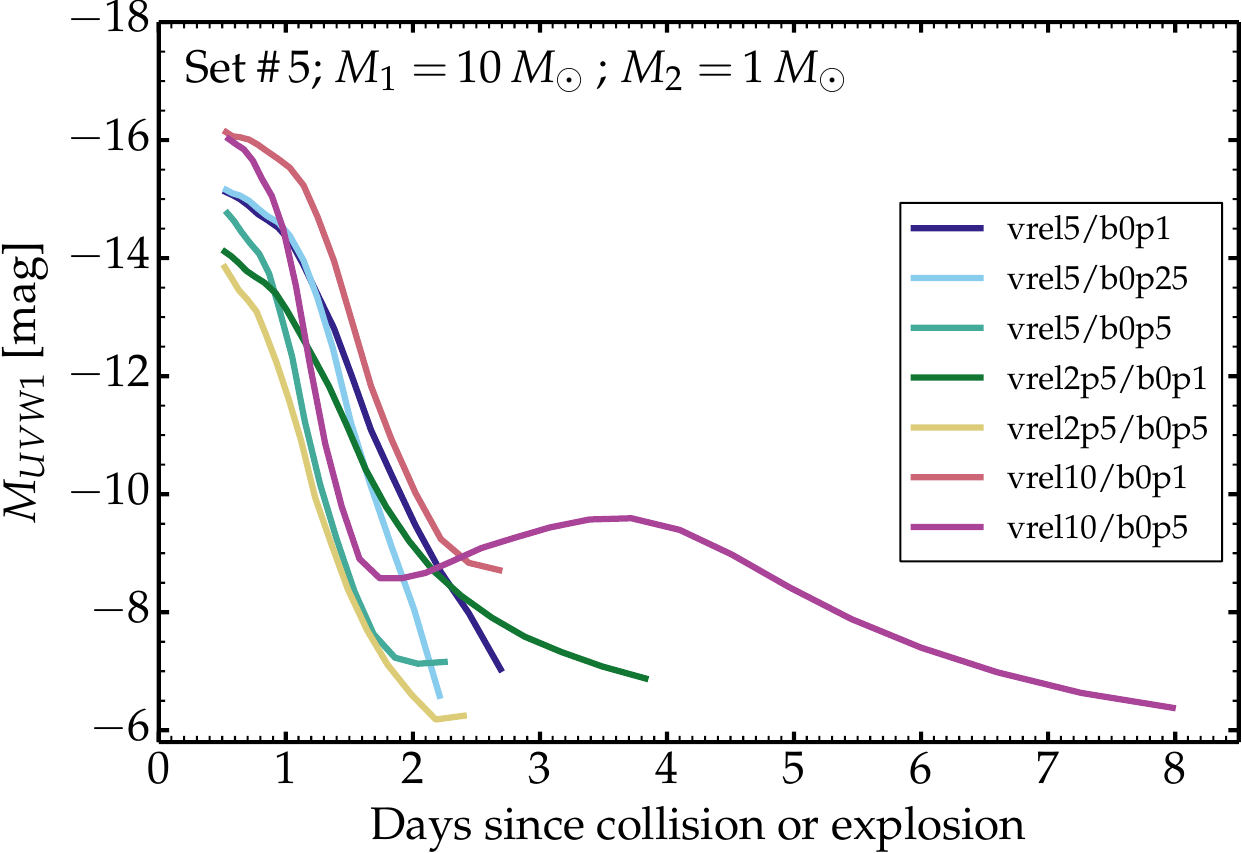}
    \end{subfigure}
     \begin{subfigure}[b]{0.48\textwidth}
       \centering
       \includegraphics[width=\textwidth]{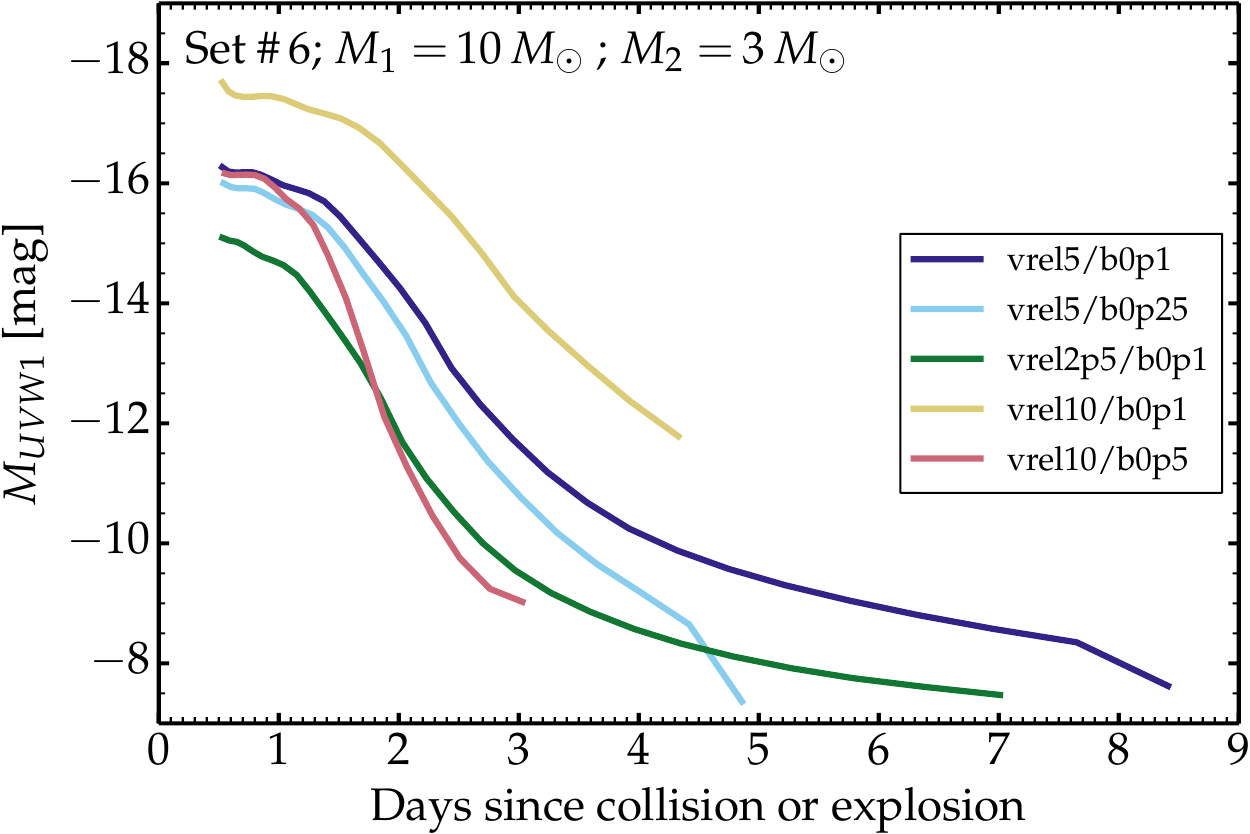}
    \end{subfigure}
\caption{$UVW1$-band light curves for our six sets of models.
\label{fig_muvw1_set}
}
\end{figure*}

\begin{figure*}
   \centering
    \begin{subfigure}[b]{0.48\textwidth}
       \centering
       \includegraphics[width=\textwidth]{Figures/RT/nlc_snobs_V_modelset_1-crop.pdf}
    \end{subfigure}
    \begin{subfigure}[b]{0.48\textwidth}
       \centering
       \includegraphics[width=\textwidth]{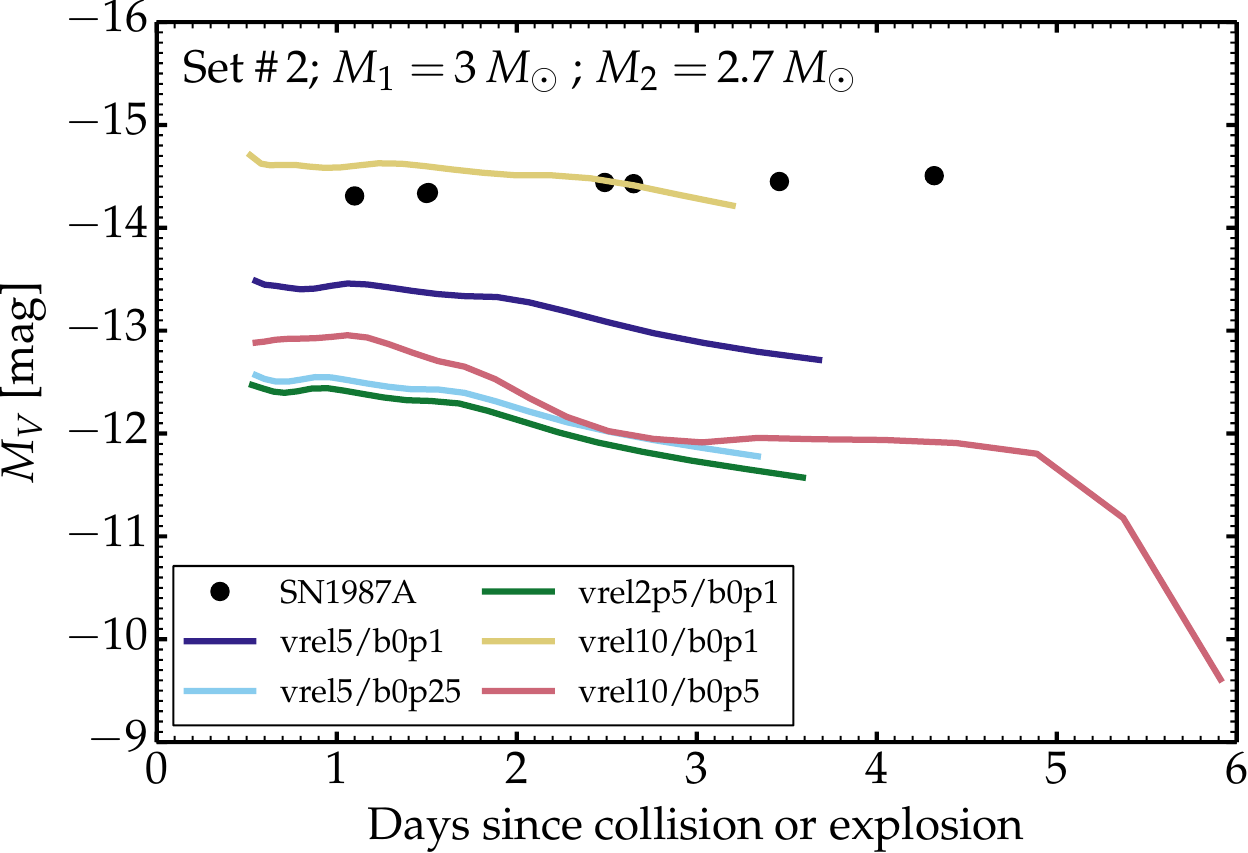}
    \end{subfigure}
    \begin{subfigure}[b]{0.48\textwidth}
       \centering
       \includegraphics[width=\textwidth]{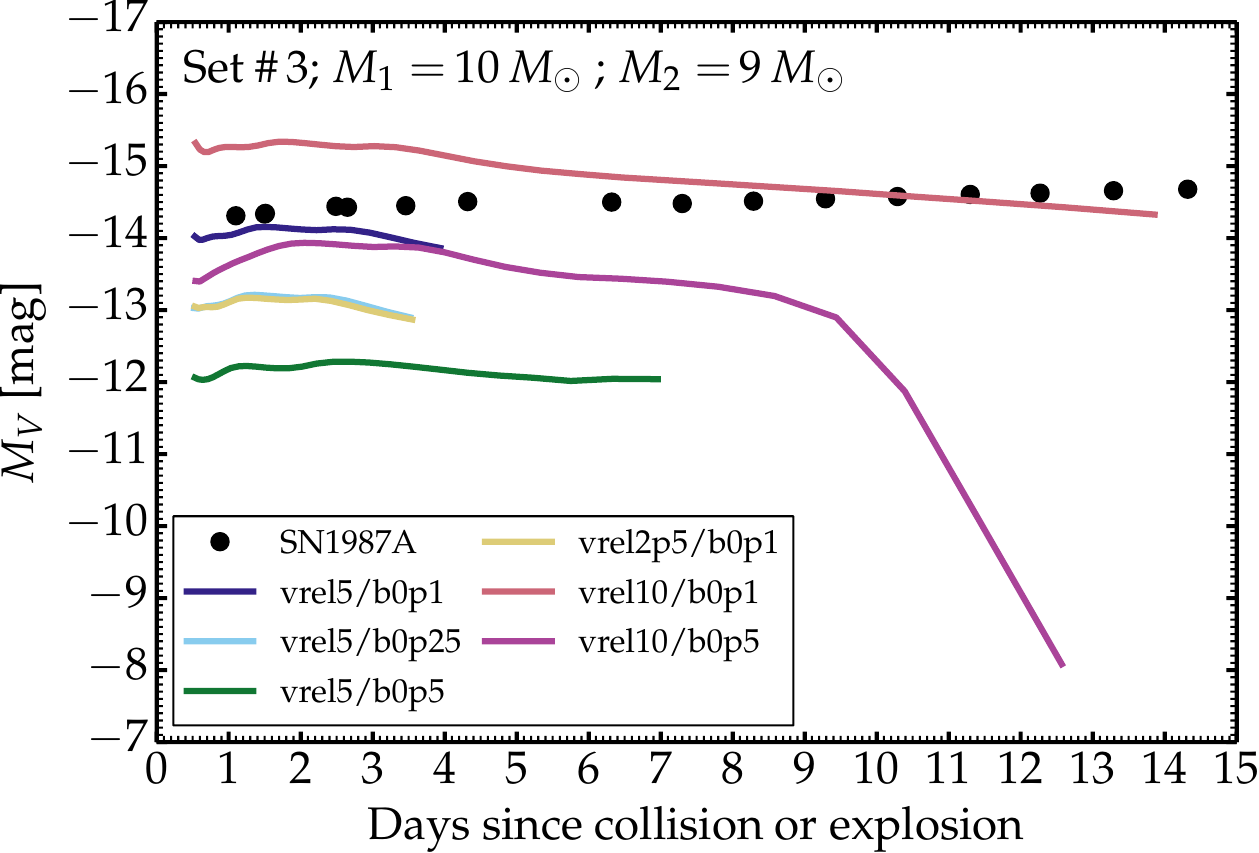}
    \end{subfigure}
    \begin{subfigure}[b]{0.48\textwidth}
       \centering
       \includegraphics[width=\textwidth]{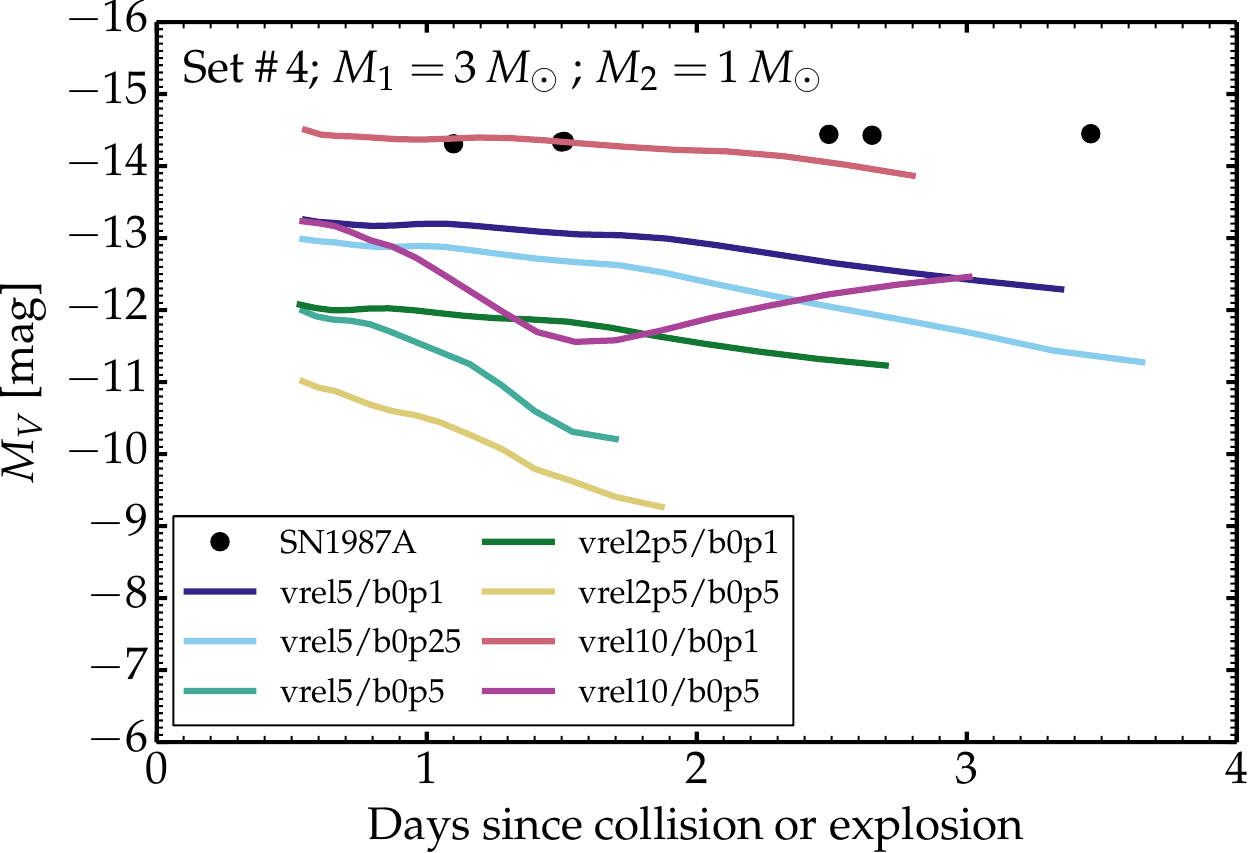}
    \end{subfigure}
     \begin{subfigure}[b]{0.48\textwidth}
       \centering
        \includegraphics[width=\textwidth]{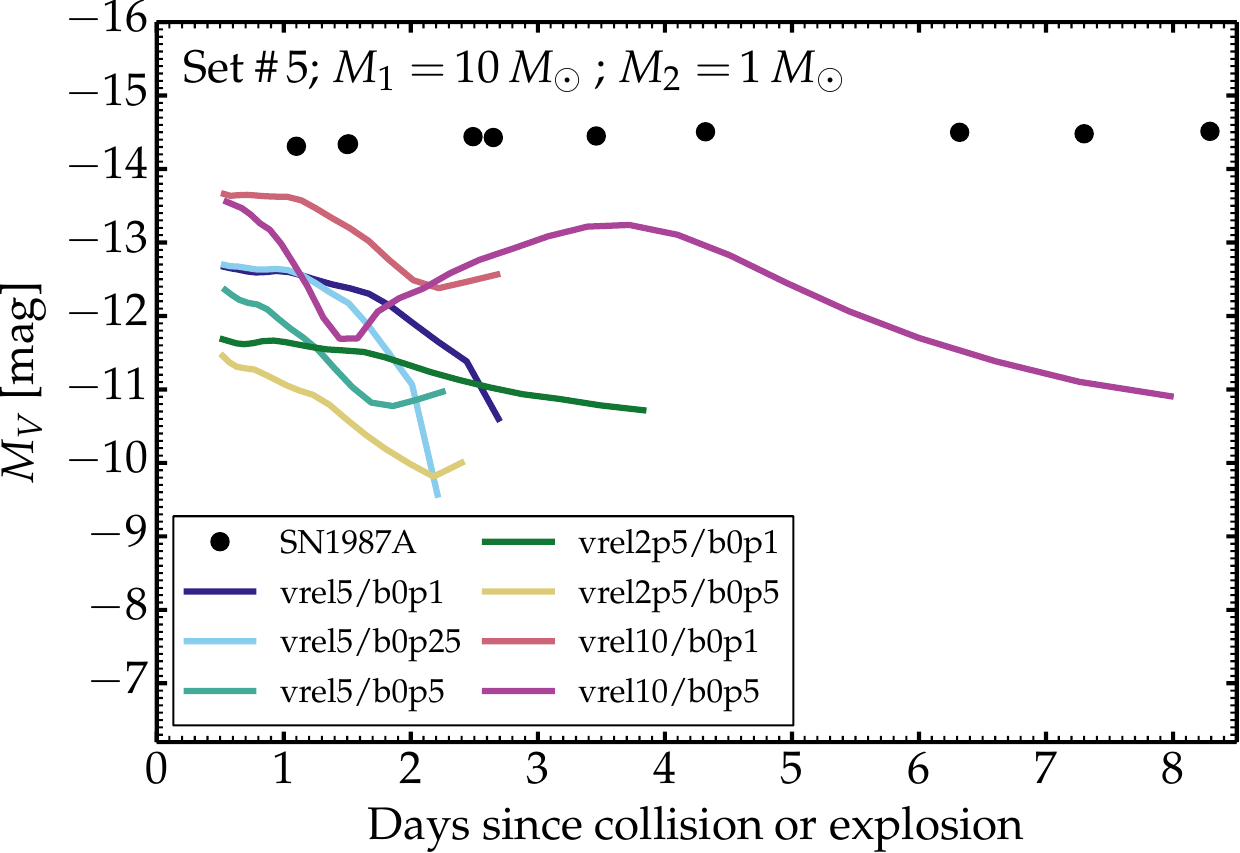}
    \end{subfigure}
     \begin{subfigure}[b]{0.48\textwidth}
       \centering
       \includegraphics[width=\textwidth]{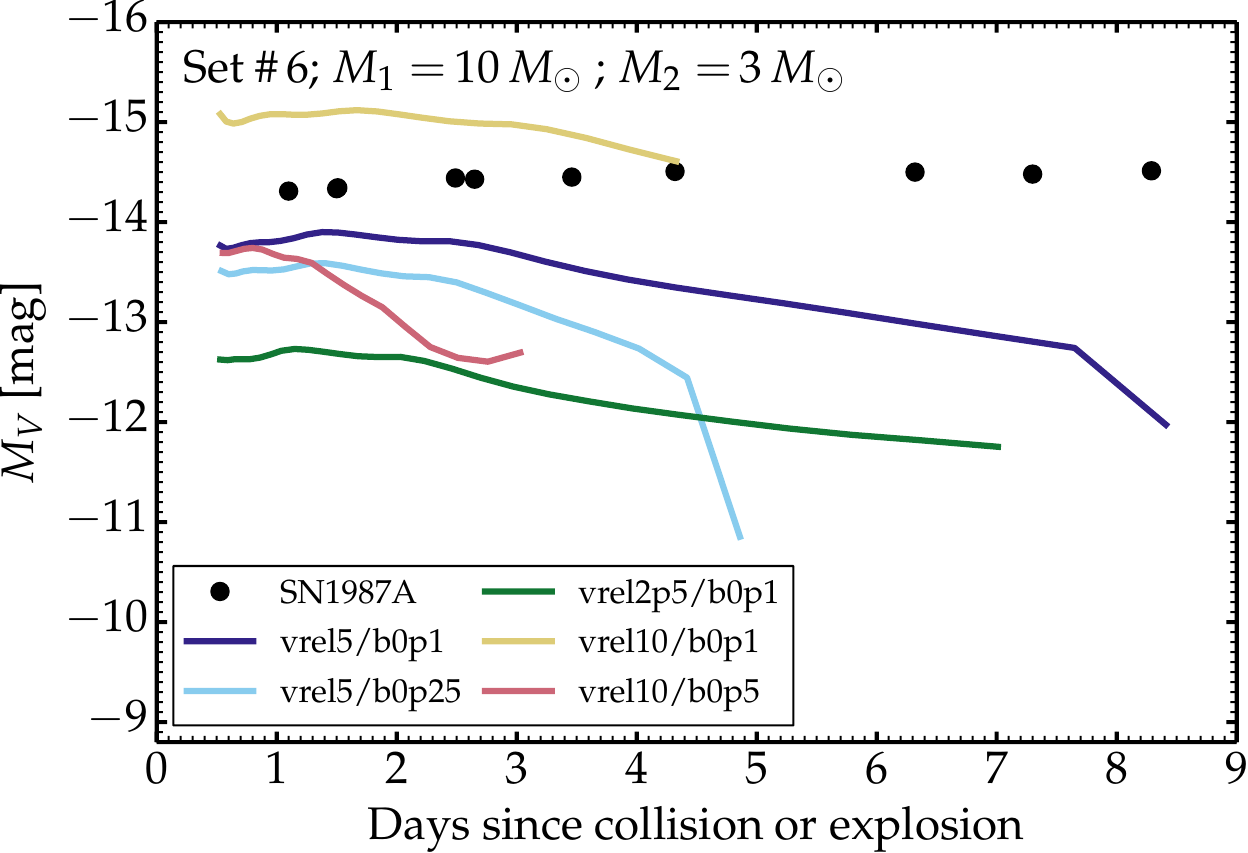}
    \end{subfigure}
\caption{$V$-band light curves for our six sets of models.
\label{fig_mv_set}
}
\end{figure*}

\begin{figure*}
   \centering
    \begin{subfigure}[b]{0.33\textwidth}
       \centering
       \includegraphics[width=\textwidth]{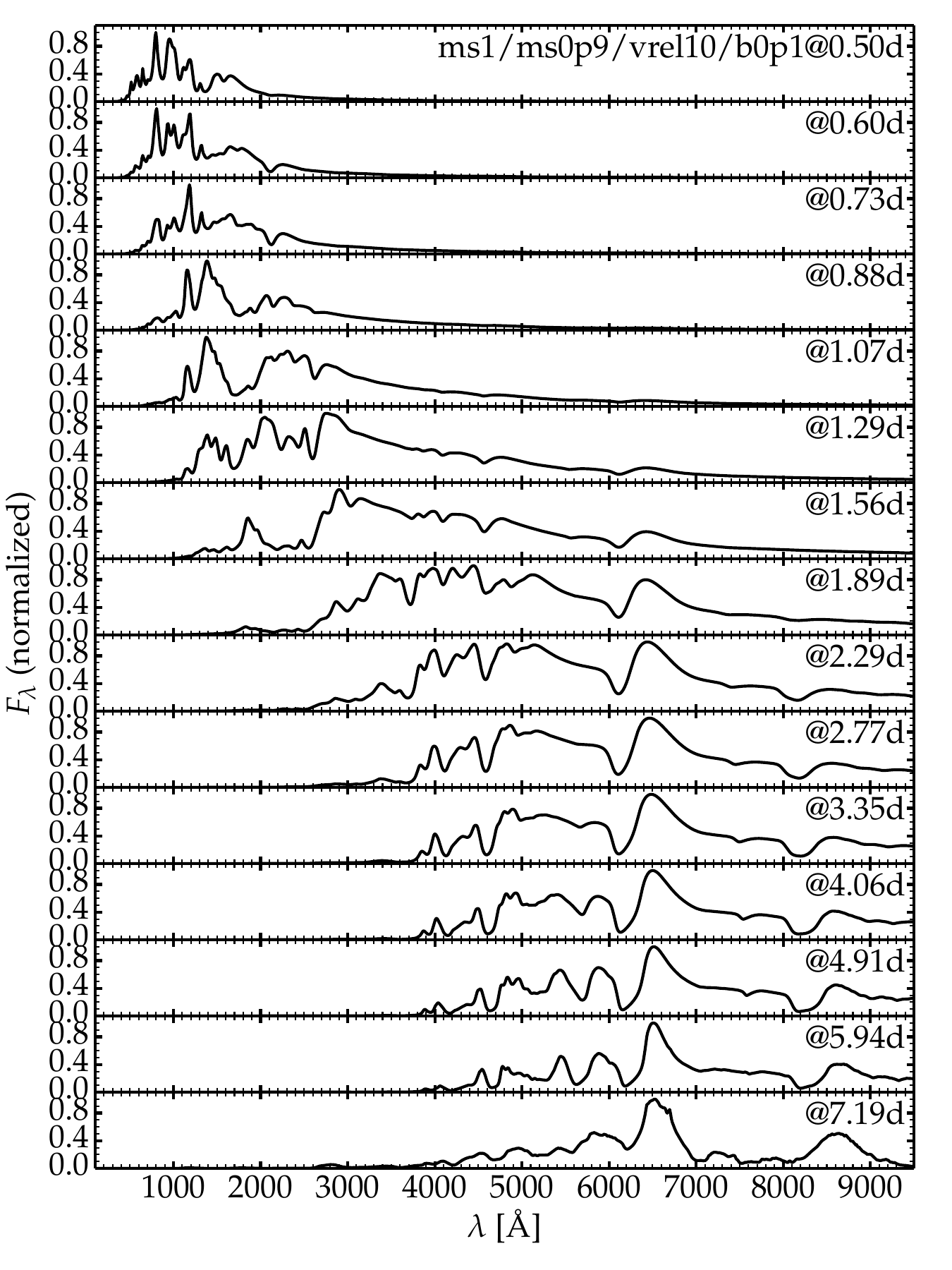}
    \end{subfigure}
    \hfill
    \begin{subfigure}[b]{0.33\textwidth}
       \centering
       \includegraphics[width=\textwidth]{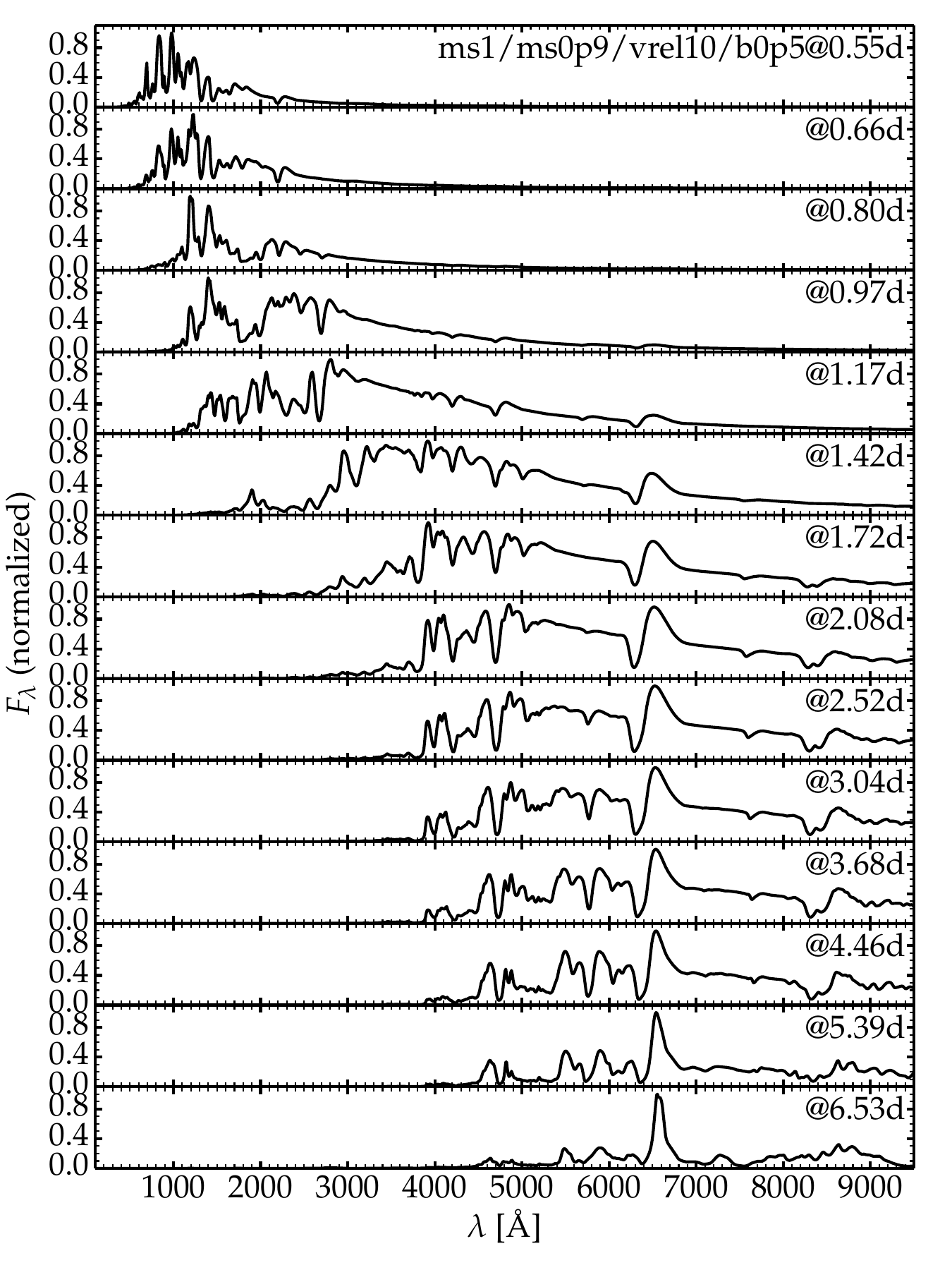}
    \end{subfigure}
     \hfill
    \begin{subfigure}[b]{0.33\textwidth}
       \centering
       \includegraphics[width=\textwidth]{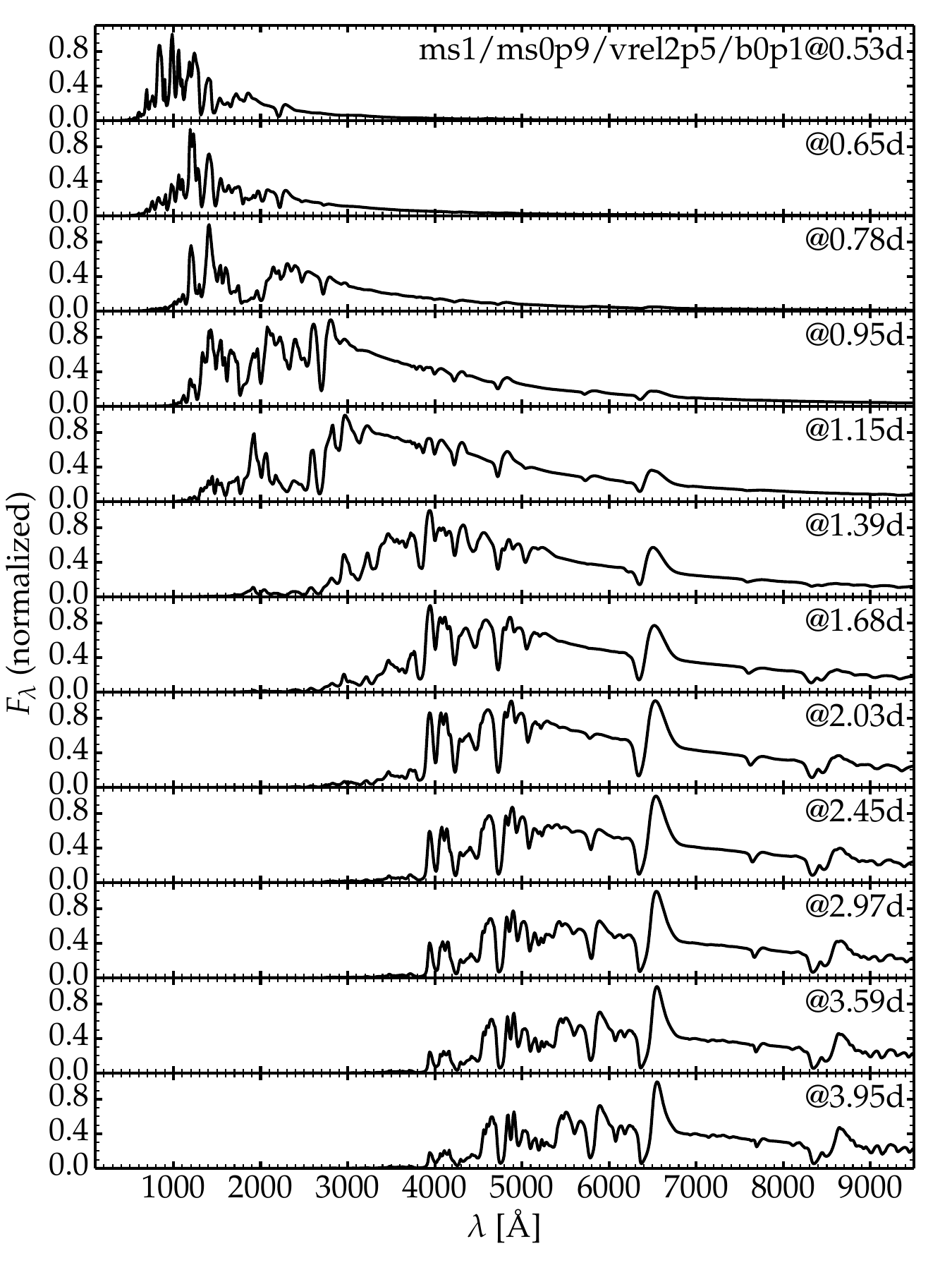}
    \end{subfigure}
     \hfill
    \begin{subfigure}[b]{0.33\textwidth}
       \centering
       \includegraphics[width=\textwidth]{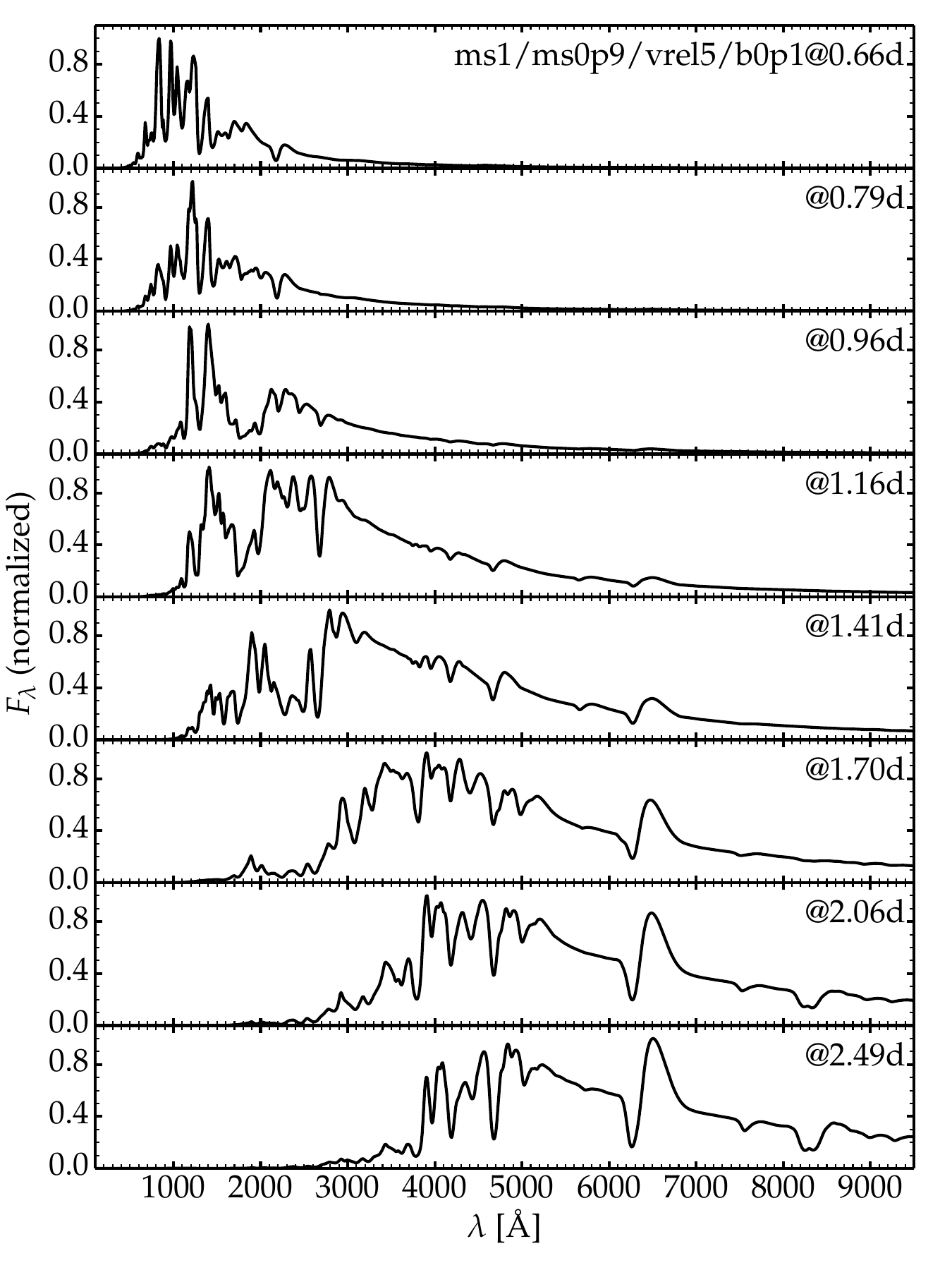}
    \end{subfigure}
     \hfill
     \begin{subfigure}[b]{0.33\textwidth}
       \centering
        \includegraphics[width=\textwidth]{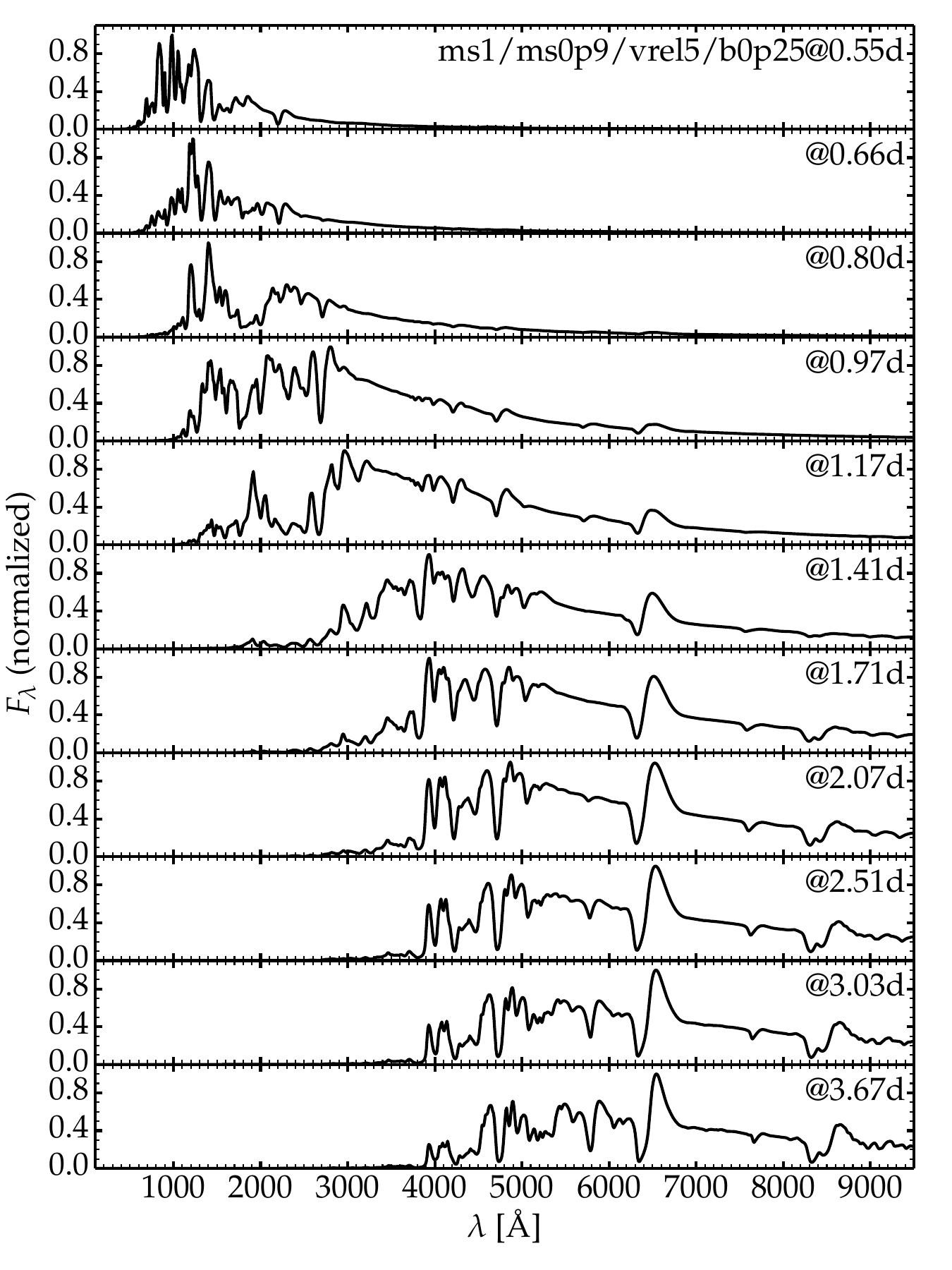}
    \end{subfigure}
\caption{Spectral montage for the modelset \# 1. 
\label{fig_montage_set_1}
}
\end{figure*}

\begin{figure*}
   \centering
    \begin{subfigure}[b]{0.33\textwidth}
       \centering
       \includegraphics[width=\textwidth]{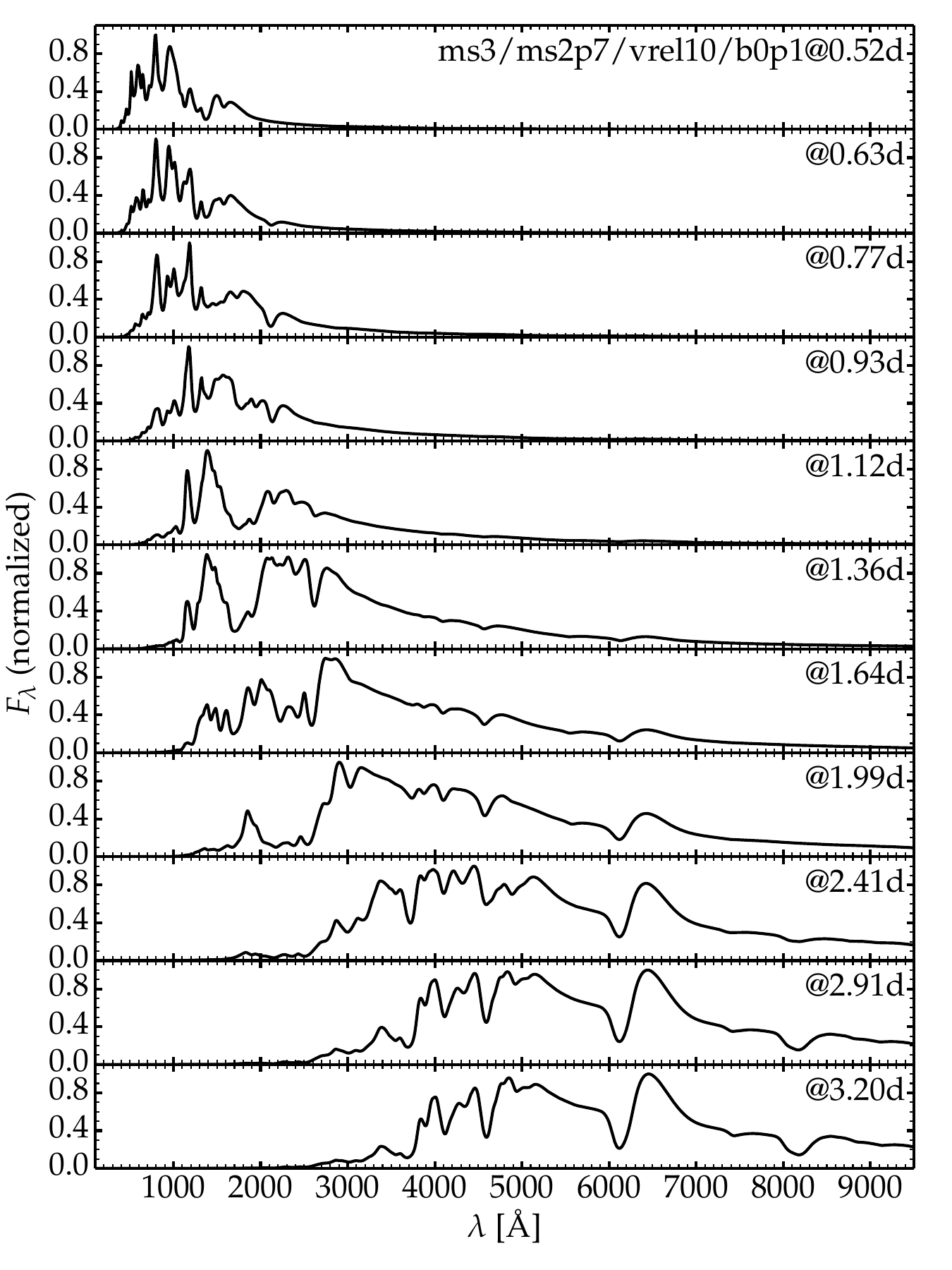}
    \end{subfigure}
    \hfill
    \begin{subfigure}[b]{0.33\textwidth}
       \centering
       \includegraphics[width=\textwidth]{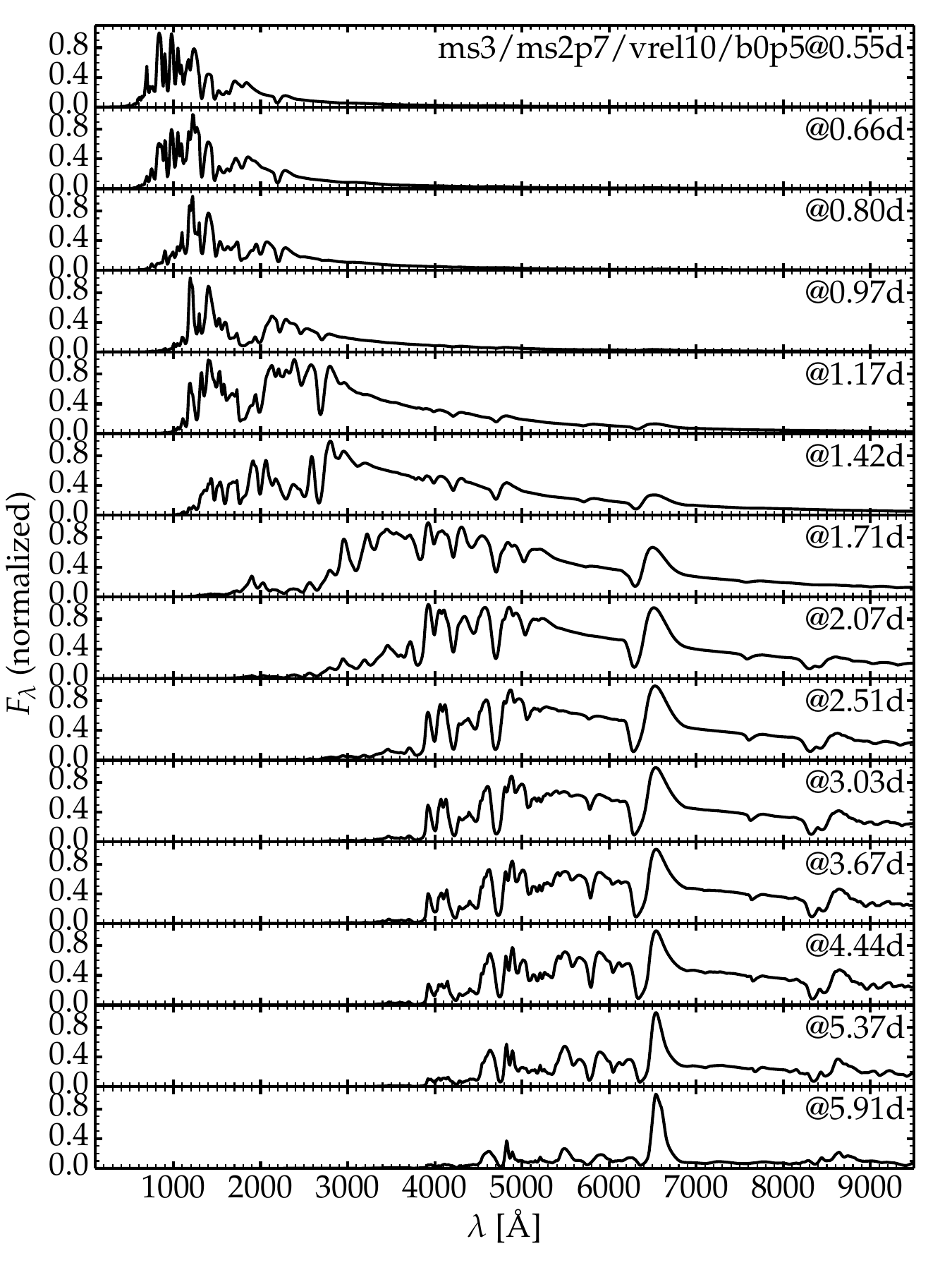}
    \end{subfigure}
     \hfill
    \begin{subfigure}[b]{0.33\textwidth}
       \centering
       \includegraphics[width=\textwidth]{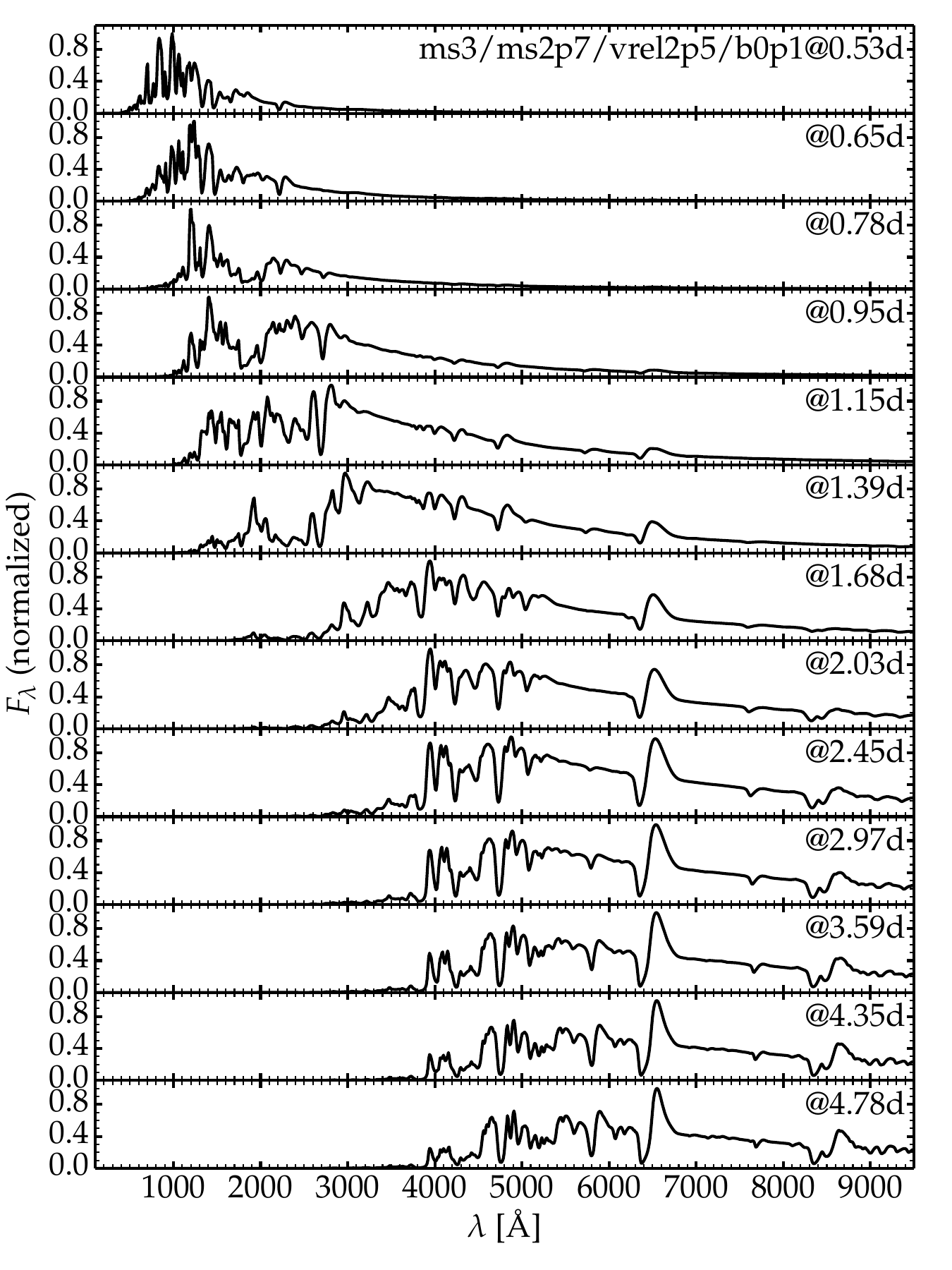}
    \end{subfigure}
     \hfill
    \begin{subfigure}[b]{0.33\textwidth}
       \centering
       \includegraphics[width=\textwidth]{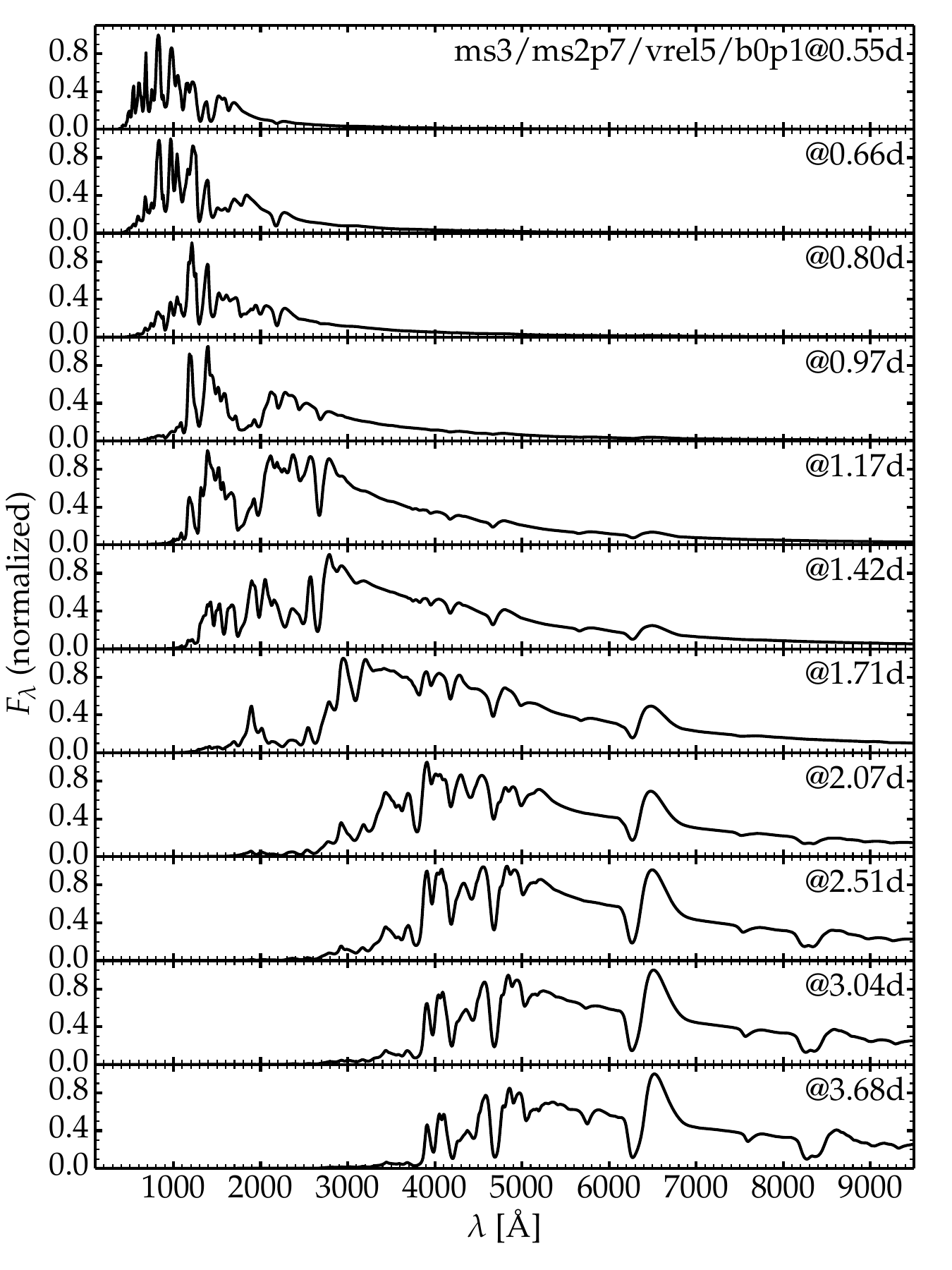}
    \end{subfigure}
     \hfill
     \begin{subfigure}[b]{0.33\textwidth}
       \centering
        \includegraphics[width=\textwidth]{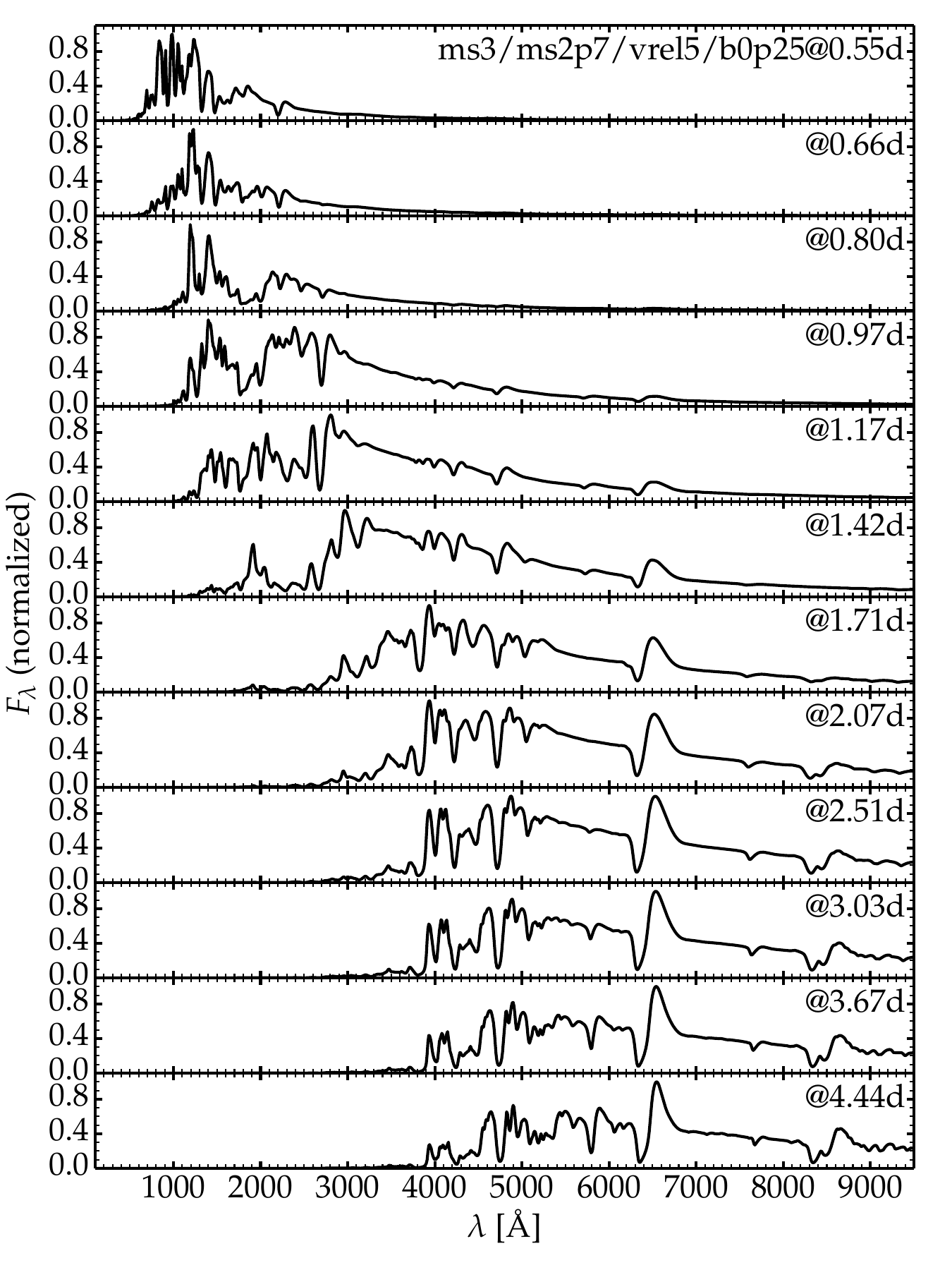}
    \end{subfigure}
\caption{Spectral montage for the modelset \# 2.
\label{fig_montage_set_2}
}
\end{figure*}

\begin{figure*}
   \centering
    \begin{subfigure}[b]{0.33\textwidth}
       \centering
       \includegraphics[width=\textwidth]{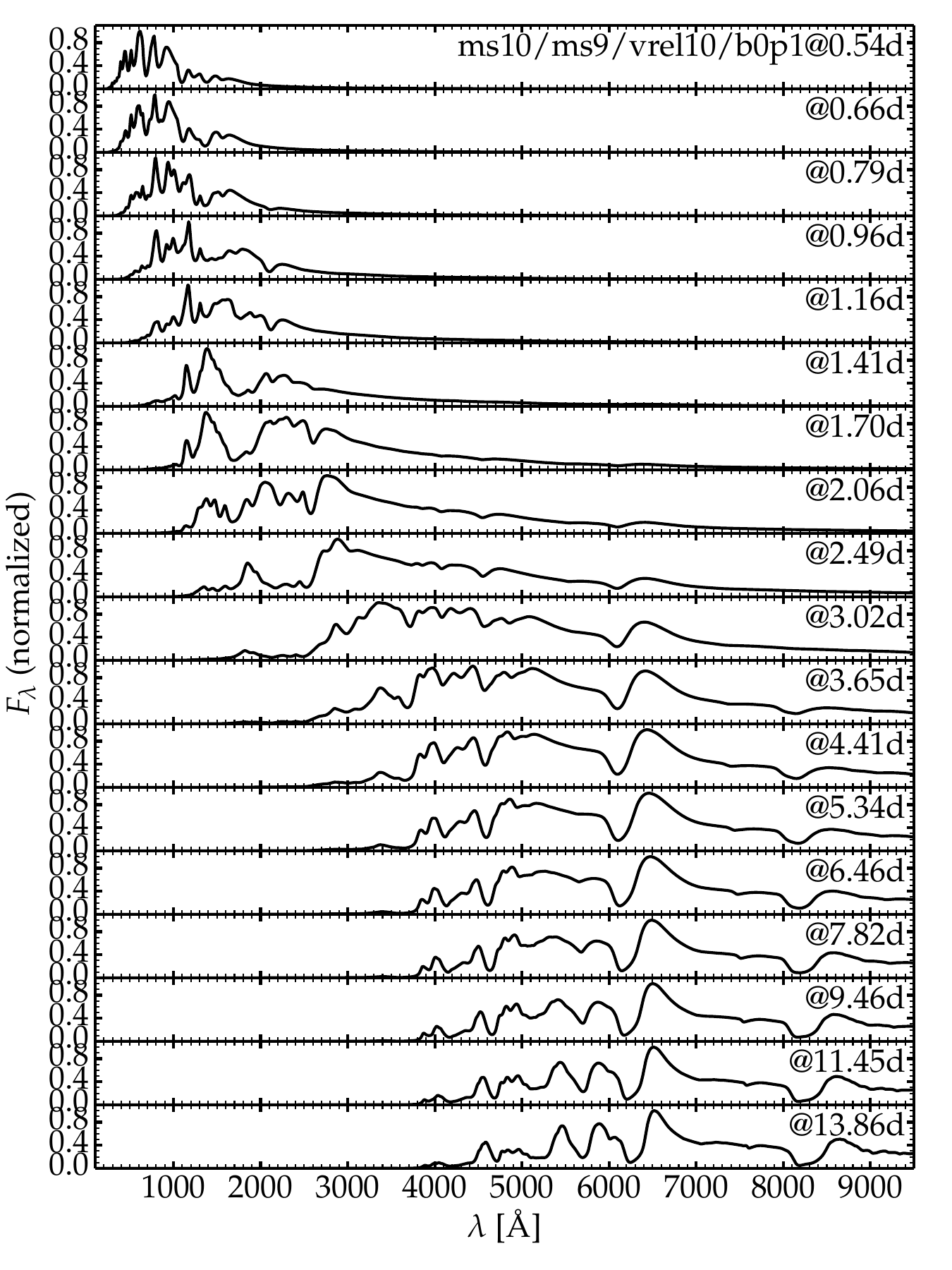}
    \end{subfigure}
    \hfill
    \begin{subfigure}[b]{0.33\textwidth}
       \centering
       \includegraphics[width=\textwidth]{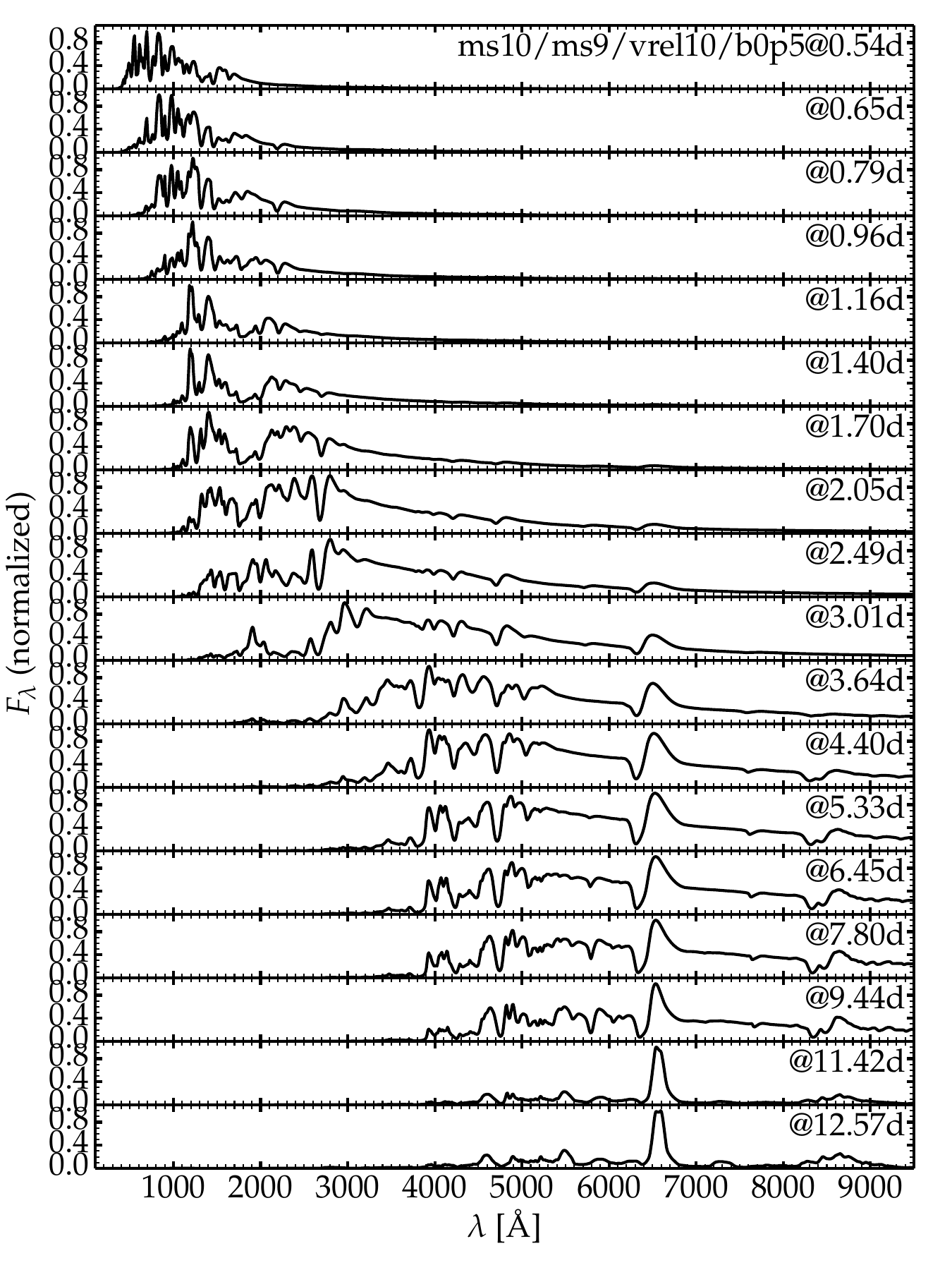}
    \end{subfigure}
     \hfill
    \begin{subfigure}[b]{0.33\textwidth}
       \centering
       \includegraphics[width=\textwidth]{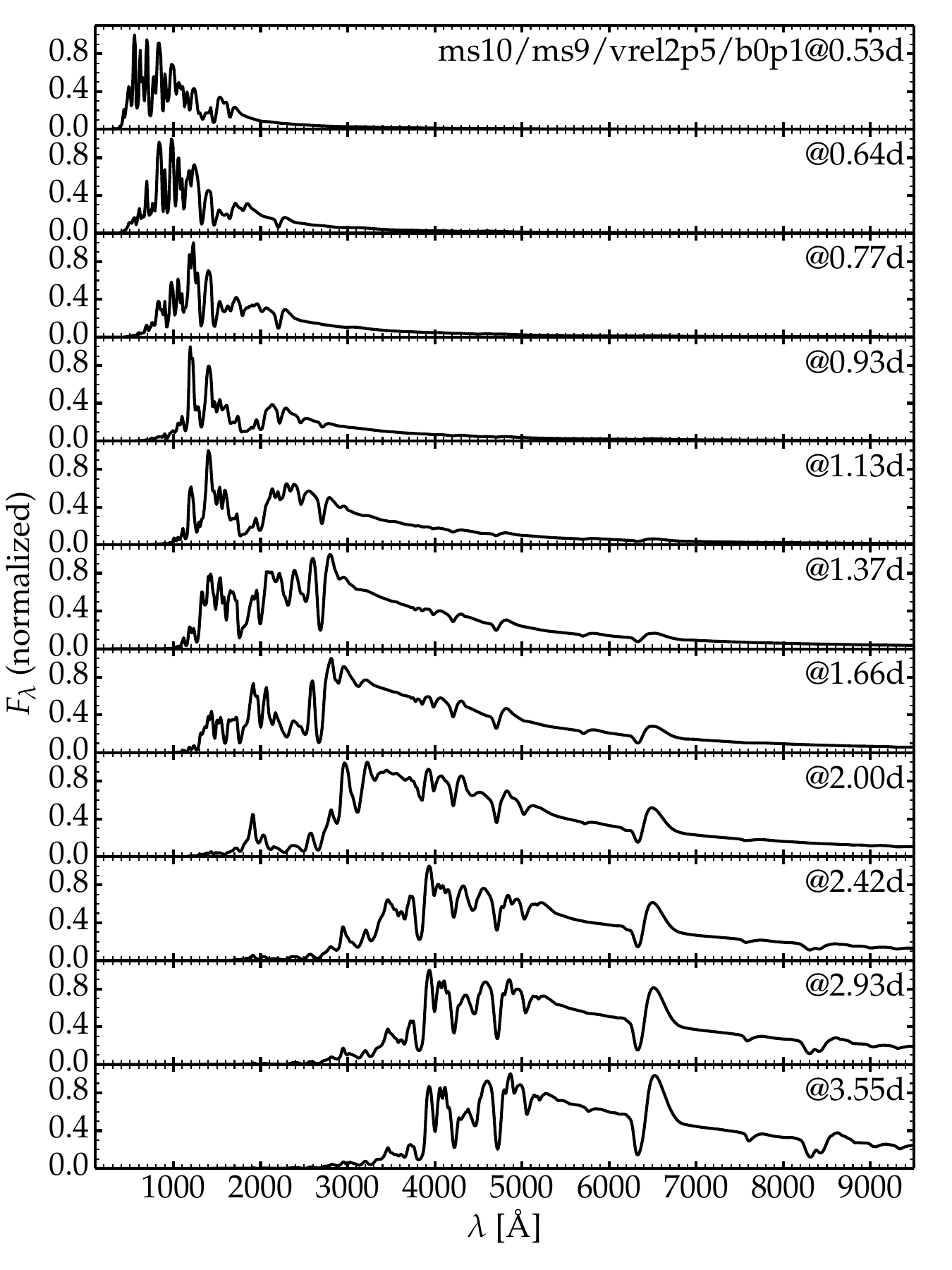}
    \end{subfigure}
     \hfill
    \begin{subfigure}[b]{0.33\textwidth}
       \centering
       \includegraphics[width=\textwidth]{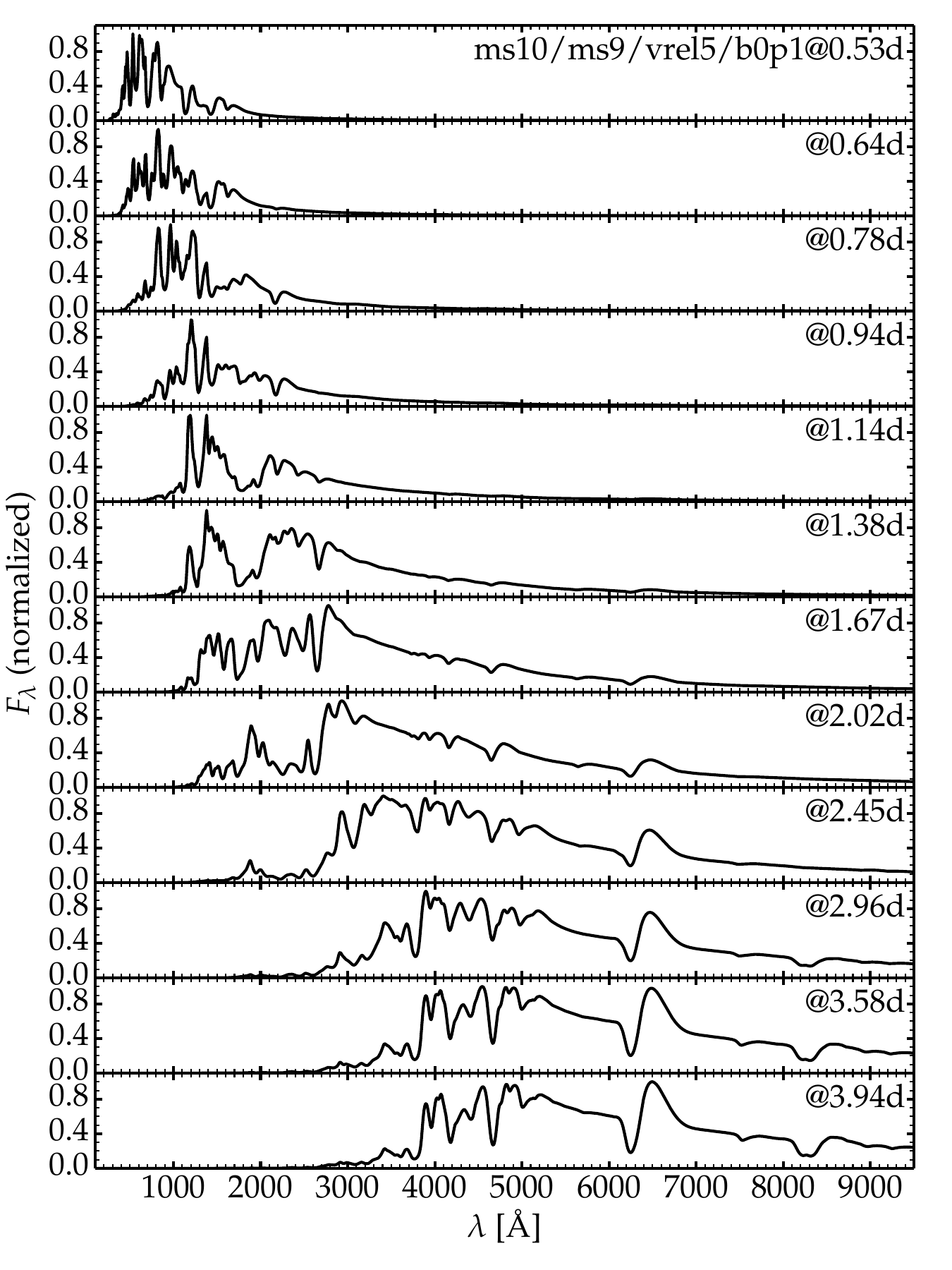}
    \end{subfigure}
     \hfill
     \begin{subfigure}[b]{0.33\textwidth}
       \centering
        \includegraphics[width=\textwidth]{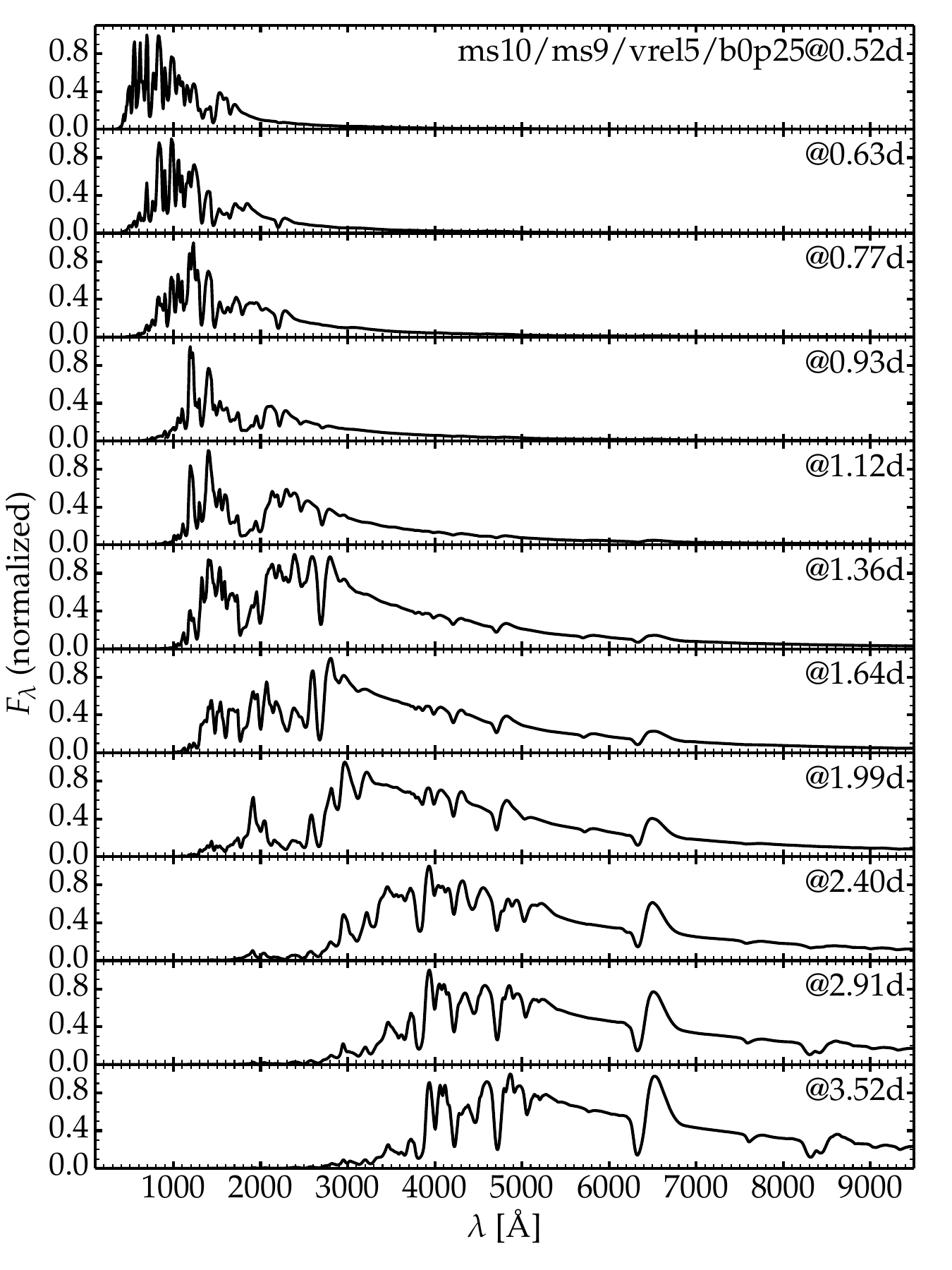}
    \end{subfigure}
     \hfill
     \begin{subfigure}[b]{0.33\textwidth}
       \centering
        \includegraphics[width=\textwidth]{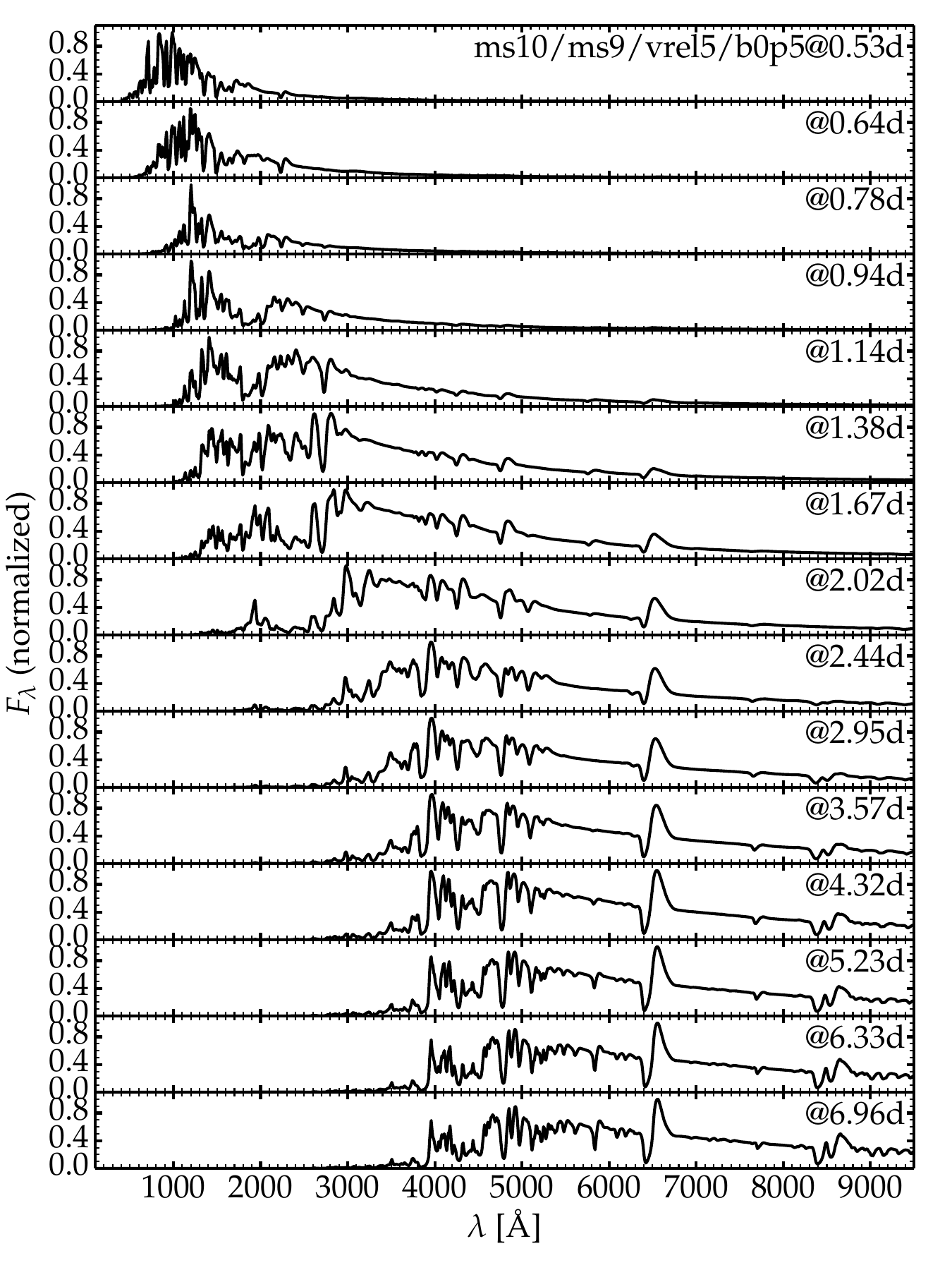}
    \end{subfigure}
\caption{Spectral montage for the modelset \# 3.
\label{fig_montage_set_3}
}
\end{figure*}

\begin{figure*}
   \centering
    \begin{subfigure}[b]{0.33\textwidth}
       \centering
       \includegraphics[width=\textwidth]{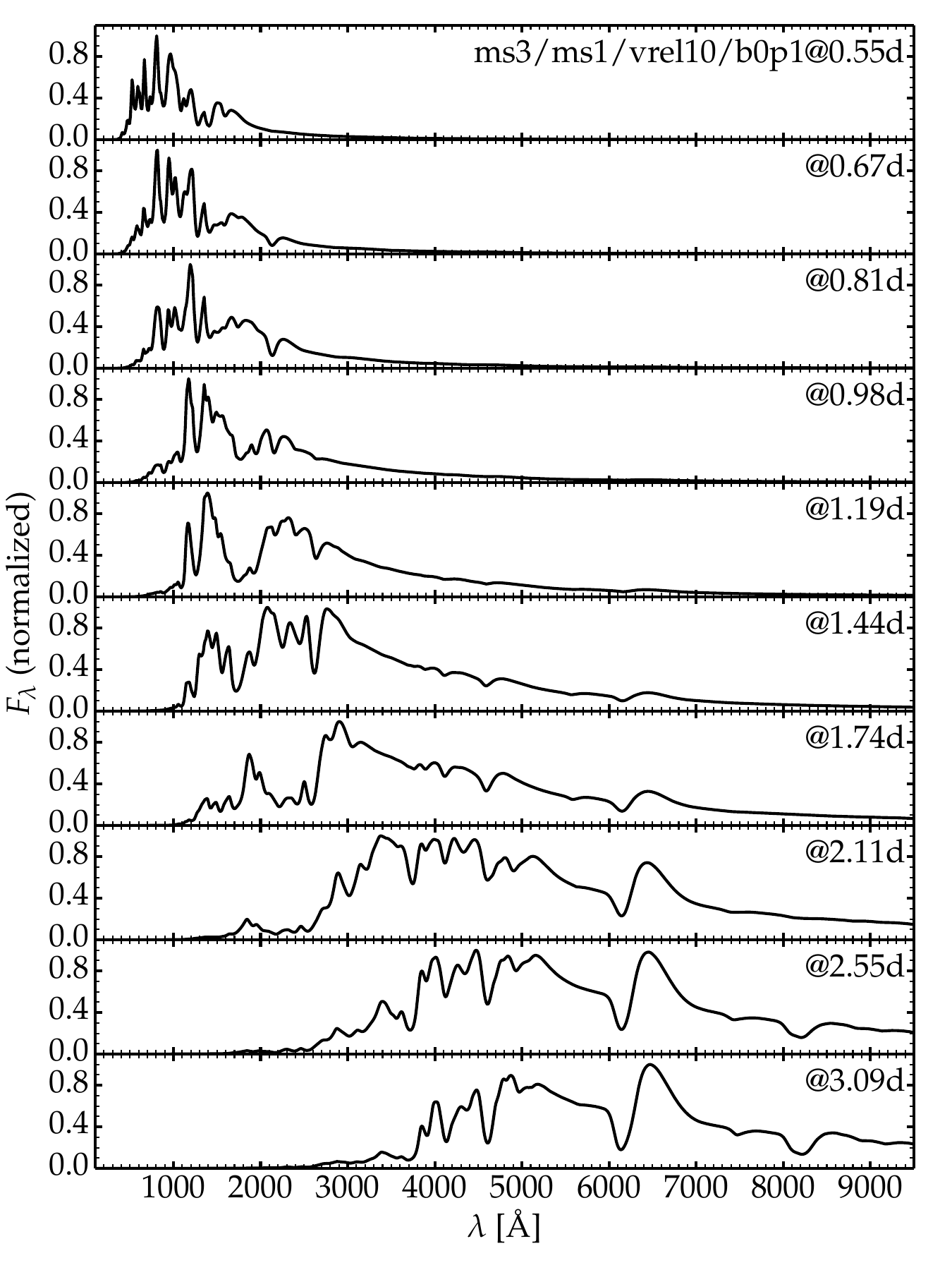}
    \end{subfigure}
    \hfill
    \begin{subfigure}[b]{0.33\textwidth}
       \centering
       \includegraphics[width=\textwidth]{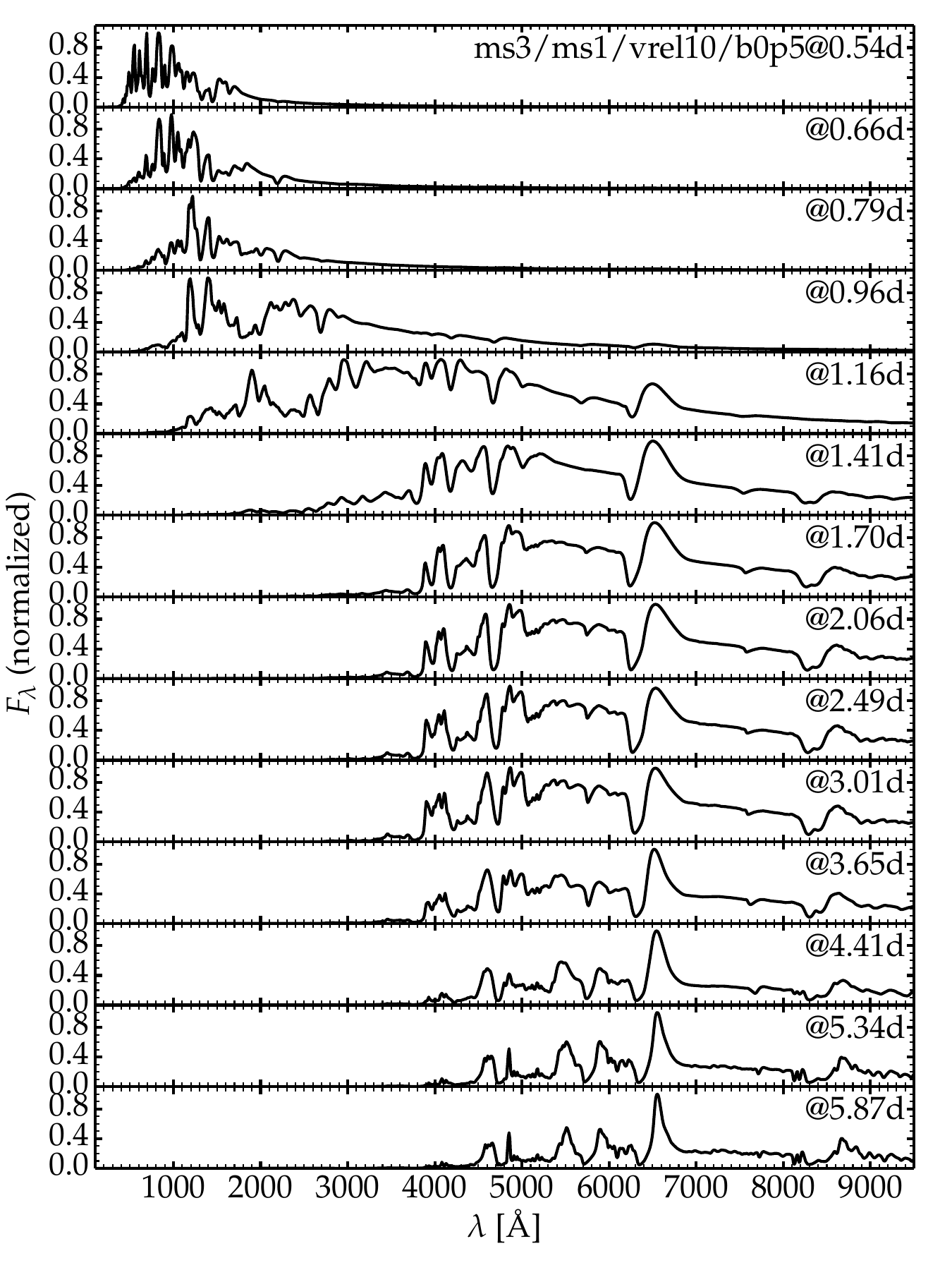}
    \end{subfigure}
     \hfill
    \begin{subfigure}[b]{0.33\textwidth}
       \centering
       \includegraphics[width=\textwidth]{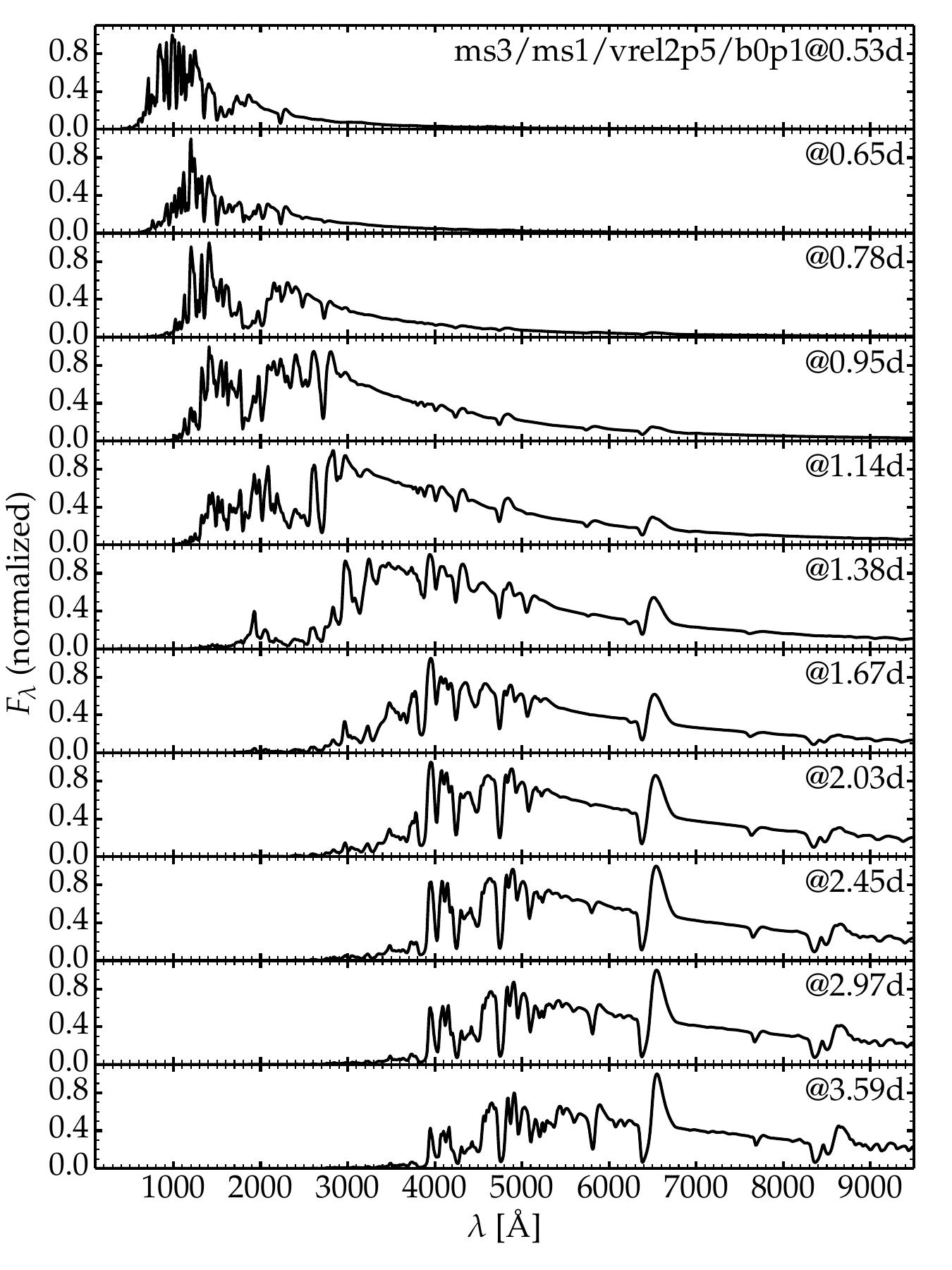}
    \end{subfigure}
     \hfill
    \begin{subfigure}[b]{0.33\textwidth}
       \centering
       \includegraphics[width=\textwidth]{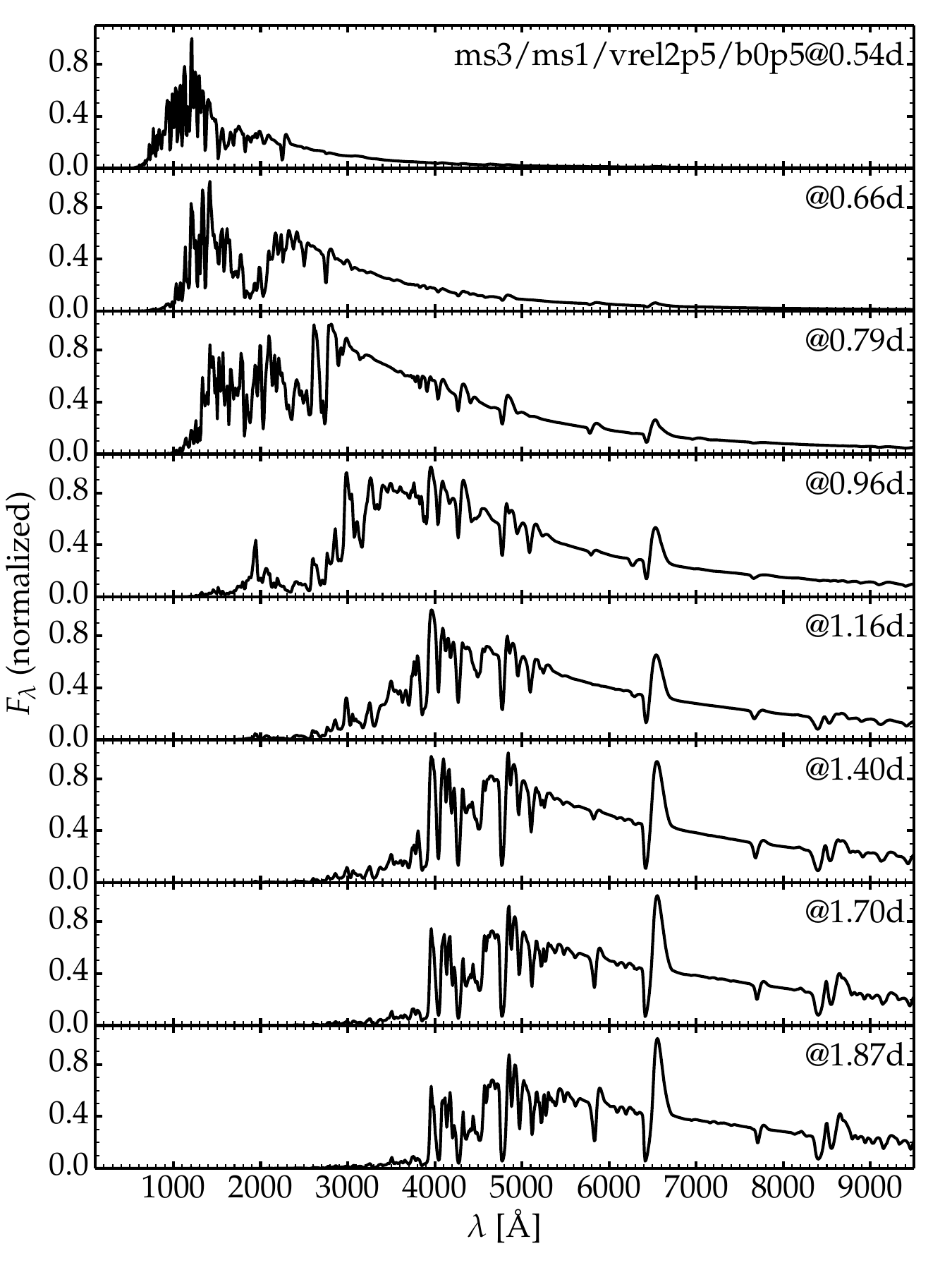}
    \end{subfigure}
     \hfill
     \begin{subfigure}[b]{0.33\textwidth}
       \centering
        \includegraphics[width=\textwidth]{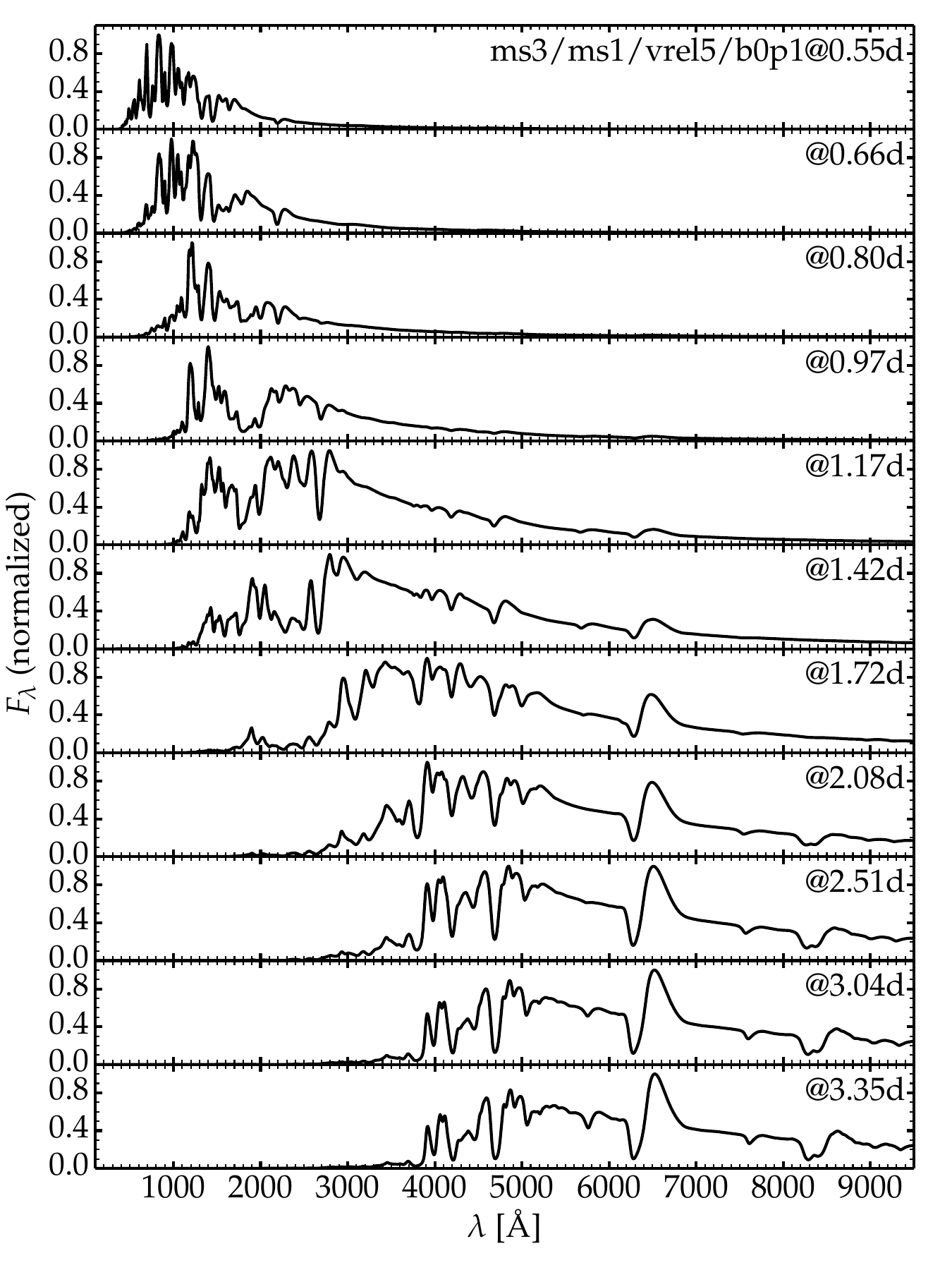}
    \end{subfigure}
     \hfill
     \begin{subfigure}[b]{0.33\textwidth}
       \centering
        \includegraphics[width=\textwidth]{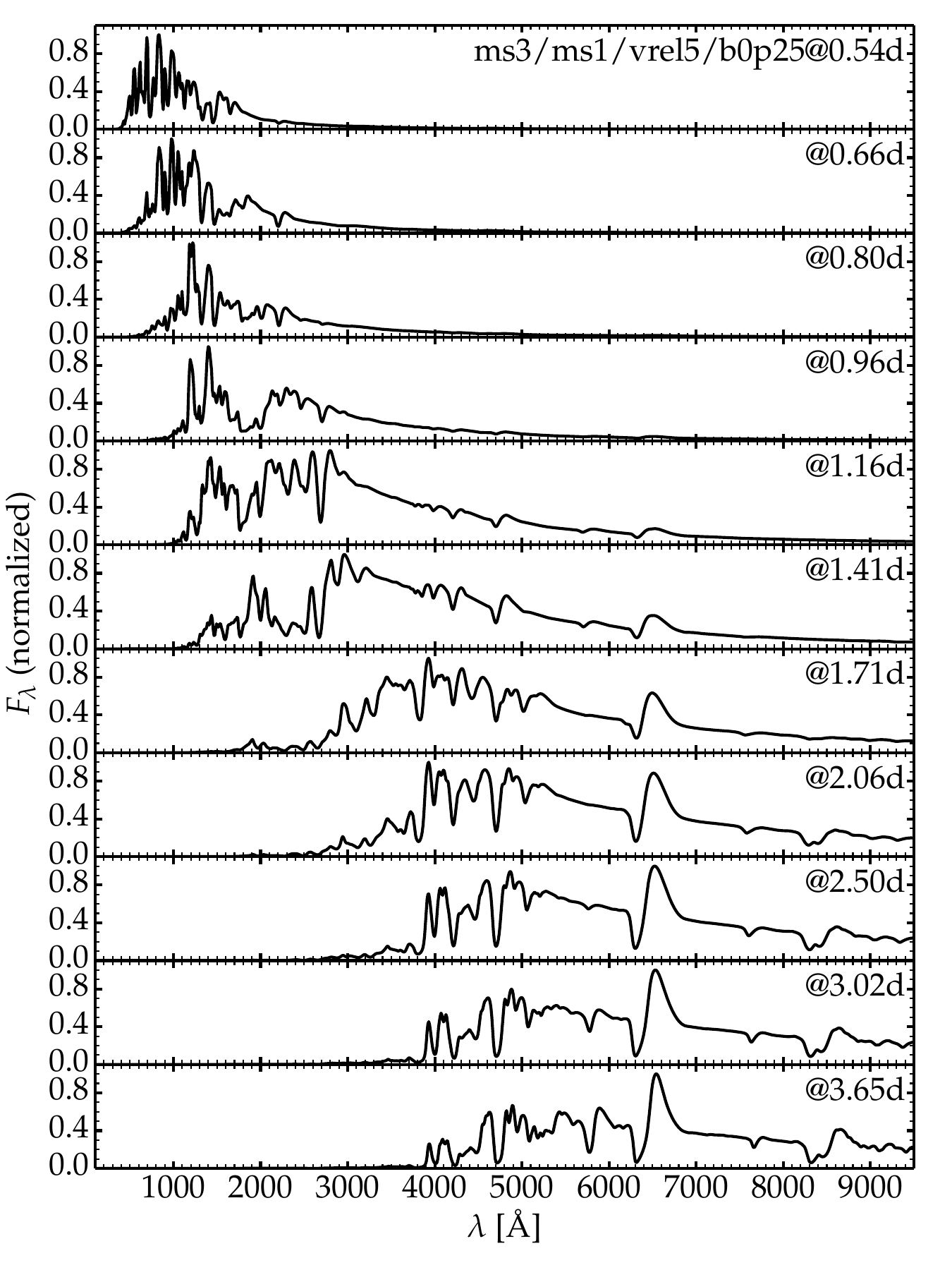}
    \end{subfigure}
\caption{Spectral montage for the modelset \# 4.
\label{fig_montage_set_4}
}
\end{figure*}

\begin{figure*}
   \centering
    \begin{subfigure}[b]{0.33\textwidth}
       \centering
       \includegraphics[width=\textwidth]{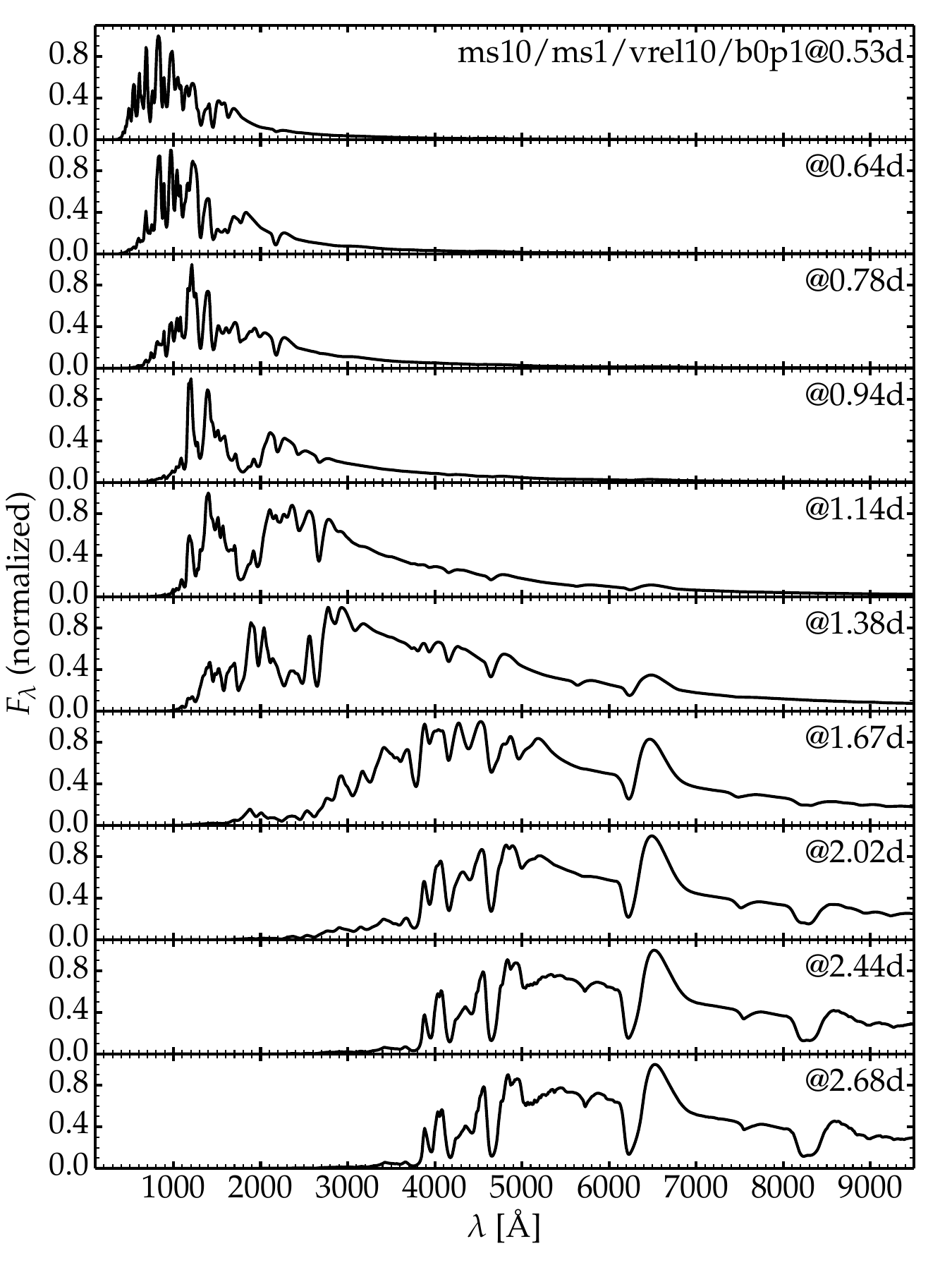}
    \end{subfigure}
    \hfill
    \begin{subfigure}[b]{0.33\textwidth}
       \centering
       \includegraphics[width=\textwidth]{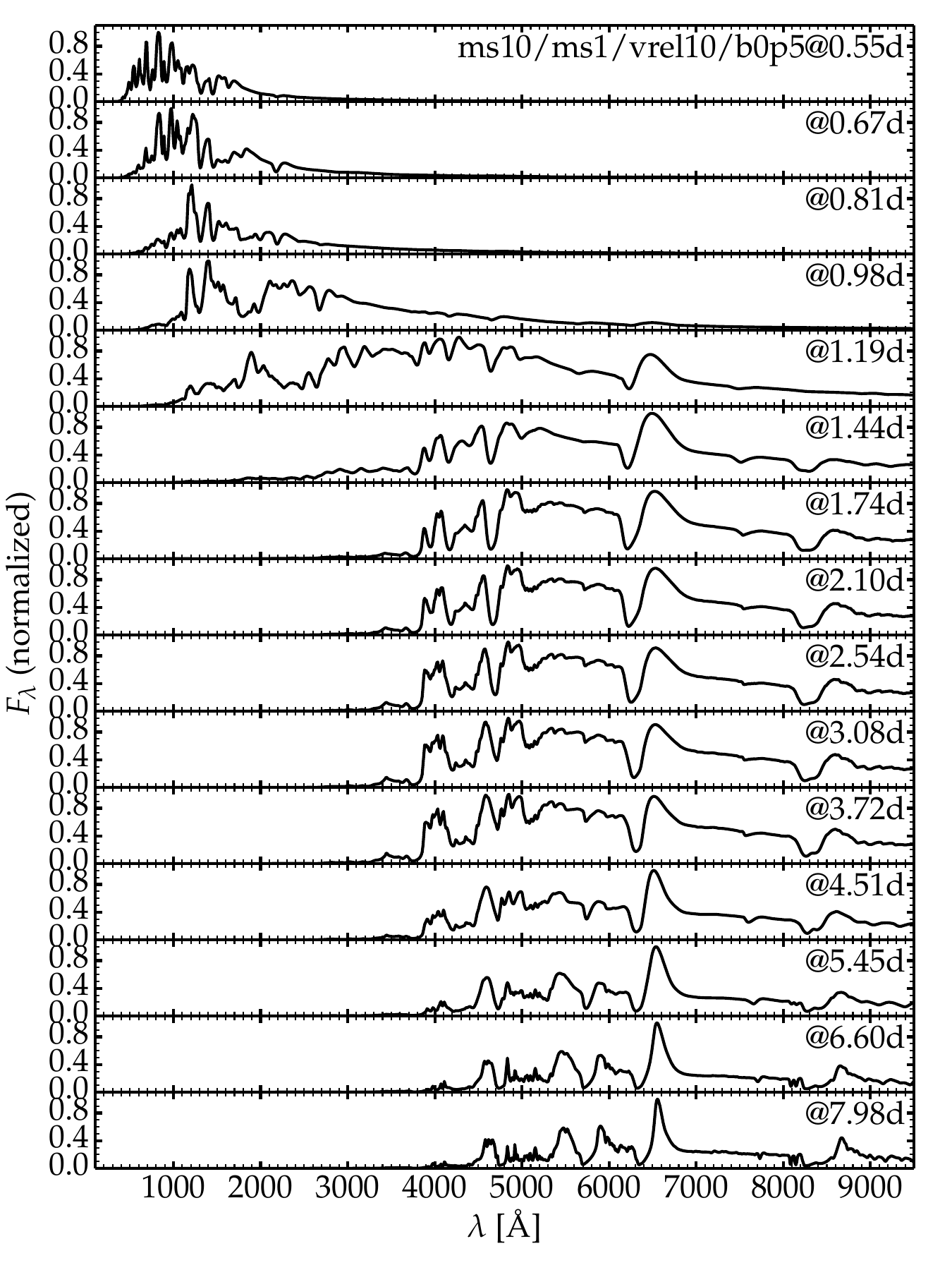}
    \end{subfigure}
     \hfill
    \begin{subfigure}[b]{0.33\textwidth}
       \centering
       \includegraphics[width=\textwidth]{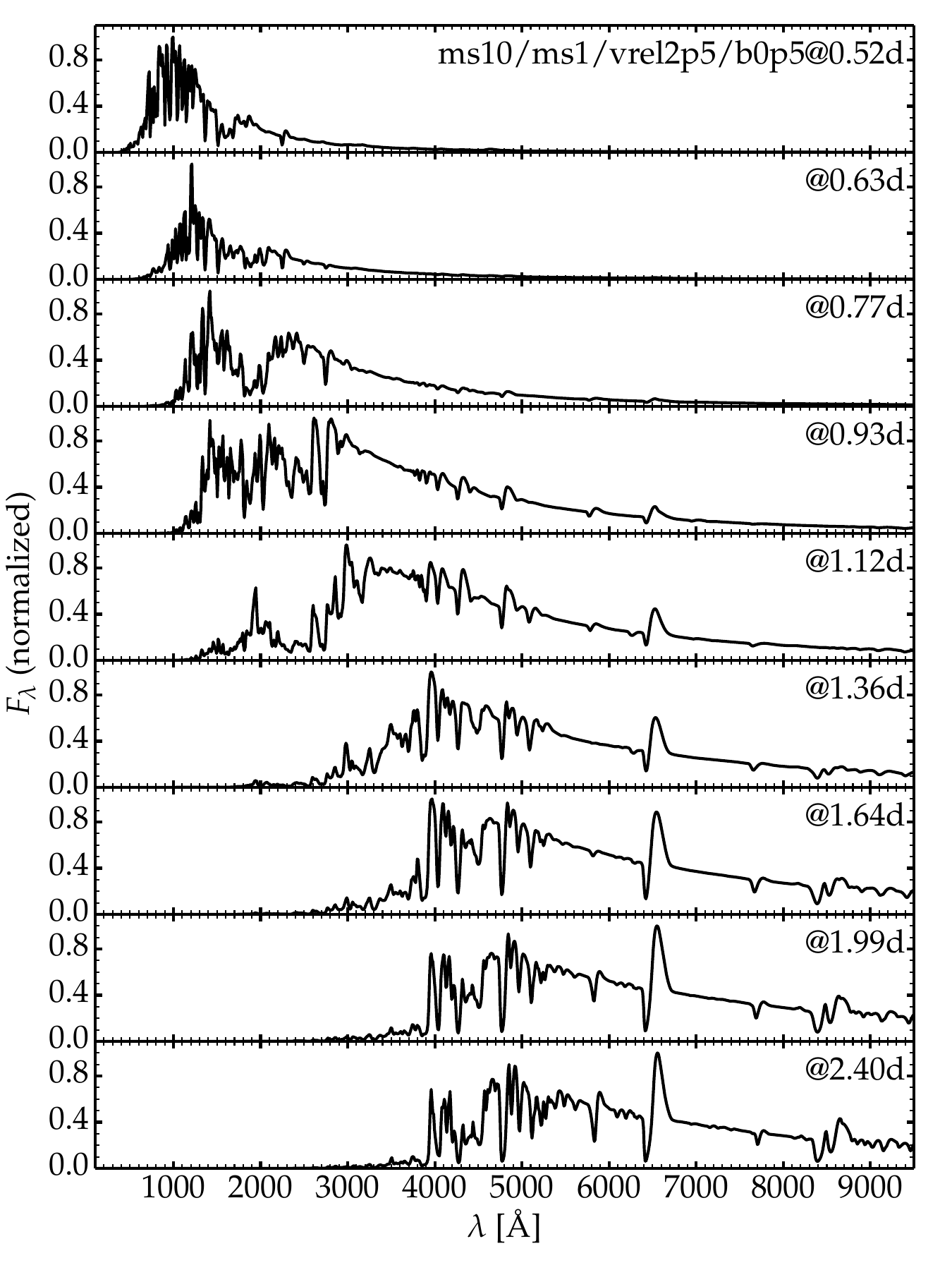}
    \end{subfigure}
     \hfill
    \begin{subfigure}[b]{0.33\textwidth}
       \centering
       \includegraphics[width=\textwidth]{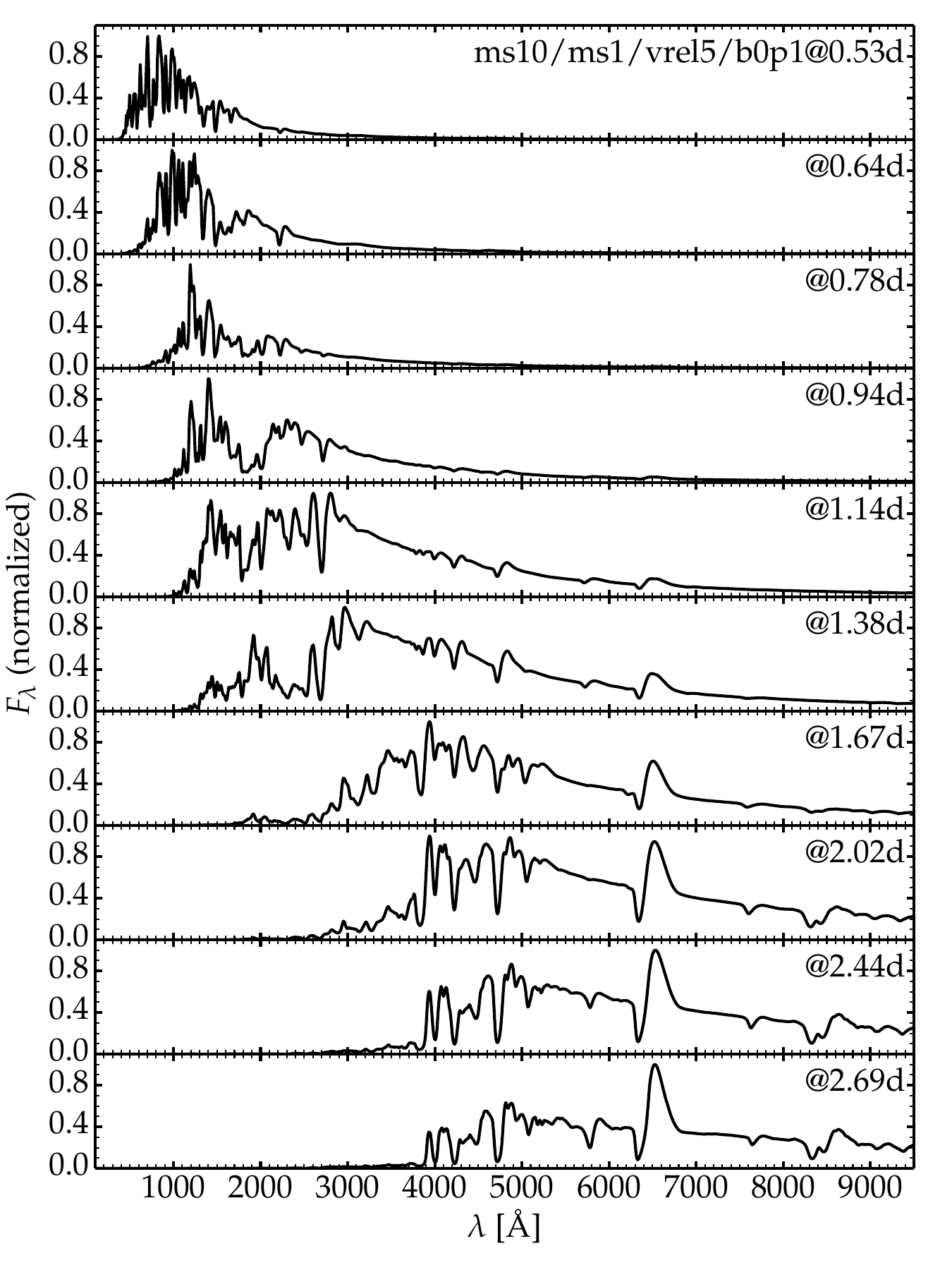}
    \end{subfigure}
     \hfill
     \begin{subfigure}[b]{0.33\textwidth}
       \centering
        \includegraphics[width=\textwidth]{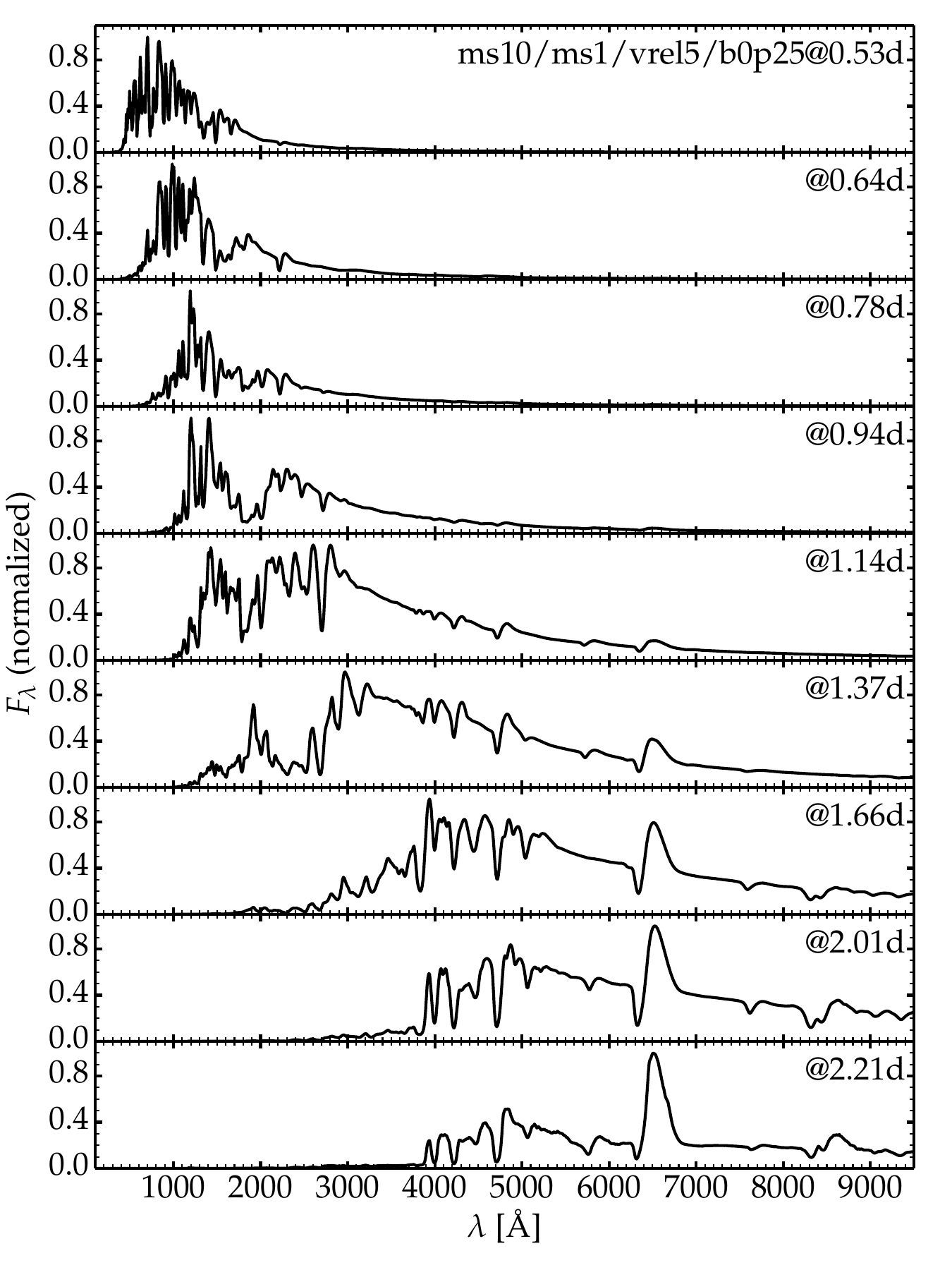}
    \end{subfigure}
     \hfill
     \begin{subfigure}[b]{0.33\textwidth}
       \centering
        \includegraphics[width=\textwidth]{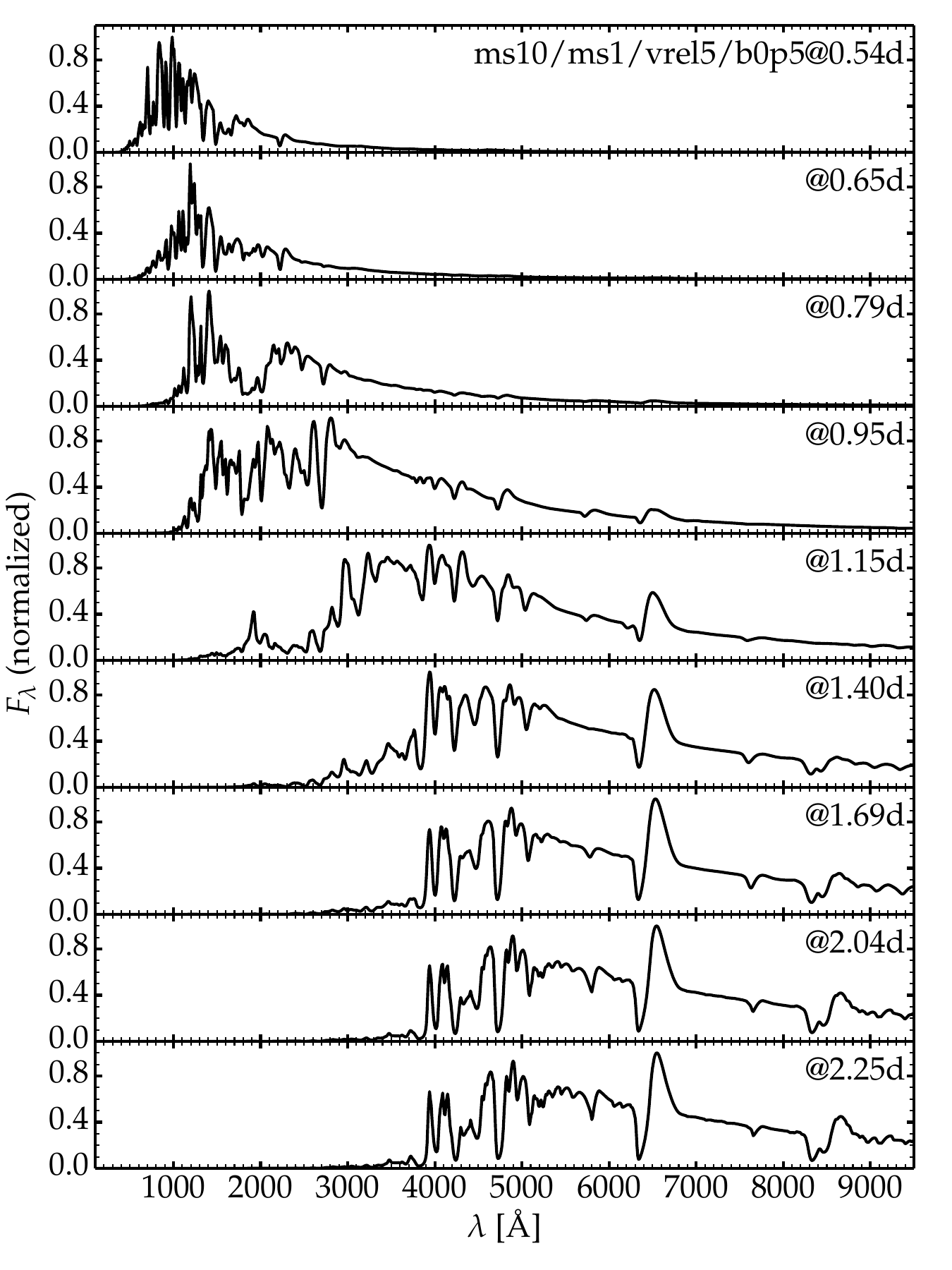}
    \end{subfigure}
\caption{Spectral montage for the modelset \# 5.
\label{fig_montage_set_5}
}
\end{figure*}

\begin{figure*}
   \centering
    \begin{subfigure}[b]{0.33\textwidth}
       \centering
       \includegraphics[width=\textwidth]{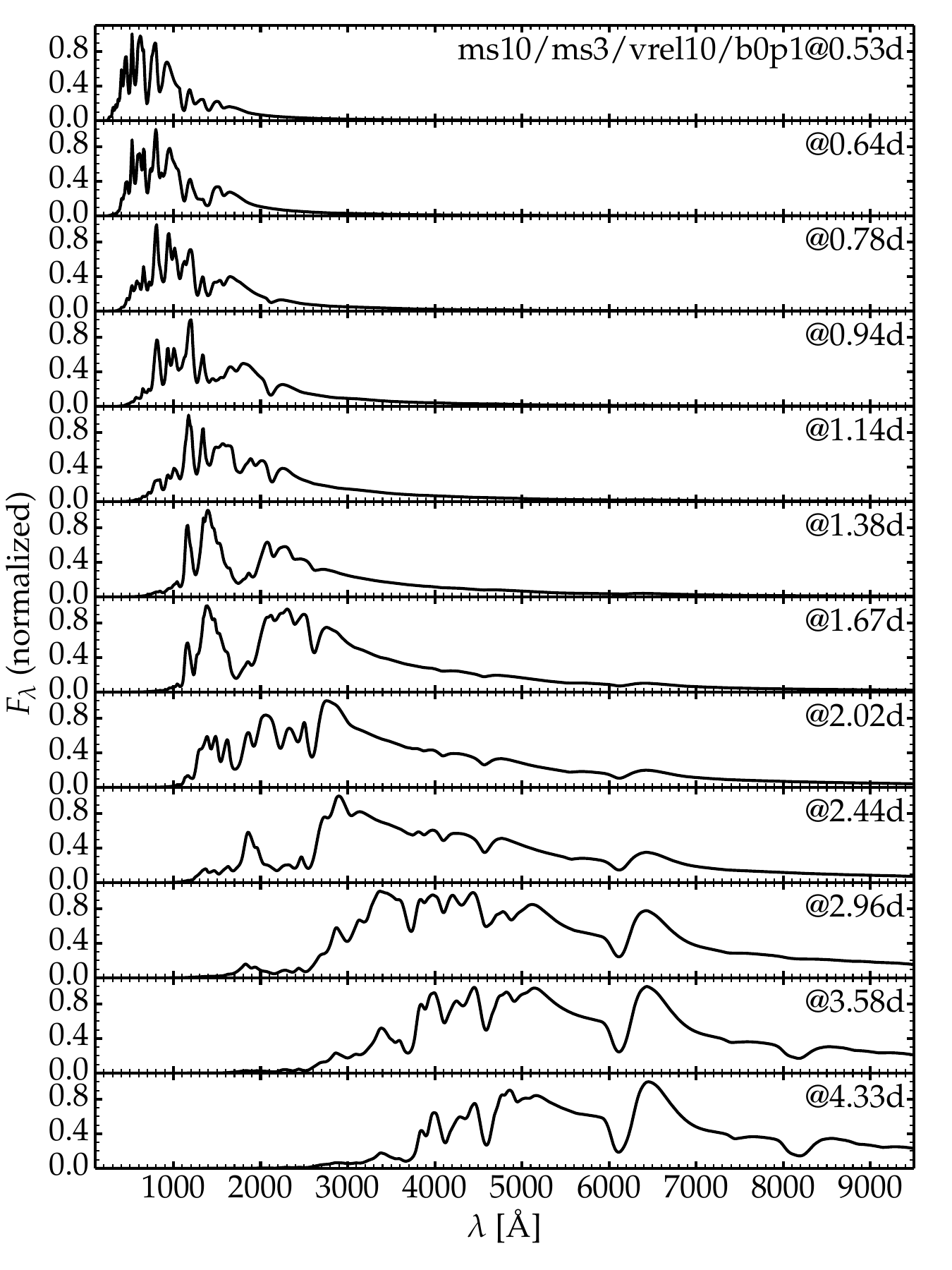}
    \end{subfigure}
    \hfill
    \begin{subfigure}[b]{0.33\textwidth}
       \centering
       \includegraphics[width=\textwidth]{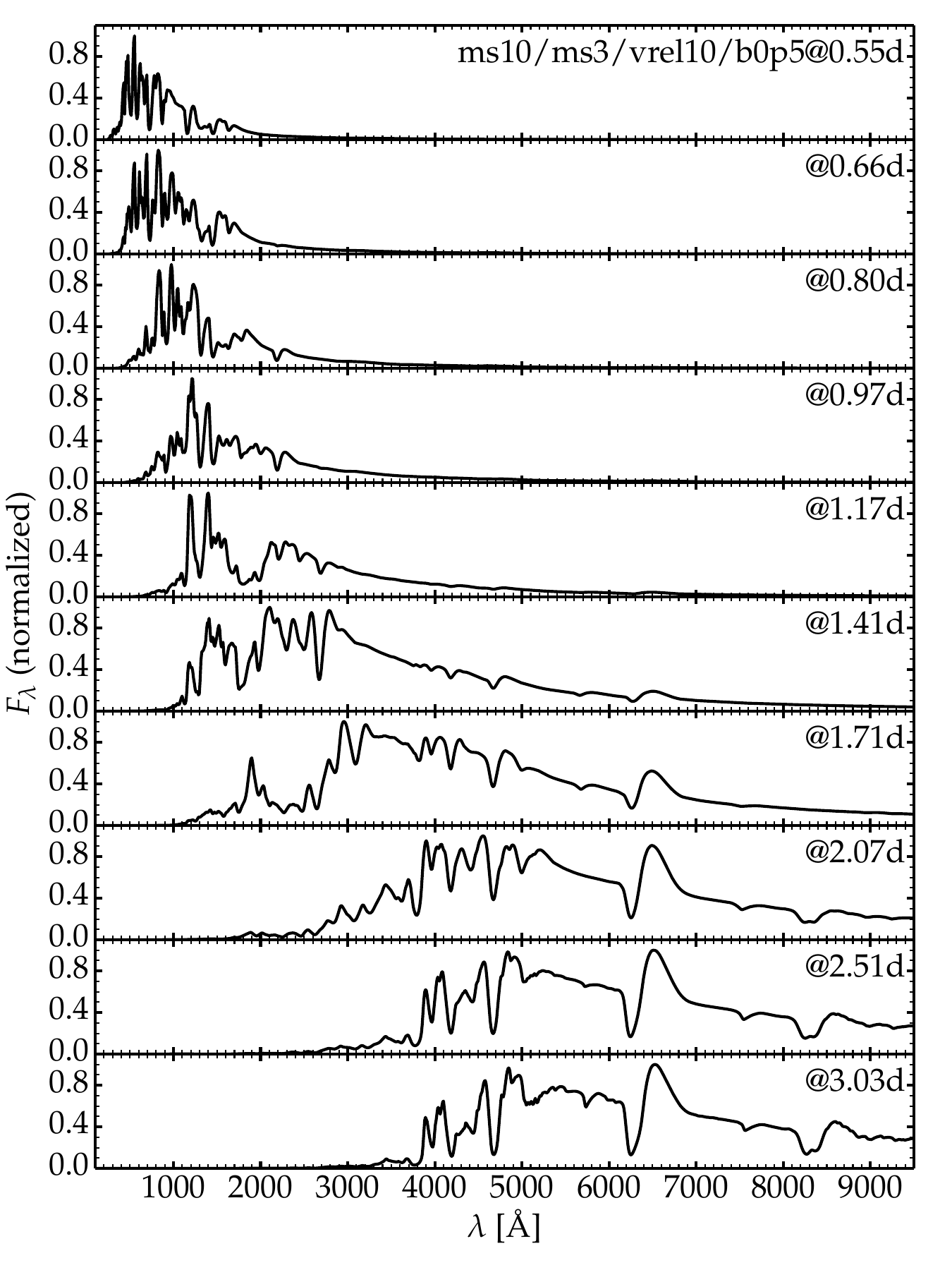}
    \end{subfigure}
     \hfill
    \begin{subfigure}[b]{0.33\textwidth}
       \centering
       \includegraphics[width=\textwidth]{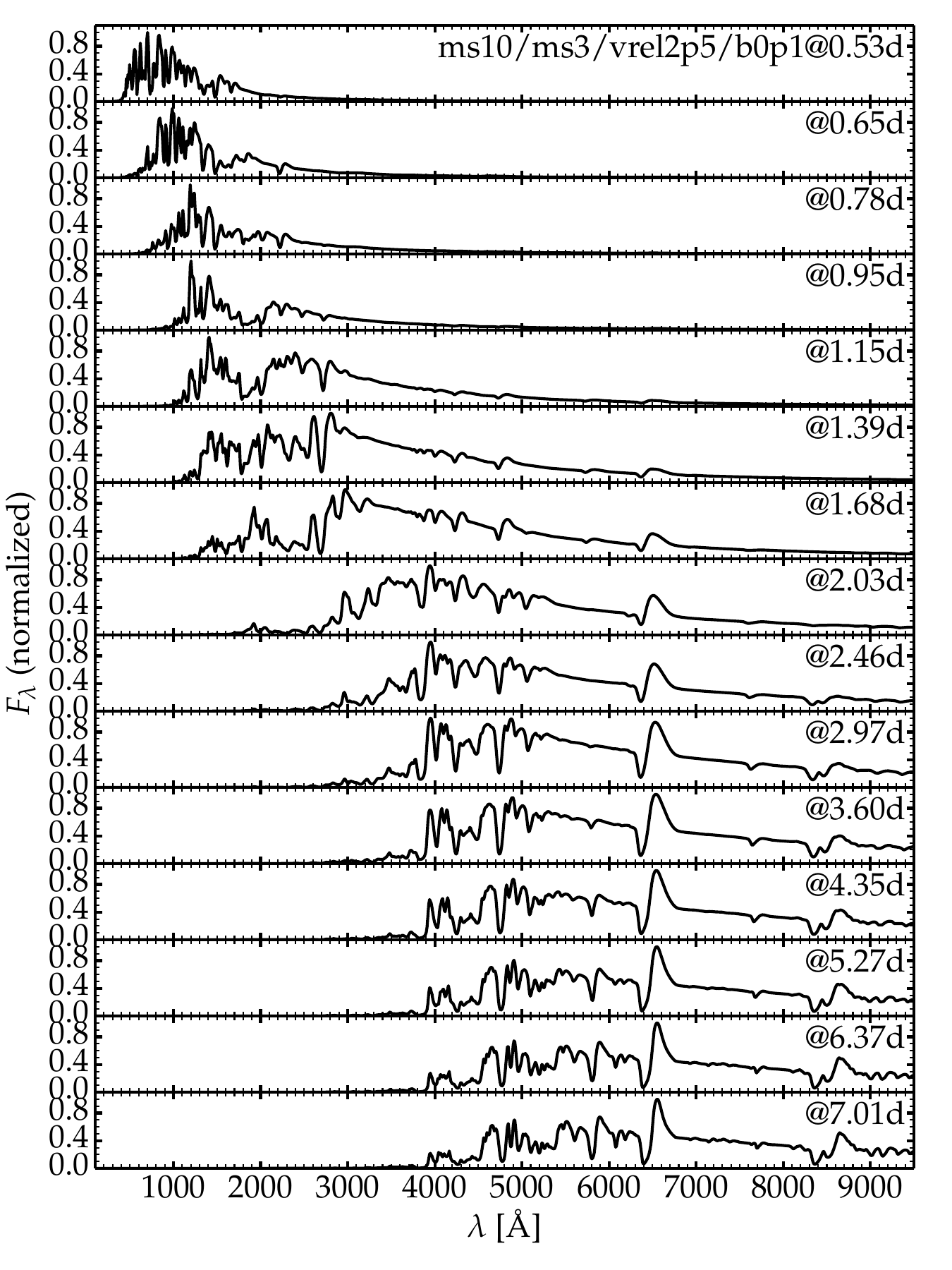}
    \end{subfigure}
     \hfill
    \begin{subfigure}[b]{0.33\textwidth}
       \centering
       \includegraphics[width=\textwidth]{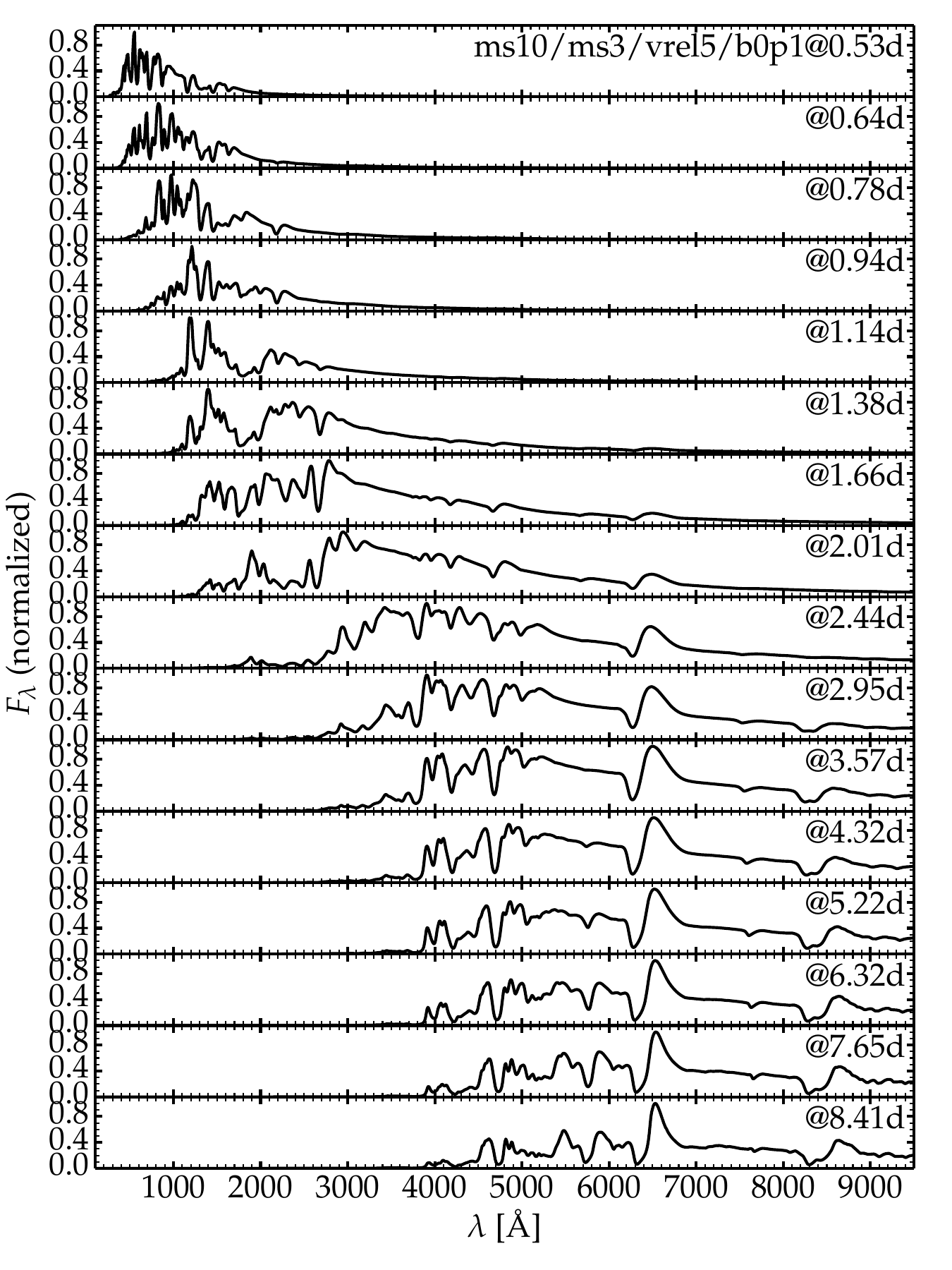}
    \end{subfigure}
     \hfill
     \begin{subfigure}[b]{0.33\textwidth}
       \centering
        \includegraphics[width=\textwidth]{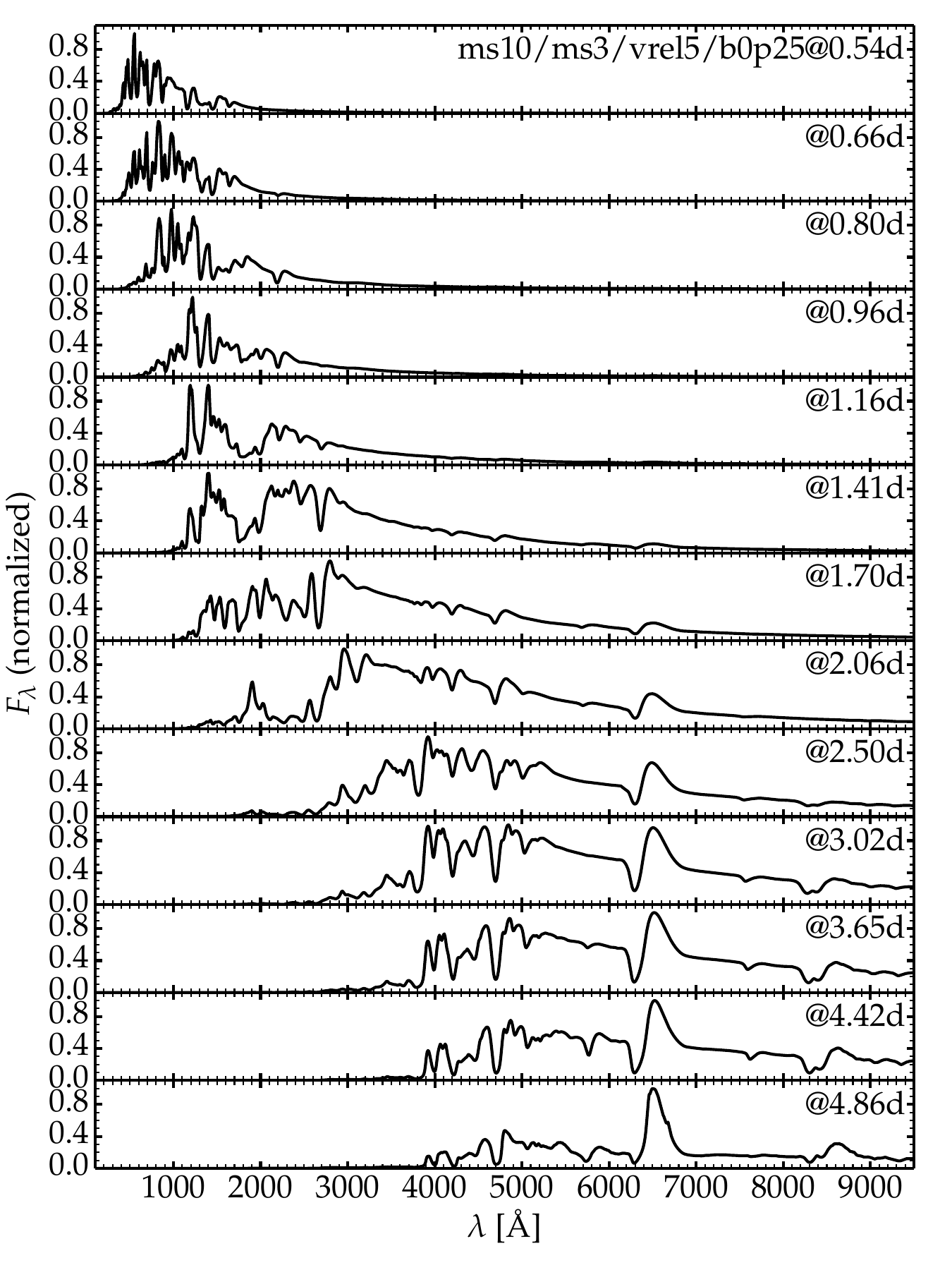}
    \end{subfigure}
\caption{Spectral montage for the modelset \# 6.
\label{fig_montage_set_6}
}
\end{figure*}

\end{appendix}
\end{document}